\documentclass[11pt]{article}
\pdfoutput=1
\usepackage{jheppub}
\usepackage{bbm}
\usepackage{booktabs}
\usepackage{tikz}
\usetikzlibrary{calc}
\usepackage{graphicx,subcaption}
\usepackage{mathtools}
\usepackage{bm}
\usepackage{enumitem}
\usepackage{verbatim}
\usepackage[makeroom]{cancel}

\renewcommand{\Re}{\operatorname{Re}}
\renewcommand{\Im}{\operatorname{Im}}
\DeclareMathOperator{\tr}{tr}
\DeclareMathOperator{\Tr}{Tr}
\DeclareMathOperator{\Ric}{Ric}

\DeclareMathOperator{\sgn}{sgn}
\DeclareMathAlphabet{\mathbfsf}{OT1}{cmss}{bx}{n}

\newcommand{\CA}{\mathcal{A}}
\newcommand{\CB}{\mathcal{B}}
\newcommand{\CC}{\mathcal{C}}

\newcommand{\CF}{\mathcal{F}}

\newcommand{\CG}{\mathcal{G}}

\newcommand{\CL}{\mathcal{L}}

\newcommand{\CN}{\mathcal{N}}
\newcommand{\CO}{\mathcal{O}}
\newcommand{\CX}{\mathcal{X}}
\newcommand{\CR}{\mathcal{R}}

\newcommand{\CH}{\mathcal{H}}
\newcommand{\CV}{\mathcal{V}}
\newcommand{\CI}{\mathcal I}
\newcommand{\CQ}{\mathcal Q}
\newcommand{\CW}{\mathcal{W}}

\newcommand{\IR}{\mathbb{R}}
\newcommand{\IC}{\mathbb{C}}
\newcommand{\IZ}{\mathbb{Z}}
\newcommand{\IH}{\mathbb{H}}

\newcommand{\Z}{\mathbb{Z}}
\newcommand{\C}{\mathbb{C}}
\newcommand{\R}{\mathbb{R}}
\newcommand{\Q}{\mathbb{Q}}

\newcommand{\mb}[1]{\mathbf{#1}}
\newcommand{\mc}[1]{\mathcal{#1}}

\newcommand{\mf}[1]{\mathfrak{#1}}

\renewcommand{\=}{\;= \;}

\newcommand\be{\begin{equation}}
\newcommand\ee{\end{equation}}
\newcommand\beq{\begin{equation}}
\newcommand\eeq{\end{equation}}
\newcommand\bea{\begin{eqnarray}}
\newcommand\eea{\end{eqnarray}}

\renewcommand{\b}{\beta}

\renewcommand{\t}{\tau}

\newcommand{\wt}{\widetilde}

\newcommand{\ve}{\varepsilon}

\newcommand{\ndt}{\noindent}

\renewcommand{\i}{{\rm i}}

\newcommand{\z}{\zeta}

\newcommand{\defeq}{\; \coloneqq \;} 

\newcommand{\uu}{\underline{u}}

\newcommand{\us}{\underline{s}}

\newcommand{\ustl}{\underline{\widetilde{s}}}

\newcommand{\rme}{{\rm e}}

\newcommand{\rd}{{\rm d}}
\newcommand{\dd}{{\rm d}}

\newcommand{\zbar}{\overline{z}}
\newcommand{\vol}{{\rm vol}}
\newcommand{\ph}[1]{\phantom{#1}}
\newcommand{\identity}{\mathbbm{1}}
\renewcommand{\j}{\varphi}
\newcommand{\ii}{{\rm i}}

\renewcommand{\tilde}{\widetilde}
\renewcommand{\hat}{\widehat}

\newcommand{\bulkF}{\CF}
\newcommand{\bulkg}{\CG}
\newcommand{\bulkA}{\CA}
\newcommand{\bulkR}{\CR}
\newcommand{\bdryA}{A}
\newcommand{\bdryg}{g}
\newcommand{\bdryF}{F}

\newcommand{\nonrotatingt}{t}
\newcommand{\nonrotatingphi}{\phi}
\newcommand{\nonrotatingpsi}{\psi}
\newcommand{\nonrotatingtau}{t_\text{E}}
\newcommand{\bdryphi}{\hat{\phi}}
\newcommand{\bdryt}{t}
\newcommand{\bdrytau}{t_\text{E}}
\newcommand{\bdrytheta}{\theta}
\newcommand{\bdrypsi}{\hat{\psi}}

\newcommand{\QFTtau}{\tau}

\newcommand{\Introtau}{T}
\newcommand{\Introq}{Q}
\newcommand{\Introc}{C}
\newcommand{\Introd}{D}

\newcommand{\SumInt}{\mathclap{\displaystyle\int}\mathclap{\textstyle\sum}}
\newcommand{\mm}{\mathfrak{m}}

\begin{document}

\title{Supersymmetric phases of~AdS$_4$/CFT$_3$}

\author[a]{Pietro Benetti Genolini,}
\emailAdd{pietro.benetti\_genolini}
\author[a,b]{Alejandro Cabo-Bizet,}
\emailAdd{alejandro.cabo\_bizet} 
\author[a]{Sameer Murthy}
\emailAdd{sameer.murthy  @kcl.ac.uk}
\affiliation[a]{Department of Mathematics, King's College London,\\
The Strand, London WC2R 2LS, U.K.}
\affiliation[b]{Università del Salento, Dipartimento di Matematica e Fisica \textit{Ennio De Giorgi},
and I.N.F.N. - sezione di Lecce, Via Arnesano, I-73100 Lecce, Italy}

\abstract{ 
We exhibit an infinite family of supersymmetric phases in the three-dimensional ABJM superconformal field theory and the dual asymptotically~AdS$_4$ gravity.
They are interpreted as partially deconfined phases which generalize the confined/pure AdS phase and deconfined/supersymmetric black hole phase.
Our analysis involves finding a family of saddle-points of the superconformal index labelled by rational points (equivalently, roots of unity), separately in the bulk and boundary theories.
In the ABJM theory we calculate the free energy of each saddle by the large-$N$ asymptotic expansion of the superconformal index to all orders in perturbation theory near the saddle-point. We find that this expansion terminates at finite order. 
In the gravitational theory we show that there is a corresponding family of solutions, constructed by orbifolding the eleven-dimensional uplift of the supersymmetric black hole.
The on-shell gravitational action of each orbifold agrees with the free energy of the corresponding saddle in the SCFT.  
We find that there are two saddles in the ABJM theory with the same entropy as the supersymmetric black hole, corresponding to the two primitive fourth-roots of unity, which causes macroscopic oscillations in the microcanonical index.
}

\maketitle \flushbottom

\section{Introduction and summary of results}
\label{sec:Intro}

A fruitful way to learn about the collective behavior of a quantum statistical system is to study the thermodynamic behavior of the theory as a function of 
external macroscopic sources (temperature, angular velocities, chemical potentials) 
which couple to conserved charges (energy, spin, global charges). 
Of particular interest are phase transitions that occur upon changing these external parameters.
The profound insight provided by AdS/CFT is that 
phase transitions  
in the bulk and boundary theories---which, a priori, have completely different mechanisms---map
to one another. This insight has led to dramatic discoveries such as the relation between the deconfinement 
transition in large-$N$ gauge theory and the Hawking--Page transition~\cite{Hawking:1982dh}  
in asymptotically AdS gravity~\cite{Witten:1998zw}. 

Recent progress on the superconformal index, initiated in~\cite{Cabo-Bizet:2018ehj,Choi:2018hmj,Benini:2018ywd}, has allowed us to revisit this line of thought  
in the supersymmetric setting, which provides better quantitative control. 
Recall that the superconformal index is defined as the Witten index of a certain (complex) supercharge 
of an SCFT, refined by chemical potentials which commute with the given supercharge. 
By the usual argument of pairing of bosonic and fermionic states, the index is invariant under small changes of the coupling \cite{Witten:1982df}. Therefore, assuming no wall-crossing, the values at weak
and strong coupling are exactly equal, and we can compare the weakly-coupled field theory index 
with the strongly-coupled gravitational answer. 
These investigations have led to 
the discovery of a rich phase structure in the supersymmetric AdS$_5$/CFT$_4$ context---from the point of 
view of gauge theory as well 
as gravity. 

On the gauge theory side, one finds complex saddle points of the large-$N$ index when a background chemical potential~$\Introtau \in \IH$ coupling to a certain combination of charges approaches rational points~\cite{Cabo-Bizet:2019eaf, ArabiArdehali:2019orz, Cabo-Bizet:2020nkr, Cabo-Bizet:2020ewf, ArabiArdehali:2021nsx, Jejjala:2021hlt,Cabo-Bizet:2021plf, Cabo-Bizet:2021jar, Jejjala:2022lrm}.  
On the gravitational side, one has a set of solutions with horizons in asymptotically AdS$_5$ space,
whose free energy agrees with that of the corresponding microscopic saddles~\cite{Aharony:2021zkr}.
Together, these results are interpreted as supersymmetric phases of the AdS/CFT, 
generalizing the  
AdS black hole/deconfined phase and pure AdS/confined phase to partially confined phases~\cite{Cabo-Bizet:2019eaf, ArabiArdehali:2019orz, Cabo-Bizet:2021jar}. 
The perturbative series for the free energy near 
each saddle terminates after a finite number of terms. Therefore, as~$N \to \infty$, the phase boundaries become sharp and the various transitions generalize the 
Hawking--Page/deconfinement transition.

In this paper, we develop a similar picture of supersymmetric phases in the setting of AdS$_4$/CFT$_3$.
The main observable that we focus on is the superconformal index of a $3d$ SCFT on~$S^2$, with independent analyses of the bulk and boundary theories. 
In the boundary theory we study the ABJM theory~\cite{Aharony:2008ug} with $U(N)_1\times U(N)_{-1}$ 
gauge group and $\CN=8$ superconformal symmetry,
and in the bulk we study $4d$ gauged supergravity. 
We find an infinite set of saddles labelled by rational points in large-$N$ ABJM theory, 
and a corresponding infinite set of asymptotically AdS$_4$ gravitational orbifold solutions in the dual supergravity.
The expansion of the statistical free energy of the field theory around the saddles agrees precisely with the corresponding  gravitational free energy 
of the~AdS$_4$ orbifolds. 

\smallskip

In the rest of the introduction we summarize the main points of the paper, drawing parallels 
and contrasts with the AdS$_5$/CFT$_4$ situation wherever possible. 

\medskip

\ndt {\bf CFT analysis of the superconformal index}

The most general superconformal index in ABJM theory receives contributions from states that preserve 
two real supersymmetries of the theory, and can be expressed as a Witten index over the Hilbert space~$\CH_{\rm BPS}(N)$
refined by four chemical potentials. 
The simplest such $\frac1{16}$-BPS index has one chemical potential~$\Introtau$ coupling to a combination~$4J + 2r$ of angular momentum $J$
and a certain $R$-symmetry $r$ rotating the preserved supercharge.
Since both angular momentum and $R$-symmetry charges are quantized in this theory, 
the index is a single-valued function of $\Introq\coloneqq \rme^{2\pi\ii \Introtau}$, and we have the Fourier expansion 
\be 
\label{INdNrel}
\CI_N(\Introtau) \= {\rm Tr}_{\CH_{\rm BPS}(N)} \, (-1)^{2J} \, \Introq^{\, 4J + 2r}  \= \sum_{\ell} d_N(\ell) \, \Introq^{\,\ell} \, .
\ee
We expect that the indexed degeneracy of states~$d_N(\ell)$ has an exponential growth reflecting the 
existence of supersymmetric black hole solutions in the holographic dual theory.
Indeed, this expectation is borne 
out by numerical studies of four-dimensional $\CN=4$ 
SYM~\cite{Murthy:2020rbd, Agarwal:2020zwm}.

At an analytic level, the problem is best solved in the grand canonical ensemble.  
It is clear from inverting~\eqref{INdNrel} that an exponential growth of~$\CI_N(\Introtau)$
as~$\Introtau \to 0$ implies an exponential growth of~$d_N(\ell)$ as~$\ell \to \infty$.  
One therefore looks for saddle points of~$\CI_N(\Introtau)$ near~$\Introtau \to 0$ whose free energy is a negative power of~$\Introtau$. 
Looking more carefully at the inversion of \eqref{INdNrel}, one realizes that an 
exponential growth of $d_N(\ell)$ as $\ell\to \infty$ only implies that $\Im(\Introtau)\to 0$. Moreover, 
in order to have a coherent addition, the Fourier series should split into a finite number of congruence 
classes with the same phase, which implies that $\Re(\Introtau)\in \Q$~\cite{Cabo-Bizet:2019eaf}. 
We thus conclude that the limits of interest are the
\begin{equation}
\label{eq:GeneralizedCardyLimits:Def}
\text{generalized Cardy limits}: \qquad    \Introtau \to \frac{\Introd}{\Introc} \in \Q \,. 
\end{equation}

For four-dimensional $\CN=4$ SYM, 
the saddle points of $\CI_N(\Introtau)$ in the generalized limits have been investigated using various approaches, including 
asymptotic analysis \cite{Kim:2019yrz, Cabo-Bizet:2019osg, Honda:2019cio, ArabiArdehali:2019tdm, GonzalezLezcano:2020yeb, Copetti:2020dil, Amariti:2020jyx, ArabiArdehali:2021nsx, Ardehali:2021irq}, contour integrals and Bethe ansatz \cite{Closset:2017bse, Benini:2018mlo, Goldstein:2020yvj}, and relations to special elliptic functions coming from number 
theory~\cite{Cabo-Bizet:2019eaf, Cabo-Bizet:2020ewf, Cabo-Bizet:2020nkr,Jejjala:2021hlt,Cabo-Bizet:2021plf,Jejjala:2022lrm}.
The results of these analyses are the following. 
The leading contribution to the microcanonical growth of states comes from the saddles~$\Introtau \to \pm \frac13$, i.e.~when $\Introq$ approaches a primitive third root of unity. (The quantization~$\frac13$ comes from the quantization of~$R$-charge 
in $\CN=4$ SYM.) 
The magnitude of this contribution 
grows as~$\exp(S_{\rm BH})$, where $S_{\rm BH}$ is the entropy of the BPS black hole in five-dimensional minimal gauged 
supergravity~\cite{Gutowski:2004ez}.
The contributions of these two leading saddles are complex conjugates of each other, leading to macroscopic oscillations in the microcanonical growth of states. 
More generally, the saddles~$(\Introc,\Introd)$ with gcd$(\Introc,\Introd)=1$ contribute to an exponential growth of states if and only if~$\Introc$ is a multiple of~3. 
For such saddles, 
the free energy is 
$\mathcal{F}_{\Introc,\Introd} (\Introtau) \sim 
 \pm \ii\pi N^2/9 \Introc \, (\Introc\,\Introtau-\Introd)^2$ if $\Introd = \pm 1 \mod 3$.

In this paper, we consider the generalized Cardy 
limits of the microscopic index of ABJM theory. 
In the large-$N$ limit, we find that 
the leading contribution to the microcanonical growth of states comes from the 
saddles~$\Introtau\to \pm 1/4$, i.e.~when~$\Introq$ approaches a primitive fourth root of unity. 
(The quantization~$\frac14$ comes from the quantization of~$R$-charge 
in ABJM theory.) 
The magnitude of the growth of this contribution grows as the exponential of the entropy of the 
supersymmetric Kerr--Newman-AdS black hole in four-dimensional minimal gauged supergravity. 
The contributions of these two leading saddles are complex conjugates of each other, 
leading to macroscopic oscillations in the microcanonical growth of states. 
More generally, the saddles~$(\Introc,\Introd)$ with gcd$(\Introc,\Introd)=1$ contribute to an exponential growth of states if and only if~$\Introc$ is a multiple of~4 (the leading saddles being~$(\Introc,\Introd)=(4,\pm 1)$). 
For such saddles, 
the free energy is 
\be \label{logIgrowth}
\mathcal{F}_{\Introc,\Introd} (\Introtau) \= \pm \frac{4}{\Introc} \, \mathcal{F}_\text{BH}(\Introc \,\Introtau -\Introd) \,,
\qquad 
\mathcal{F}_\text{BH} (\varepsilon)  \, \sim \, \frac{\pi}{3\sqrt{2}} \frac{N^{\frac{3}{2}}}{\varepsilon} \,,
\ee
if $\Introd=\pm 1$ mod 4.

\medskip

In order to analyze the phase transitions, we 
summarize the above facts by 
(as~$N \to \infty$),  
\be \label{indphases}
\mc{I}_N (\Introtau) \; \sim \; \sum_{(\Introc,\Introd)} \exp\bigl(-\CF_{\Introc,\Introd} (\Introtau) \bigr) \,,
\ee
where the free energy of the saddle~$(\Introc,\Introd)$
is given by~\eqref{logIgrowth}.
We thus reach a picture of the $\Introtau$ plane divided into regions bounded by co-dimension-one walls. These regions are identified with the phases and the phase boundaries occur at the walls where the neighboring entropies become equal. 
In particular, since $\CF_{4,1}$ is precisely the BH free energy, it is clear that 
the transition between the BH phase and the pure AdS phase is precisely the Hawking--Page phase transition.

In fact, we consider an all-order perturbation expansion in the small parameter~$\ve = \Introc\Introtau-\Introd$ 
and show that the asymptotic expansion of the free energy terminates at~$\text{O}(\ve)$. 
This implies that as~$N\to \infty$, the phase boundaries become sharp, similar to the phenomenon observed in 4d SYM theory~\cite{ArabiArdehali:2021nsx}. It is worth noting, though, that  
the saddle point equations in 4d SYM theory are typically algebraic and can be solved easily at finite~$N$, 
while in~SCFT$_3$ they involve transcendental functions which can be solved 
in the~$N \to \infty$ limit using techniques introduced in \cite{Herzog:2010hf, Benini:2015eyy}.

\bigskip

\ndt {\bf Gravitational analysis of the superconformal index}

In the gravitational theory we do not have a good Hilbert space interpretation and, instead, we try to 
interpret the index as a sum over saddle points. 
The parameter~$N$ is interpreted according to the standard AdS/CFT dictionary as proportional to
the inverse gravitational coupling and, 
similarly, other chemical potentials in the most general index 
 are interpreted as  geometrical parameters in~AdS space.
The saddles are defined in the limit~$N \to \infty$ as solutions to the equations of motion of the 
semiclassical Euclidean gravitational theory. 
The natural idea is to interpret the microscopic sum over saddles~\eqref{indphases} as a sum over gravitational solutions of the Euclidean bulk theory like the supersymmetric black hole. 
We now summarize how this understanding is reached purely from bulk considerations.

Firstly, there is the question:~``(How) does a supersymmetric black hole contribute to the supersymmetric index?"  
The saddle points of the gravitational theory are solutions of the effective low-energy  
supergravity with boundary conditions fixed to be asymptotically AdS with conformal boundary~$S^2 \times S^1$.
A first puzzle is that the~$(-1)^{2J}$ in the trace~\eqref{INdNrel} naively implies that the fermions should have periodic boundary conditions around the~$S^1$, but this is in apparent tension with the fact that the~$S^1$ is 
contractible in the Euclidean black hole geometry, which requires the spinor to be anti-periodic around the circle. 
A second issue is that supersymmetric black holes are extremal, so they have an infinite throat at the horizon, leading to an infinite on-shell action. These tensions were resolved in~\cite{Cabo-Bizet:2018ehj} in the context of the supersymmetric AdS$_5$ 
black holes. 
The same idea has been applied to 
black holes asymptotically AdS$_4$~\cite{Cassani:2019mms, Bobev:2019zmz}  (which are relevant for the current paper), as well as  to five-dimensional supersymmetric black holes in 
asymptotically locally flat 
space~\cite{Iliesiu:2021are}. 

The idea at the heart of the resolution is to allow \textit{complex} gravitational solutions. 
The first point is that the supersymmetric periodicity condition can be absorbed by an imaginary shift of a chemical potential. 
In the concrete example of \eqref{INdNrel}, we have 
\be
\Tr_{\CH_{\rm BPS}(N)} \, (-1)^{2J} \rme^{2\pi\ii \Introtau \, (4 J + 2r)} \= \Tr_{\CH_{\rm BPS}(N)} \,  \rme^{2\pi\ii \left( 1 + 4\Introtau \right) \, J + 4\pi\ii \Introtau \, r} \, .
\ee
(One can equivalently shift the potential for the $R$-charge.)
The shifted chemical potential on the right-hand side  is naturally interpreted in the gravitational black hole 
background as \emph{complex} boundary values of \emph{bosonic} fields.
Interpreting the index trace as a functional integral, 
it is clear that only those configurations that extend the boundary Killing spinor to the bulk contribute 
to the index, as otherwise there would be extra fermion zero modes coming from broken supersymmetry 
which would kill the corresponding contribution of such solutions to the path integral.\footnote{This has been demonstrated by an explicit calcuation in the asymptotically flat space in~\cite{Iliesiu:2021are}.}

The second point is the definition of the contribution of the supersymmetric black hole to the functional integral via a regularization procedure. One starts with a one-parameter 
family of deformations of the Euclidean supersymmetric black hole with the above complex boundary conditions, 
consisting of geometries that are not extremal but are supersymmetric. 
Importantly, these configurations are inherently complex and they do not admit regular real Lorentzian continuations.
They have a real Euclidean section with the topology of the product of a disc and a sphere, and are labelled by a continuous 
parameter~$\beta>0$ corresponding to the boundary size of the Euclidean circle. 
In the limit~$\beta \to \infty$ we recover the Euclidean supersymmetric extremal black hole with an infinite throat, 
which is the Wick-rotation of the Lorentzian supersymmetric black hole solution. 
These complex supersymmetric solutions also support a non-trivial gauge field, which vanishes at the origin of the disc, thus preserving smoothness, and has a non-trivial holonomy around the boundary of the disc. 
Since the Killing spinor is charged under the gauge field, parallel transport around a circle in the disc in the chosen gauge gives an anti-periodic spinor, consistently with topology, whereas parallel transport with the fully covariant derivative does indeed lead to a periodic spinor, as consistent with the naive expectation from supersymmetry.

One  then calculates the holographically renormalized on-shell action of the complex supersymmetric solutions.  
Notably, this action, as a function of appropriately reduced chemical potentials, is (a) finite, (b) independent of~$\beta$, and (c) exactly equal to the logarithm of the functional form of the microscopic 
grand canonical index near the~$(4,\pm 1)$ saddle-point.
In order to obtain the entropy of the supersymmetric extremal black hole, one does a Legendre transform with respect to the reduced chemical potentials, and finds agreement with the Bekenstein--Hawking area law for the black hole entropy \cite{Hosseini:2017mds}. This extends the prescriptions of Euclidean quantum gravity \cite{Gibbons:1976ue} to supersymmetric black holes \cite{Cabo-Bizet:2018ehj}.

\medskip

The above procedure allows one to trace the relation between the Wick-rotated BPS black hole and the $(4,\pm 1)$ saddle of the field theory index. One then  expects that the $(\Introc, \Introd)$ saddle of the index with $\gcd(C,4)=4$ corresponds to supersymmetric solutions (regularized as above) with the same conformal boundary conditions as the black hole, and on-shell action equal to $\CF_{\rm BH}/(\Introc/4)$. In the context of the AdS/CFT correspondence applied to four-dimensional $\CN=4$ SYM, these solutions were described in \cite{Aharony:2021zkr}: they are supersymmetry-preserving quotients of the ten-dimensional uplift on $S^5$ to IIB of the Wick-rotated black hole solution. 
These quotients crucially involve the Euclidean time circle and thus their Lorentzian interpretation is not transparent. 

In this paper, we construct the analogous solutions dual to the $(\Introc, \Introd)$ saddles in the large-$N$ limit of the index of ABJM. We start with solutions to gauged supergravity with the same conformal boundary conditions as the supersymmetric electrically charged black hole, uplift them on $S^7$ to eleven-dimensional supergravity, and then perform a $\Z_{C/4}$ quotient while preserving supersymmetry. The solutions in this infinite family generalize the two Hawking--Page-like phases, namely the supersymmetric Kerr--Newman-AdS black hole and pure AdS$_4$, and have on-shell action equal to the microscopic free energy~\eqref{logIgrowth} at that rational point.

\medskip

\ndt {\bf Open questions and further directions}

The structure of the sum over saddles~\eqref{indphases}, which we derive independently from field theory and from gravity, is very interesting from
the point of view of the underlying symmetries. 
Such a sum over $(\Introc,\Introd)$ saddles has previously appeared in the context of supersymmetric AdS$_3$/CFT$_2$ \cite{Maldacena:1998bw, Dijkgraaf:2000fq}, and in the context of~AdS$_2$/CFT$_1$ \cite{Banerjee:2008ky, Murthy:2009dq, Dabholkar:2014ema, Iliesiu:2022kny}. 
In both cases, there is an underlying $SL(2,\IZ)$ modular symmetry which is closely tied to the existence of such an exact convergent
formula.\footnote{The sum over~$(\Introc,\Introd)$ here is really a sum over the equivalence classes $\Gamma_\infty \backslash \, \Gamma \,/\Gamma_\infty$ where~$\Gamma=SL(2,\mathbb{Z})$ and $\Gamma_\infty$ is its  subgroup stabilizing the point~$\tau=\ii \infty$.}
In contrast, the formula~\eqref{indphases} for SCFT$_3$ and the analogue in the SCFT$_4$ problem is not an exact formula: 
firstly, we do not know if this is an exhaustive set of saddles (although there are indications that this may be so 
at large~$N$); secondly, although we have control over the all-order perturbation theory around each saddle,  
there is no claim of convergence of the sum. 
Nevertheless, the observation that the index is arranged as a sum over such saddles with a number-theoretic constraint on~$(\Introc,\Introd)$ 
is remarkable since the superconformal index in~$4d$ and in~$3d$ are not  
$SL(2,\IZ)$ modular forms. 
Perhaps it hints to an approximate modular symmetry in these systems. 

The phase structure of the~$3d$ as well as the~$4d$ theory and, in particular, the dominant phase as one moves along the real axis in~$\Introtau$ has an erratic behavior. It would be very interesting if this fundamentally related to 
the appearance of randomness in gravitational systems~\cite{Saad:2018bqo}. 

The analysis of ABJM theory at level $k>1$ is an interesting problem: in the gravity dual, the $\Z_{\Introc/4}$ quotient described above would be woven with the $\Z_k$ quotient of the internal $S^7$. 
More generally, one could generalize the uplifts discussed here to uplifts on different seven-dimensional Sasaki--Einstein spaces, where the quotient would identify points along the circle orbit of the Reeb vector. These constructions would be dual to three-dimensional $\CN=2$ SCFTs obtained from arrangements of $M2$-branes. 

In another direction, it would be interesting to study the $\CN=2$ SCFTs obtained by wrapping $M5$-branes on hyperbolic three-manifolds. In this case, the $R$-symmetry charge of the states on $S^2$ is quantized since it is a compact subgroup of the $\mf{so}(5)$ $R$-symmetry group of the original six-dimensional theory, so the superconformal index has a similar structure as that described around \eqref{INdNrel}. The canonical Cardy limit describing the leading growth of states has been calculated, and the large-$N$ limit agrees with the gravity dual~\cite{Bobev:2019zmz, Benini:2019dyp}. The gravity duals to the generalized Cardy limits would be constructed by quotienting the eleven-dimensional geometry uplifting the supersymmetric Kerr--Newman-AdS black hole to eleven dimensions on a fibration of $S^4$ over the hyperbolic manifold \cite{Gauntlett:2006ux}.

Another issue is the overabundance of eleven-dimensional geometries with boundary conditions appropriate to describe the dual to the large-$N$ saddles of the generalized Cardy limits of ABJM. As pointed out in \cite{Aharony:2021zkr}, their presence would make the gravitational grand canonical partition function diverge, but one can argue for their absence from the sum studying the action of wrapped branes in the geometry. 
In this paper we include some comments on this subject, and we plan to report on this in more detail in the near future.

\medskip

\ndt {\bf Plan of the paper}

In Section \ref{sec:ABJMIndex} we review the construction of the superconformal index for ABJM, highlighting the interpretation as a functional integral over a complex background and the conditions imposed on the chemical potentials in order to preserve supersymmetry. 
We also comment on the subtleties arising from the fact that the index as usually defined is a multi-valued function, and identify two choices of refinements that will be matched in gravity. 
In Section \ref{sec:RationalPoints} we start from the expression of the index as a matrix model integral, and obtain the free energy of the saddle points in the large-$N$ limit, including subleading 
effects in the generalized Cardy limit.

We then move to the dual gravity side. In Section \ref{sec:BlackHole} we begin by reviewing aspects of the Euclidean Kerr--Newman-AdS black hole and its holographic renormalization, 
highlighting the importance of the gauge choice for the Abelian gauge field.
We then construct a family of complex supersymmetric solutions deforming the Wick-rotation of the supersymmetric black hole away from extremality and compute their on-shell action.
In Section \ref{sec:Uplift} we uplift these solutions of four-dimensional minimal gauged supergravity  on $S^7$ to eleven dimensions, describing the conformal boundary conditions and the matching with the supersymmetric background of the boundary field theory. 
This leads us to the construction of the quotient solutions, whose free energy matches the large-$N$ limit of the $(\Introc, \Introd)$ saddle of the unrefined index of ABJM. Finally, in Section~\ref{sec:U12} we consider black holes electrically charged under two gauge fields, 
which are dual to saddles of the index of ABJM with $R$-symmetry fugacities being pairwise equal. 
Tracing the same path as in the previous sections, we review the construction of complex solutions, regularize the action of the BPS black hole, uplift the non-minimal gauged supergravity to eleven dimensions, studying the conformal boundary conditions, and finally take the $\Z_{\Introc/4}$ quotients.

In multiple appendices we explain some details of the computations in the main text, and review the analogy between the large-$N$ behavior of the superconformal index of ABJM and the superconformal index of $4d$ $\CN=4$ SYM.

\section{The superconformal index of ABJM theory}
\label{sec:ABJMIndex}

In this section we introduce the superconformal index of ABJM theory in the Hamiltonian as well as functional integral formalism, and discuss its behavior under 
shifts of various chemical potentials.

\medskip

ABJM theory \cite{Aharony:2008ug} is a $U(N)_k\times U(N)_{-k}$ Chern--Simons-matter theory with $\CN=6$ superconformal symmetry for generic~$k$.
The superconformal algebra  is~$\mf{osp}(6|4)$ whose bosonic subalgebra is $\mf{so}(3,2)\times \mf{so}(6)_R$. In addition the theory has a~$\mf{u}(1)_b$ symmetry. 
The field content in~$\CN=2$ language is as follows. There are two vector supermultiplets $\CV, \tilde{\CV}$ corresponding,
respectively, to the two gauge groups, two chiral supermultiplets $\CA^{1,2}$ in the $(\mb{N}, \mb{\overline{N}})$ of $U(N)\times U(N)$ and two chiral supermultiplets $\CB_{1,2}$ in the anti-bifundamental $(\mb{\overline{N}}, \mb{N})$.
The matter multiplets  
interact via a superpotential
\beq
W \= \frac{2\pi}{k} \epsilon_{ab}\epsilon^{cd}\tr\left( \CA^a \CB_c \CA^b \CB_d \right) \, .
\eeq
The $\CN=2$ formalism makes the bosonic subalgebra~$\mf{su}(2)_A\times \mf{su}(2)_B\times \mf{u}(1)_b\times \mf{u}(1)_R$ of the global symmetry manifest. Here $\mf{su}(2)_{A,B}$ rotates, respectively, the two pairs of chiral multiplets~$\CA$ and~$\CB$, $\mf{u}(1)_R$ is the $R$-symmetry of the $\CN=2$ supercharges (under which the chiral multiplets have charge $\frac{1}{2}$) and $\mf{u}(1)_b$ is the ``baryonic symmetry'' under which $\CA^a$ have charge $\frac{1}{2}$ and $\CB_a$ have charge $- \frac{1}{2}$.\footnote{This symmetry is gauged, but it is related by the Chern--Simons term to the topological symmetry with current $J \propto * \tr (F + \tilde{F})$. One can also construct the index refined by the topological symmetry, and then perform a change of variables in the resulting matrix integral to obtain the same formula we have, see for instance \cite{Benini:2015eyy, Bobev:2022wem} for some related comments.\label{footnote:DetailsIndex}}
It is convenient to group the components of the chiral fields~$\CA = \{ A, \zeta \}$, $\CB = \{ B, \omega \}$ as
\begin{equation}
\begin{aligned}
    Y^A &\= \{ A^a, B^{\dagger a} \} \,, &\qquad Y^\dagger_A &\= \{ A^\dagger_a , B_a \} \, , \\
    \psi_A &\= \{ \epsilon_{ab}\zeta^b \rme^{-\ii \pi/4} , - \epsilon_{ab}\omega^{\dagger b} \rme^{\ii\pi/4} \} \, , &\qquad \psi^{\dagger A} &\= \{ - \epsilon^{ab} \zeta_b^\dagger \rme^{\ii\pi/4} , \epsilon^{ab}\omega_b \rme^{- \ii \pi/4} \} \, .
\end{aligned}
\end{equation}
The potential of the theory when written in terms of these combinations is invariant 
under the full $\mf{so}(6)_R\times \mf{u}(1)_b$ symmetry
provided $Y^A, \psi^{\dagger A}$ transform in the complex representation $\mb{4}$, and $Y^\dagger_A, \psi_A$ transform in the $\mb{\overline{4}}$~\cite{Aharony:2008ug, Benna:2008zy}.
We focus on the $k=1$ case, when the supersymmetry is enhanced to $\CN=8$ by non-perturbative effects \cite{Benna:2009xd, Bashkirov:2010kz}, and the symmetry algebra  $\mf{so}(6)_R\times \mf{u}(1)_b$ combines into $\mf{so}(8)_R$. 

\medskip

In this theory we consider the superconformal index based on the $\CN=6$ superconformal algebra 
refined by the symmetry~$\mf{u}(1)_b$.
In order to define the index, we quantize the theory on $S^2$ obtaining the Hilbert space $\CH_{S^2}$. 
States in $\CH_{S^2}$ are labelled by the eigenvalues of the Hamiltonian~$H$ and the angular momentum~$J$ (quantized as a half-integer), the weights of the $\mf{so}(6)_R$ $R$-symmetry, 
and the half-integer eigenvalues of the generator $B$ of $\mf{u}(1)_b$. For the $\mf{u}(1)^3$ Cartan subalgebra of $\mf{so}(6)_R$, we pick the ``orthogonal'' basis, consisting of the generators of the rotations in the three orthogonal planes of $\R^6$, which we label $H_1, H_2, H_3$, having half-integer eigenvalues.
Following \cite{Bhattacharya:2008zy, Bhattacharya:2008bja}, we pick a supercharge $\CQ$ with eigenvalues $\left( \frac{1}{2}, -\frac{1}{2}, 1,0,0,0 \right)$ under $(H,J,H_1,H_2,H_3,B)$, with the anticommutation relation
\begin{equation} 
\label{eq:QQbar}
 \{ \CQ, \CQ^\dagger \} \= H - J - H_1 \, .
\end{equation}
On states annihilated by~$\CQ$,
which define the subspace~$\CH_{\rm BPS}$, 
the right-hand side of the above equation clearly vanishes, 
i.e.~the energy~$H$ is determined by the values of~$H_1$ and~$J$. In the remaining five-dimensional subspace of bosonic charges, the subalgebra commuting with $\CQ$ and $\CQ^\dagger$ is generated by $J+\frac12 H_1$,~$H_2$,~$H_3$,~$B$. Therefore, the refined index of ABJM theory of interest here is
\begin{equation}
\label{eq:ABJMSCI_v1}
\begin{split}
   I(\QFTtau, \xi_2, \xi_3, \xi_B) &\= \Tr_{\CH_{S^2}} (-1)^{2J} \rme^{- \beta \{ \CQ, \CQ^\dagger \} + 2\pi \ii \QFTtau \left(J+\frac{1}{2} H_1 \right) + 2\pi\ii \left( \xi_2 H_2 + \xi_3 H_3 + \xi_B B \right) } \\
    &\= \Tr_{\CH_{\rm BPS}} \, (-1)^{2J} \rme^{ 2 \pi \ii \QFTtau \left( J + \frac{1}{2} H_1 \right) + 2\pi\ii \left( \xi_2 H_2 + \xi_3 H_3 + \xi_B B \right) } \,.
\end{split}
\end{equation}
The second equality follows, as usual,
from the fact that the index only receives contributions from states in $\CH_{\rm BPS}$~\cite{Witten:1982df}. Because of the half-integer quantization of the generators, 
the index is invariant under the
identifications~$\QFTtau \sim \QFTtau + 4$, and $\xi_{2,3,B} \sim \xi_{2,3,B} + 2$.

\medskip

It is useful to introduce a more symmetric basis of generators (and the corresponding chemical potentials)
\begin{equation}
\label{eq:ChangeVariablesGenerators_ABJM}
\begin{aligned}
    H_1 &\= \frac{R_1 + R_2 + R_3 + R_4}{2} \, , &\qquad H_2 &\= \frac{- R_1 + R_2 - R_3 + R_4}{2} \, , \\
    H_3 &\= \frac{- R_1 + R_2 + R_3 - R_4}{2} \, , &\qquad B &\= \frac{- R_1 - R_2 + R_3 + R_4}{2} \, ; \\[5pt]
    \xi_2 &\= - \lambda_1 - \lambda_3 \, , &\qquad \xi_3 &\= \lambda_2 + \lambda_3 \, , \\
    \xi_B &\= - \lambda_1 - \lambda_2 \, . 
\end{aligned}
\end{equation}
In terms of these, the refined index \eqref{eq:ABJMSCI_v1} takes the form
\begin{equation}
\label{eq:ABJMSCI_v2}
\begin{split}
    I &\= \Tr_{\CH_{S^2}} (-1)^{2J} \rme^{ - \beta \{ \CQ, \CQ^\dagger \} + 2\pi\ii \tau \left( J + \frac{1}{4}\sum_{a=1}^4 R_a \right) + 2\pi\ii \sum_{i=1}^3 \lambda_i (R_i - R_4) } \\
    &\= \Tr_{\CH_{S^2}} \rme^{ - \beta H + \beta \Omega J + \sum_{a=1}^4 \beta \Phi_a R_a}
\end{split}
\end{equation}
with
\begin{equation}
\label{eq:SCIOmega_v0}
    \Omega \= 1 + \frac{2\pi\ii}{\beta}(\tau + n_0) \,,
\end{equation}
and
\begin{equation}
\label{eq:SCIPhia_v0}
    \Phi_i \= \frac{1}{2} + \frac{2\pi\ii}{\beta} \left( \frac{\tau}{4} + \lambda_i \right) \, , \quad i\= 1, 2, 3 \, ,  \qquad \Phi_4 \= \frac{1}{2} + \frac{2\pi\ii}{\beta} \left( \frac{\QFTtau}{4} - \sum_{i=1}^3 \lambda_i \right)  \, .
\end{equation}
Here, we have used the expression \eqref{eq:QQbar} for $\{\CQ, \CQ^\dagger\}$
and written~$(-1)^{2J}=\rme^{2\pi \ii n_0 J}$ for an arbitrary \emph{odd} integer~$n_0$.
The charges of the fields (and the supercharge) under the spacetime and global symmetry used to define the refined index are summarised in Table~\ref{tab:ChargesABJM}. It is clear from this table that all states satisfy $J=R_a \mod 1$ for all $a=1, \dots, 4$, so that $\lambda_i \sim \lambda_i + 1$.

\begin{table}[t]
    \centering
    \begin{tabular}{ccccc}
    \toprule
    Fields & $J$ & $H$ & $\left( H_1,H_2,H_3,B \right)$ & $\left( R_1, R_2, R_3, R_4 \right)$ \\
    \midrule
    $Y^1$ & $0$ & $\frac{1}{2}$ & $\left( \frac{1}{2}, \frac{1}{2}, - \frac{1}{2}, \frac{1}{2}\right) $ & $\left( 0,0,0,1 \right)$ \\
    $Y^2$ & $0$ & $\frac{1}{2}$ & $\left( \frac{1}{2}, -\frac{1}{2}, \frac{1}{2}, \frac{1}{2}\right) $ & $\left( 0,0,1,0 \right)$ \\
    $Y^3$ & $0$ & $\frac{1}{2}$ & $\left( - \frac{1}{2}, \frac{1}{2}, \frac{1}{2}, \frac{1}{2}\right) $ & $\left( -1,0,0,0\right)$ \\
    $Y^4$ & $0$ & $\frac{1}{2}$ & $\left( - \frac{1}{2}, - \frac{1}{2}, - \frac{1}{2}, \frac{1}{2}\right) $ & $\left( 0,-1,0,0 \right)$ \\
    \midrule
    $\psi_{1\pm}$ & $\pm \frac{1}{2}$ & $1$ & $\left( - \frac{1}{2}, - \frac{1}{2}, \frac{1}{2}, \frac{1}{2}\right)$ & $\left( - \frac{1}{2}, -\frac{1}{2}, \frac{1}{2}, - \frac{1}{2} \right)$ \\
    $\psi_{2\pm}$ & $\pm \frac{1}{2}$ & $1$ & $\left( - \frac{1}{2}, \frac{1}{2}, - \frac{1}{2}, \frac{1}{2}\right)$ & $\left( - \frac{1}{2}, -\frac{1}{2}, -\frac{1}{2}, \frac{1}{2} \right)$ \\
    $\psi_{3\pm}$ & $\pm \frac{1}{2}$ & $1$ & $\left( \frac{1}{2}, - \frac{1}{2}, - \frac{1}{2}, \frac{1}{2}\right)$ & $\left( \frac{1}{2}, -\frac{1}{2}, \frac{1}{2}, \frac{1}{2}\right)$ \\
    $\psi_{4\pm}$ & $\pm \frac{1}{2}$ & $1$ & $\left( \frac{1}{2}, \frac{1}{2}, \frac{1}{2}, \frac{1}{2}\right)$ & $\left( - \frac{1}{2}, \frac{1}{2}, \frac{1}{2}, \frac{1}{2} \right)$ \\
    \midrule
    $Y^\dagger_1$ & $0$ & $\frac{1}{2}$ & $\left( - \frac{1}{2}, - \frac{1}{2}, \frac{1}{2}, - \frac{1}{2}\right)$ & $\left(0,0,0,-1\right)$ \\
    $Y^\dagger_2$ & $0$ & $\frac{1}{2}$ & $\left( - \frac{1}{2}, \frac{1}{2}, -\frac{1}{2}, - \frac{1}{2}\right)$ & $\left(0,0,-1,0\right)$ \\
    $Y^\dagger_3$ & $0$ & $\frac{1}{2}$ & $\left( \frac{1}{2}, - \frac{1}{2}, -\frac{1}{2}, - \frac{1}{2}\right)$ & $\left(1,0,0,0\right)$ \\
    $Y^\dagger_4$ & $0$ & $\frac{1}{2}$ & $\left( \frac{1}{2}, \frac{1}{2}, \frac{1}{2}, - \frac{1}{2}\right)$ & $\left(0,1,0,0\right)$ \\
    \midrule
    $\psi^{\dagger 1 \pm}$ & $\pm \frac{1}{2}$ & $1$ & $ \left( \frac{1}{2}, \frac{1}{2}, - \frac{1}{2}, - \frac{1}{2}\right)$ & $\left( \frac{1}{2}, \frac{1}{2}, -\frac{1}{2}, \frac{1}{2} \right)$ \\
    $\psi^{\dagger 2 \pm}$ & $\pm \frac{1}{2}$ & $1$ & $ \left( \frac{1}{2}, -\frac{1}{2}, \frac{1}{2}, - \frac{1}{2}\right)$ & $\left( \frac{1}{2}, \frac{1}{2}, \frac{1}{2}, -\frac{1}{2}  \right)$ \\
    $\psi^{\dagger 3 \pm}$ & $\pm \frac{1}{2}$ & $1$ & $ \left( -\frac{1}{2}, \frac{1}{2}, \frac{1}{2}, - \frac{1}{2}\right)$ & $\left( -\frac{1}{2}, \frac{1}{2}, -\frac{1}{2}, -\frac{1}{2} \right)$ \\
    $\psi^{\dagger 4 \pm}$ & $\pm \frac{1}{2}$ & $1$ & $ \left( -\frac{1}{2}, -\frac{1}{2}, - \frac{1}{2}, - \frac{1}{2}\right)$ & $\left( \frac{1}{2}, -\frac{1}{2}, -\frac{1}{2}, -\frac{1}{2}  \right)$ \\
    \midrule
    $\CQ$ & $- \frac{1}{2}$ & $\frac{1}{2}$ & $\left( 1,0,0,0 \right)$ & $\left( \frac{1}{2}, \frac{1}{2}, \frac{1}{2}, \frac{1}{2}\right)$ \\
    \bottomrule
    \end{tabular}
    \caption{Weights of the fields after the change of basis in the Cartan.}
    \label{tab:ChargesABJM}
\end{table}

\medskip

By the usual relation between the Hamiltonian and the path integral quantization, we can write the index~\eqref{eq:ABJMSCI_v2} 
as a functional integral on the background $S^1_\beta \times S^2$ with metric
\begin{equation}
    \rd s^2 \= \rd \nonrotatingtau^2 + \rd\bdrytheta^2 + \sin^2\bdrytheta \, \rd\nonrotatingphi^2 \, , 
\end{equation}
where $\bdrytheta$ has the canonical periodicity $\pi$, and
\begin{equation}
    (\nonrotatingtau, \nonrotatingphi) \; \sim \; (\nonrotatingtau, \nonrotatingphi + 2\pi) \; \sim \; (\nonrotatingtau + \beta, \nonrotatingphi - \ii \Omega \beta) \, .  
\end{equation}
The fields $f$ satisfy the following twisted boundary conditions around $S^1_\beta$,
\begin{equation}
    f \left( \, \bdrytau  + \beta , \bm{x} \right) \= (-1)^F \rme^{\beta \Omega J +  \sum_{i=a}^4 \beta \Phi_a R_a } f \left( \, \bdrytau , \bm{x} \right)
\end{equation}
Equivalently, we can write the index as the same functional integral, but  over the fibered background metric 
\begin{equation}
\label{eq:FiberedBdryMetric}
    \rd s^2 \= \rd {\bdrytau}^2 + \rd \bdrytheta^2 + \sin^2\bdrytheta \, \bigl( \rd \bdryphi - \ii \Omega \, \rd \bdrytau \bigr)^2
\end{equation}
and with background gauge fields coupling to the currents for the symmetries generated by~$R_a$ 
\begin{equation}
\label{eq:BackgroundGaugeFields_v0}
    A_a \= \ii \Phi_a \, \rd \bdrytau \,.
\end{equation}
In this case, the coordinates have periodicities
\begin{equation}
\label{eq:BdryCoordsPeriodicities}
    \left( \bdrytau, \bdryphi \right) \; \sim \; \left( \bdrytau + \beta, \bdryphi \right) \; \sim \; \left( \bdrytau, \bdryphi + 2\pi \right) \, ,
\end{equation}
and the fields satisfy standard thermal boundary conditions. The metric \eqref{eq:FiberedBdryMetric} is real if $\Omega$ is pure imaginary, but is otherwise complex-valued.

The background gauge field holonomies and the fibration parameter satisfy the following constraint
\begin{equation}
\label{eq:ChemicalPotentialsConstraint_v0}
    \beta \Bigl( 1 - \sum_a \Phi_a + \Omega \Bigr) \= 2\pi\ii n_0 \, .
\end{equation}
Due to the relation $J=R_a$ mod $1$, the index is invariant under the following shifts,
\begin{equation}
\label{eq:PeriodicitiesABJM}
    \Phi_a \; \to \; \Phi_a + \frac{2\pi\ii}{\beta} n_a \, , \qquad \Omega \; \to \; \Omega + \frac{2\pi\ii}{\beta} 2n_\Omega + \frac{2\pi\ii}{\beta} \sum_a n_a \, , \qquad n_a, n_\Omega \in \Z \, .
\end{equation}
Of course, these transformations preserve the parity of $n_0$ in the right-hand side of \eqref{eq:ChemicalPotentialsConstraint_v0}, since their effect is $n_0 \to n_0 - 2n_\Omega$. 

\medskip

It will simplify part of the following analysis to include an additional chemical potential $\lambda_4$ defined by the constraint
\begin{equation}
\label{eq:ConstraintFT}
    \sum_{a=1}^4 \lambda_a \= n_1 \in\Z \, .
\end{equation}
This leads to the following expression for the index, obtained from \eqref{eq:ABJMSCI_v2}
\begin{equation}
\label{eq:ABJMSCI_v3}
\begin{split}
    \CI(\tau; \lambda) &\= \Tr_{\CH_{\rm BPS}} (-1)^{2J} \rme^{2\pi\ii\tau \left( J + \frac{1}{4}\sum_a Q_a \right) + 2\pi\ii \sum_a \lambda_a (J+R_a)} \\
    &\= \Tr_{\CH_{\rm BPS}} (-1)^{2J} \rme^{2\pi\ii \sum_a \left( \frac{\tau}{4} + \lambda_a \right)(J + R_a) }
\end{split}
\end{equation}
In these variables we can identify 
the fibration parameter~$\Omega$ and the holonomies~$\Phi_a$ (which now have a symmetric form) for the background gauge fields 
\begin{equation}
\label{eq:SymmetricBackgroundsABJM}
    \Omega \= 1 + \frac{2\pi\ii}{\beta}(\tau + n_0 + n_1) \, , \qquad \Phi_a \= \frac{1}{2} + \frac{2\pi\ii}{\beta}\left( \frac{\QFTtau}{4} + \lambda_a \right) \, , \qquad a = 1, \dots, 4, , 
\end{equation}
which is related to \eqref{eq:SCIOmega_v0} and \eqref{eq:SCIPhia_v0} by a shift of the type \eqref{eq:PeriodicitiesABJM}.
The advantage of this basis is that the chemical potentials couple to orthogonal generators of the Cartan of the $\mf{so}(3)\times \mf{so}(8)_R$ subalgebra of the $\CN=8$ superalgebra. Thus the fact that $J=R_a$ mod $1$ is seen as a consequence of the $\CN=8$ superalgebra. Indeed, both in the free $\CN=8$ superconformal theory and in the interacting BLG theory, we can find a triality frame for the $R$-symmetry such that the supercharge sits in the $\mb{8_s^*}$, the scalars in the $\mb{8_v}$, and the spinors in the $\mb{8_s}$ \cite{Bandres:2008vf}.\footnote{This is different from the frame used in \cite{Bhattacharya:2008bja}, where the supercharge is taken in the vector of $\mf{so}(8)_R$.}

\medskip

The rewriting \eqref{eq:ABJMSCI_v3} shows that the index can be effectively rewritten as a function of four variables $\tau/4 + \lambda_a$, each defined modulo $1$, since $J=R_a$ mod $1$. However, it will be useful in later sections to consider shifts of $\tau$ alone, which lead to:
\begin{equation}
\label{eq:ABJMSCI_Shifts}
\begin{split}
    \CI(\tau;\lambda) &\= \Tr_{\CH_{\rm BPS}} (-1)^{2J} \rme^{2\pi\ii \sum_a \left( \frac{\tau}{4} + \lambda_a \right)(J + R_a) } \, , \\
    \CI(\tau+1;\lambda) &\= \Tr_{\CH_{\rm BPS}} (-1)^{H_1} \rme^{2\pi\ii \sum_a \left( \frac{\tau}{4} + \lambda_a \right)(J + R_a) } \; \equiv \; \CI_R(\tau;\lambda) \, , \\
    \CI(\tau+2;\lambda) &\= \Tr_{\CH_{\rm BPS}} (-1)^{2(J + H_1)} \rme^{2\pi\ii \sum_a \left( \frac{\tau}{4} + \lambda_a \right)(J + R_a) } \, , \\
    \CI(\tau + 3 ;\lambda) &\= \CI_R(\tau;\lambda)
\end{split}
\end{equation}
In particular, the first and the third indices are graded by~$(-1)^{2J}$ and~$(-1)^{2(J+H_1)}$, respectively, both of which take values~$\pm 1$ on the states of the ABJM theory, while the second and the fourth ones are indices graded by~$(-1)^{H_1}$ ($H_1= \frac{1}{2}\sum_a R_a$ is the $R$-symmetry generator) which takes values in the fourth roots of unity.\footnote{In fact, this quantization arises whenever~$\mf{u}(1)_R$ is part of a larger non-Abelian algebra.}
These phases are reflected in the contribution of the bosonic and fermionic states to the grand canonical index. 
In particular, as we see in section \ref{sec:largeN}, the $R$-charge index has saddle points with exponential growth near~$\t=0$, while the other two do not.\footnote{Of course, the Fourier coefficients of all four indices in \eqref{eq:ABJMSCI_Shifts} have an exponential growth in magnitude, with shifted phases.}
\footnote{This phenomenon is analogous to the structure of the index of four-dimensional~$\CN=4$ SYM, where the $R$-charges of the states are quantized in units of $1/3$, so that there are \textit{three} different indices, defined by shifts of~$\QFTtau$. Two of them lead to large growth as $\QFTtau\to 0$, while the third does not~\cite{ArabiArdehali:2021nsx, Cassani:2021fyv} (see Appendix \ref{app:N4Index}).}

\medskip

We can understand the constraint~\eqref{eq:ChemicalPotentialsConstraint_v0} in another equivalent manner.
Recall that the index can be computed as a functional integral over field configurations satisfying standard thermal boundary conditions around the~$ \bdrytau$ circle, in the background~\eqref{eq:FiberedBdryMetric},~\eqref{eq:SymmetricBackgroundsABJM}.  
In order to formulate a supersymmetric field theory on such a background, one couples it to off-shell supergravity and then considers its rigid limit~\cite{Festuccia:2011ws}. In this case, coupling the $\CN=8$ field theory to three-dimensional off-shell supergravity requires the existence of spinors solving the (generalised) Killing equation obtained from the vanishing of the gravitino variation (see e.g.~\cite{Nishimura:2012jh}). Noting that the spinor is charged under the potentials~$\Phi_a$ (for the $\mf{u}(1)^4 \subset \mf{so}(8)$ gauge group of the supergravity), and that the spin connection is proportional to $\Omega$, we see that the~$\bdrytau$ dependence of the Killing spinor is
\begin{equation}
\label{eq:TimeDependenceNonMinimalKS}
    \rme^{ \ii \bigl( 1 - \sum_a \Phi_a + \Omega \bigr) \bdrytau } \, .
\end{equation}
Now, from the fact that we impose anti-periodic boundary conditions for the fermions around the~$\bdrytau$ circle, we obtain 
precisely the constraint~\eqref{eq:ChemicalPotentialsConstraint_v0} with~$n_0$ being odd.

\medskip

When comparing with the gravity solutions, we shall consider less refined versions of the index, obtained by setting certain combinations of the four chemical potentials associated to the Cartan generators of $\mf{so}(8)_R$ to be equal. We may first set the four chemical potentials to be pairwise equal, that is $\lambda_{1,2} \equiv \sigma_1$, $\lambda_{3,4} \equiv \sigma_2$. Since the chemical potentials $\lambda_a$ are constrained to satisfy $\sum_a \lambda_a = n_1$, we have $2 (\sigma_1 + \sigma_2) = n_1$. In this case, we find that the index has the form
\begin{equation}
\label{eq:Index_PairwiseEqual}
    \CI(\QFTtau;\lambda_{1,2} = \sigma_1, \lambda_{3,4} = \sigma_2) \= \Tr_{\CH_{\rm BPS}} (-1)^{2J} \rme^{2\pi\ii \QFTtau \left( J + \frac{1}{4}\sum_a R_a \right) + 2\pi\ii \left( \sigma_1 (R_1 + R_2) + \sigma_2 (R_3 + R_4) \right) } \, .
\end{equation}
We can identify two $\mf{u}(1)$ generators $2 Q_{1} \equiv R_1 + R_2$ and $2 Q_{2} \equiv R_3 + R_4$. In terms of these generators, the interpretation as a functional integral over a fibered background requires the following parameters
\begin{equation}
\label{eq:FibrationParameters_PairwiseEqual}
    \Omega \= 1 + \frac{2\pi\ii}{\beta} (\tau+n_0 + n_1) \, , \qquad \Phi^{(Q_A)} \= 1 + \frac{2\pi\ii}{\beta} \left( \frac{\QFTtau}{2} + 2 \sigma_A \right) \, , \quad A\=1, 2 \, ,
\end{equation}
which are constrained by
\begin{equation}
\label{eq:Constraint_PairwiseEqual}
    \beta \, \biggl( 1 - \Phi^{(Q_1)} - \Phi^{(Q_2)} + \Omega \biggr) \= 2\pi\ii n_0 \, .
\end{equation}
The shifts of the fibration parameters that leave the partition function invariant are
\begin{equation}
\label{eq:PeriodicitiesABJM_PairwiseEqual}
    \Phi^{(Q_A)} \; \to \; \Phi^{(Q_A)} + \frac{2\pi\ii}{\beta} 2 n_{Q_A} \, , \qquad \Omega \; \to \; \Omega + \frac{2\pi\ii}{\beta} 2n_\Omega \, , \qquad n_{(R_A)}, n_\Omega \in \Z \, .
\end{equation}
Unlike~\eqref{eq:PeriodicitiesABJM}, these two shifts are now independent because the constraint between the charges of the states in the theory, namely $2Q_A + 2J = 0 \mod 1$, is now trivially satisfied. Notice that these two shifts preserve the parity of $n_0$ in the right-hand side of \eqref{eq:Constraint_PairwiseEqual}.

Finally, we can further simplify this by setting all the generators equal: $\lambda_a \equiv \lambda$ for all $1\leq a \leq 4$. Because of the constraint between the chemical potentials, we have $4 \lambda = n_1$. In this case, we find that the index has the form
\begin{equation}
\label{eq:N2Index}
\begin{split}
    \CI(\QFTtau;\lambda_{1,2,3,4}=\lambda) &\= \Tr_{\CH_{\rm BPS}} (-1)^{2J} \rme^{2\pi\ii \left(\QFTtau + n_1 \right) \left( J + \frac{1}{2}H_1 \right) } \, .
\end{split}
\end{equation}
In this form it is clear that, as observed in~\eqref{eq:ABJMSCI_Shifts},  $n_1 = \sum_a \lambda_a$ appears as a shift of $\tau$, leading to indices graded by different operators and thus with different asymptotic behaviors.

When we view the unrefined index as a thermal partition function over a fibered background, we find the parameters
\begin{equation}
\label{eq:FibrationParameters_AllEqual}
    \Omega \= 1 + \frac{2\pi\ii}{\beta} (\tau + n_0 + n_1)  \, , \qquad \Phi^{(H_1)} \= 1 + \frac{2\pi\ii}{\beta} \frac{\QFTtau + n_1}{2} \, , 
\end{equation}
constrained by
\begin{equation}
\label{eq:Constraint_AllEqual}
    \beta \left( 1 - 2 \Phi^{(H_1)} + \Omega \right) \= 2\pi\ii n_0 \, .
\end{equation}
As in the previous cases, we can shift the holonomy and fibration parameter while leaving the partition function invariant
\begin{equation}
\label{eq:PeriodicitiesABJM_AllEqual}
    \Phi^{(H_1)} \; \to \; \Phi^{(H_1)} + \frac{2\pi\ii}{\beta} 2 n_{H_1} \, , \qquad \Omega \; \to \; \Omega + \frac{2\pi\ii}{\beta} 2n_\Omega \, , \qquad n_{(H_1)}, n_\Omega \in \Z \, .
\end{equation}
The shifts~\eqref{eq:PeriodicitiesABJM_AllEqual}
and~\eqref{eq:PeriodicitiesABJM_PairwiseEqual} will play an important role in the gravitational interpretation in the later sections.

\medskip

As we shall see in Section~\ref{sec:BlackHole}, the background with fibration parameters \eqref{eq:FibrationParameters_AllEqual} is what is found at the boundary of the supersymmetric electrically charged black hole in minimal gauged supergravity, and so it is the correct background in order to match the gravity computation in the bulk. However, it has a generically complex metric, which is not standard from the field theory viewpoint. 
On the other hand, one can also define the index using a supersymmetric background consisting of a real metric and background gauge field on $S^1\times S^2$ (see for instance~\cite[Sec.~7.2]{Closset:2013vra}). 
Since both backgrounds are used to compute the same supersymmetric observable, which should only depend on the moduli of the transversely holomorphic foliation, it is natural to conjecture that they are related by a $\CQ$-exact deformation.\footnote{One should be able to formulate the analogous discussion in extended supergravity for the background to the index refined according to the $\CN=8$ superalgebra.} This is similar to the discussion about four-dimensional backgrounds on $S^1\times S^3$ \cite{Cassani:2021fyv, ArabiArdehali:2021nsx, Cabo-Bizet:2021jar}.

\section{ABJM index near rational points}
\label{sec:RationalPoints}

In this section we investigate the behavior of the superconformal index near rational points. We consider the refined index~\eqref{eq:ABJMSCI_v3}, which is a periodic function of~$\tau$ with period~4. In terms of the variable~$q=\exp(2 \pi \ii \t)$, it is 
defined on a 4-sheeted cover of the complex plane. The variable~$\exp(2 \pi \ii \t/4)$ lives on~$\IC$ and we consider the behavior of the index as this approaches a primitive root of unity, i.e.~$\frac14 \tau \to \frac{d}{c}$,  $c,d\in \Z$, $c>0$ and $\gcd(c,d)=1$, further taking the large-$N$ limit of the result.

\medskip

Recall from~\eqref{eq:ABJMSCI_v3} that~$\QFTtau$ couples to $J+\frac{1}{4}\sum_a R_a$, and $\lambda = (\lambda_1, \dots, \lambda_4)$, with  $\sum_a\lambda \in\Z$, couple to the orthogonal generators of the Cartan of $\mf{so}(3)\times \mf{so}(8)$ in the definition of the index. 
The index $\CI$ is the $\CN=6$ superconformal index further refined by the~$\mf{u}(1)_b$. The index without this refinement was computed 
in~\cite{Bhattacharya:2008zy} using the free field theory in the background of zero magnetic fluxes for the gauge groups, 
and in~\cite{Kim:2009wb} using localization allowing for fluxes. The refinement by the baryonic symmetry can be introduced referring to Table~\ref{tab:ChargesABJM} (see also footnote \ref{footnote:DetailsIndex}). An analogous expression for the refined index has been considered in \cite{Choi:2019zpz, Bobev:2022wem}. 
As a matrix integral, the index \eqref{eq:ABJMSCI_v3} for ABJM at level $k=1$ takes the form
\be
\label{eq:IndexABJM}
\begin{split}
& \CI(\t;\lambda)  \= \; \SumInt_{\; \; \, \mathfrak{m}} \, [D\uu] \quad 
 \SumInt_{\; \; \,  \mathfrak{\wt m}} \, [D \wt \uu] \;  q^{ \frac{1}{2}\sum_{i,j} |\mf{m}_i - \tilde{\mf{m}}_j| - \frac{1}{4} \sum_{i,j} |\mf{m}_i - \mf{m}_j| - \frac{1}{4} \sum_{i,j} |\tilde{\mf{m}}_i - \tilde{\mf{m}}_j| } \\
 & \qquad \qquad \qquad \qquad \qquad \times Z_\text{class}  (\uu, \mm, \wt \uu, \mathfrak{\wt m})
\;  Z_\text{vec} (\uu, \mm, \wt \uu, \mathfrak{\wt m};\t)   \; 
 \prod_{a=1}^4 Z^a_\text{chi} (\uu, \mm, \wt \uu, \mathfrak{\wt m};\t,\lambda)  \,.
\end{split}
\ee
Here the notation 
\be \SumInt_{\; \; \, \mm } \, [D\uu]  \; \equiv \;  \frac{1}{N!}\, \sum_{\mathfrak{m} \,\in\,\mathbb{Z}^N} \, 
\prod_{i=1}^N \, \int_{\IR/\IZ} \dd u_i 
\ee
denotes the sum over the magnetic fluxes~$\mm = (\mm_1,\dots \mm_N)$ and the integral over the gauge holonomies~$\uu = (u_1, \dots , u_N)$ for each gauge group. 
We introduce the fugacities~$q = \rme^{2 \pi \ii \t}$, $\zeta_a = \rme^{2 \pi \ii \lambda_a}$, $a=1,\dots,4$, and~$x_i= \rme^{2 \pi \ii u_i}$, $\wt x_i = \rme^{2 \pi \ii \wt u_i}$.
In all the products, $i, j$ run from $1$ to~$N$. 

The various pieces in the integrand are the following. The contribution of the classical action is 
\be \label{zclass}
Z_\text{class} (\uu, \mm, \wt \uu, \mathfrak{\wt m}) \= 
\prod_i  \;   \exp \Bigl(2 \pi \ii \bigl( \mathfrak{m}_i  \, u_i - \tilde{\mf{m}}_i \, \wt u_i  \bigr) \Bigr) \,,
\ee
the contribution of the two $\CN=2$ vector multiplets for the $U(N)\times U(N)$ gauge group is
\be \label{vec1loop}
\begin{split}
Z_\text{vec} (\uu, \mm, \wt \uu, \mathfrak{\wt m};\t) \= &
\prod_{i\neq j} \bigl( 1 - x_i \, x_j^{-1} q^{\frac{1}{2} |\mf{m}_i - \mf{m}_j|} \bigr) \, \prod_{i\neq j} \bigl( 1 - \tilde{x}_i \, \tilde{x}_j^{-1} q^{\frac{1}{2} |\tilde{\mf{m}}_i - \tilde{\mf{m}}_j|} \bigr) 
\end{split}
\ee
and the contributions of the four $\CN=2$ chiral multiplets are
\be  \label{chi1loop}
\begin{split}
Z^a_\text{chi} (\uu, \mm, \wt \uu, \mathfrak{\wt m};\t,\lambda) &\= 
\prod_{i,j} \; \dfrac{ \bigl( x_i^{-1}\, \tilde{x}_j \, \zeta_a^{-1} \, q^{\frac{3}{4} + \frac{1}{2} | \mf{m}_i - \tilde{\mf{m}}_j | } \, ; 
\, q \bigr)_\infty }{ \bigl( x_i \, \tilde{x}_j^{-1} \, \zeta_a \; q^{\frac{1}{4} + \frac{1}{2} | \mf{m}_i - \tilde{\mf{m}}_j | } \, ; 
\, q \bigr)_\infty } \,, \qquad a \= 1,2 \,, \\
Z^a_\text{chi} (\uu, \mm, \wt \uu, \mathfrak{\wt m};\t,\lambda) &\= 
\prod_{i,j} \; \dfrac{ \bigl( x_i \, \tilde{x}_j^{-1} \, \zeta_a^{-1} \, q^{\frac{3}{4} + \frac{1}{2} | \mf{m}_i - \tilde{\mf{m}}_j | } \, ; \, 
q \bigr)_\infty }{ \bigl( x_i^{-1} \, \tilde{x}_j \, \zeta_a \; q^{\frac{1}{4} + \frac{1}{2} | \mf{m}_i - \tilde{\mf{m}}_j | } \, ; 
\, q  \bigr)_\infty }  \,, \qquad a \= 3,4 \,. 
\end{split}
\ee
It is manifest from the above expressions that~$\CI(\t;\lambda)$ is invariant under the separate shifts~$\lambda_a \to \lambda_a + 1$, $a=1,\dots, 4$, and $\QFTtau \to \QFTtau + 4$. 

\medskip

We are interested in the behavior of $\CI(\t;\lambda)$ for
\be
\tau \= \frac{4d}{c} + \ii \frac{\varepsilon}{2\pi c} \; \Rightarrow \; q\=\rme^{2 \pi \ii \frac{4 d}{c}} \,\rme^{-\frac{\ve}{c}} \,, \qquad \gcd(c,d) =1 \,,  \quad \ve \searrow 0 \,,  
\ee
Although we begin our analysis for~$\ve \in \IR^+$, all the statements that we make below
hold for~$\ve$ tending to zero from any direction in the upper 
half-plane, and we continue to use the symbol~$\varepsilon \searrow 0$ to denote this more general limit.

\medskip

In this limit we expect that the integral over~$\uu$ and the sum over~$\mf{m}$ are both dominated by saddle-points. 
In order to implement the saddle-point analysis, it is useful to change variables so that the integrand factorizes into what are essentially holomorphic and anti-holomorphic pieces. This is similar to the $\tau\to 0$ treatment in~\cite{Choi:2019zpz} but as we see below the~$\t \to \mathbb{Q}$ limit 
has additional subtleties.
It is useful to define the following 
variables
\begin{equation}
\label{eq:VariablesForIntegration}
\begin{aligned}
    s_i  &\= u_i + \ii \mf{m}_i \frac{\varepsilon}{4\pi c} \, , 
    &\qquad \overline{s}_i &\= - u_i + \ii \mf{m}_i \frac{\varepsilon}{4\pi c} \, , \\ 
    \tilde{s}_i &\= \tilde{u}_i + \ii \tilde{\mf{m}}_i \frac{\varepsilon}{4\pi c} \, , 
    &\qquad \overline{\tilde{s}}_i &\= - \tilde{u}_i + \ii \tilde{\mf{m}}_i \frac{\varepsilon}{4\pi c} \, ,
\end{aligned}
\end{equation}
and the corresponding exponentiated 
variables~$z_i = \rme^{2 \pi \ii s_i}$, $\overline{z}_i= \rme^{2 \pi \ii \overline{s}_i}$, 
and~$\wt z_i = \rme^{2 \pi \ii \wt s_i}$,  $\wt{\overline{z}}_i= \rme^{2 \pi \ii \overline{\wt s}_i}$, 
with~$\overline{z}_i=z_i^*$,~${\overline{\wt z}}_i=\wt z_i^*$.
We set~$\mf{m}_{ij} = \mf{m}_{i}- \mf{m}_{j}$, $\mf{\wt m}_{ij} = \mf{\wt m}_{i}- \mf{\wt m}_{j}$, and introduce
\begin{equation}
\begin{split}
    \xi_{ij} \; &\equiv \; \exp \left(-2 \pi \ii \frac{4d}{c} \frac{\mf{m}_{ij}}{2} \right) \,,\qquad 
   \wt \xi_{ij} \; \equiv \; \exp \left(-2 \pi \ii \frac{4d}{c} \frac{\tilde{\mf{m}}_{ij}}{2} \right) \,,\\
    \xi_{ij}' \; &\equiv \; \exp \left(-2 \pi \ii \frac{4d}{c} \frac{\mf{m}_i - \tilde{\mf{m}}_j}{2} \right) \, ,
\end{split}
\end{equation}
which are roots of unity depending on $\mf{m}_{ij}$.

As we show in Appendix~\ref{app:factorization}, in terms of these variables, the integrand of~\eqref{eq:IndexABJM} essentially factorizes into two parts\footnote{More precisely, in the case $c>1$, the presence of the terms $\xi_{ij}$, $\tilde{\xi}_{ij}$ and $\xi'_{ij}$ forbids us from declaring that $Z_{\rm hol}$ is a holomorphic function of $z$ (and $Z_{\rm anti-hol}$ of $\zbar$). However, as we shall see in the next section, at the leading order in the Cardy-like limit $\varepsilon\searrow 0$, the dependence on $\xi_{ij}$, $\tilde{\xi}_{ij}$, $\xi'_{ij}$ drops and the integrand indeed factorizes, even for $c>1$ in the cases we are interested in. This factorization should be thought of as a simplifying manipulation, and should not be crucial to the derivation. Indeed, in a related problem, the authors of~\cite{Hosseini:2022vho} directly compute the large-$N$ limit of the twisted index, without using this manipulation.}
\be
\label{eq:ABJMFactorization}
 Z_{\rm hol}(\underline{z},\underline{\tilde{z}}; \t,\lambda) \, 
 Z_{\rm antihol}(\underline{\overline{z}}, \underline{\overline{\tilde{z}}}; \t,\lambda) \, ,
\ee
with
\begin{equation} \label{defZhol}
\begin{split}
Z_{\rm hol}(\underline{z},\underline{\tilde{z}}; \tau,\lambda)  \= 
 & \prod_i  \; \Bigg[ \exp \Bigl( 2 \pi \ii \frac{4\pi c}{4\ii \varepsilon} \left( s_i^2 - \tilde{s}_i^2 \right)  \Bigr) \, z_i^{-1/2}\, \tilde{z}_i^{1/2} \, \xi_{ii}'^{\frac{1}{2}}  \\
& \qquad \quad \times \; \frac{\prod_{a=1,2} \, \bigl( z_i^{-1} \,\tilde{z}_i \, \xi'_{ii} \, \zeta_a^{-1} \, q^{\frac{3}{4}} \, ; \, 
q \bigr)_\infty }{  \prod_{a=3,4} \, \bigl( z_i^{-1}\, \tilde{z}_i \, \xi'_{ii} \, \zeta_a \, q^{\frac{1}{4}} \, ; \, 
q \bigr)_\infty } \Bigg] \\
&  \times \;   \prod_{i > j}  \; \Bigg[ \frac{\bigl( z_i^{-1} \, z_j  \, \xi_{ij}  \, ;  \, q \bigr)_\infty}{\bigl( z_i^{-1}  \, z_j  \, \xi_{ij} 
\, q \, ; \, q \bigr)_\infty} \; 
\frac{\bigl( \tilde{z}_i^{-1} \,  \tilde{z}_j  \, \tilde{\xi}_{ij}  \, ;  \, q \bigr)_\infty}{\bigl( \tilde{z}_i^{-1}  \, \tilde{z}_j  \, 
\tilde{\xi}_{ij}  \, q  \, ; \, q \bigr)_\infty}  \\
& \qquad \qquad \times \;  \prod_{a=1,2} \,  \frac{\bigl( z_i^{-1} \, \tilde{z}_j  \, \xi'_{ij}  \, \zeta_a^{-1}  \, q^{\frac{3}{4}} \, ;  
\, q \bigr)_\infty }{ \bigl( z_j  \, \tilde{z}_i^{-1}  \, \xi_{ji}'^{-1}  \, \zeta_a  \, q^{\frac{1}{4}} \, ;  \, q \bigr)_\infty}  \\
& \qquad \qquad \times \;  \prod_{a=3,4} \,  \frac{ \bigl( z_j  \, \tilde{z}_i^{-1}  \, \xi_{ji}'^{-1}  \, \zeta_a^{-1}  \, q^{\frac{3}{4}} \, ;  
\, q \bigr)_\infty }{
\bigl( z_i^{-1} \, \tilde{z}_j  \, \xi_{ij}'  \, \zeta_a  \, q^{\frac{1}{4}} \, ;  \, q \bigr)_\infty}  \Bigg]
\end{split}
\end{equation}
and
\begin{equation}
\label{eq:ZantiHol}
    Z_{\rm antihol}(\underline{\overline{z}}, \underline{\overline{\tilde{z}}}; \t,\lambda) \= 
    Z_{\rm hol}(\underline{z},\underline{\tilde{z}}; \t ,\lambda) \big|_{k\to -k, \, z_i \to \overline{z}_i, \, 
    \tilde{z}_i \to \overline{\tilde{z}}_i, \, \zeta_1 \leftrightarrow \zeta_3, \, \zeta_2 \leftrightarrow \zeta_4} \, .
\end{equation}

Given this factorization of the integrand, we now use the idea of~\cite{Pasquetti:2019uop} to make a similar split 
in the integration measure using a change of contour. 
Recall that~$z_i = \rme^{- \mf{m}_i \varepsilon/{2 c}} \rme^{2\pi \ii u_i}$, so that we identify $\rme^{- \mf{m}_i \varepsilon/{2c}}$ as the modulus of $z_i$, and $2\pi u_i$ as its argument. 
The idea is to exchange the domain of the sum over~$\mf{m}_i$ and integration over~$u_i$ with the full complex plane, i.e., 
\begin{equation}
\begin{split}
    \frac{1}{\varepsilon} \sum_{\mf{m}_i\in \Z} \varepsilon \Delta \mf{m}_i \, \int_0^1 \rd u_i 
    & \; \xrightarrow{\varepsilon \to 0} 
    \;  \frac{1}{\varepsilon} \int_{-\infty}^{+\infty} \rd (\varepsilon \mf{m}_i) \, \int_0^1 \rd u_i \\
    &\= \frac{2c}{\varepsilon} \int_0^\infty \frac{\rd |z_i|}{|z_i|} \int_0^{2\pi} \frac{\rd {\rm Arg}(z_i)}{2\pi} 
    \= \frac{2c}{\varepsilon} \int_{\C} \frac{\rd^2 z_i}{2\pi |z_i|^2} \, , 
\end{split}
\end{equation}
where $\rd^2 z_i = |z_i| \, \rd |z_i| \rd {\rm Arg}(z_i)$ is the flat measure on the plane. 
We thus obtain
\begin{equation} 
\label{IntZZbar}
\CI(\t;\lambda)    
 \= \frac{1}{(N!)^2} \frac{(2c)^{2N}}{\varepsilon^{2N}} \; \biggl(\;  \prod_{i=1}^N \int_{\C^{2}} \frac{\rd^2 z_i}{2\pi z_i \overline{z}_i} 
    \frac{\rd^2 \tilde{z}_i}{2\pi \tilde{z}_i \overline{\tilde{z}}_i} \,  \biggr)  \; 
     Z_{\rm hol}(\underline{z},\underline{\tilde{z}}; \t,\zeta_a) \, 
     Z_{\rm antihol}(\underline{\overline{z}}, \underline{\overline{\tilde{z}}}; \t,\zeta_a) \, .
\end{equation}
One then factorizes the full integral into ``holomorphic'' and ``anti-holomorphic''
pieces as
\begin{equation} 
\label{IntZZbarsep}
\CI(\t;\lambda)    
    \= \frac{(4\pi c)^{N}}{\ve^{N}} \; \int [D \us]  \, [D\ustl] \; 
     Z_{\rm hol}(\underline{z},\underline{\tilde{z}}; \t,\lambda) \, \times
  \frac{(4 \pi c)^{N}}{\ve^{N}}\;  \int [D \overline{\us}]  \, [D \overline{\ustl}] \; 
     Z_{\rm antihol}(\underline{\overline{z}}, \underline{\overline{\tilde{z}}}; \t,\lambda)
\end{equation}
where the holomorphic variables~$\us$, $\ustl$ are integrated
\be 
\int [D\us]  \; \equiv \;  \frac{1}{N!}\, \prod_{i=1}^N \, \int \dd s_i \,, \qquad 
\int [D\ustl]  \; \equiv \;  \frac{1}{N!}\, \prod_{i=1}^N \, \int \dd \wt{s}_i 
\ee
over some contour. The corresponding anti-holomorphic variables are taken to be independent 
and run over a possibly different contour.  
The contour integrals are then evaluated by a saddle-point approximation, which we discuss now. 

\subsection{Generalized Cardy limits to roots of unity}
\label{subsec:GeneralizedCardyLimites}

Our goal is to calculate the asymptotic behavior of the integral~\eqref{IntZZbarsep}
in the generalized Cardy limit~$q=\rme^{2\pi\ii \frac{4d}{c}} \rme^{- \frac{\varepsilon}{c}}$ as~$\ve \searrow 0$. 
In the saddle-point approximation, we can analyze the holomorphic and anti-holomorphic parts separately, which are built out of Pochhammer symbols.

To study the asymptotics of these building blocks, we use a result of \cite{Garoufalidis:2018qds}, whose details we summarize in Appendix~\ref{app:Asymp}. This requires~$q=\xi_{\underline{m}} \rme^{ - \frac{\underline{\ve}}{\underline{m}}}$, where $\xi_{\underline{m}}$ is a primitive root of unity of order $\underline{m}$. Therefore, we write~$ (c,4d) = \gcd(c,4d) (\ell_c , \ell_d) $
with~$\gcd(\ell_c,\ell_d)=1 $,
so that $q=\xi_{\ell_c} \rme^{ - \frac{\ve/\gcd(c,4d)}{\ell_c}}$, 
and we can now apply the result from \cite{Garoufalidis:2018qds} directly. 
The leading order result for the holomorphic part of the integrand~\eqref{defZhol} as~$\ve \searrow 0$ is
\begin{equation} 
\label{Zholexp}
\begin{split}
   & {\rm log} \; Z_{\rm hol}(\underline{z},\underline{\tilde{z}}, q, \lambda) \; \sim \; \\
   & \qquad - \frac{\gcd(c,4d)^2}{c\varepsilon} \Biggl\{ \sum_i \, \frac{1}{2} (2\pi\ii \ell_c)^2 (s_i^2 - \tilde{s}_i^2) \\
    & \qquad \qquad \quad  + \sum_{i > j} \, \biggl[ \; \sum_{a=1,2} {\rm Li}_2 \bigl( z_i^{-\ell_c}\tilde{z}_j^{\ell_c} \zeta_a^{-\ell_c} (\xi'_{ij}\xi_{\ell_c}^{-\frac{1}{4}})^{\ell_c} \bigr) - {\rm Li}_2 \bigl( z_j^{\ell_c} \tilde{z}_i^{-\ell_c} \zeta_a^{\ell_c} (\xi_{ji}'^{-1}\xi_{\ell_c}^{\frac{1}{4}})^{\ell_c} \bigr)  \\
    & \qquad \qquad \qquad \qquad \quad + \sum_{a=3,4} {\rm Li}_2\bigl( z_j^{\ell_c} \tilde{z}_i^{-\ell_c} \zeta_a^{-\ell_c} (\xi_{ji}'^{-1}\xi_{\ell_c}^{-\frac{1}{4}})^{\ell_c} \bigr) - {\rm Li}_2\bigl( z_i^{-\ell_c}\tilde{z}_j^{\ell_c} \zeta_a^{\ell_c} (\xi'_{ij}\xi_{\ell_c}^{\frac{1}{4}})^{\ell_c} \bigr)  \biggr] \\
    & \qquad \qquad \quad + \sum_i \biggl[ \sum_{a=1,2} {\rm Li}_2 \bigl( z_i^{-\ell_c}\tilde{z}_i^{\ell_c}\zeta_a^{-\ell_c} (\xi'_{ii}\xi_{\ell_c}^{-\frac{1}{4}})^{\ell_c} \bigr) - \sum_{a=3,4} {\rm Li}_2 \bigl( z_i^{-\ell_c}\tilde{z}_i^{\ell_c} \zeta_a^{\ell_c} (\xi'_{ii}\xi_{\ell_c}^{\frac{1}{4}})^{\ell_c} \bigr) \biggr] \Biggr\} \,.
\end{split}
\end{equation}
Note that the vector multiplet does not contribute in the Cardy-like limit. Upon rescaling the integration variables $\bigl(s_i, \wt s_i, \overline{s_i} , \overline{\wt s_i} \bigr) \mapsto \frac{1}{\ell_c} \bigl(s_i, \tilde{s_i} , \overline{s}_i,  \overline{\wt s}_i \bigr)$
we obtain
\be 
\label{Zholoc}
  {\rm log} \; Z_{\rm hol}(\underline{z},\underline{\tilde{z}}, q, \lambda) \; \sim \; 
   - \frac{\gcd(c,4d)^2}{c\varepsilon} \CW(\underline{z}, \underline{\tilde{z}}, \lambda) + \text{O}(1) \,,
\ee
where
\begin{equation} 
\label{eq:Wstart}
\begin{split}
    \CW(\underline{z}, \underline{\tilde{z}}, \lambda) &\= \sum_i \, \frac{1}{2} (2\pi\ii)^2 (s_i^2 - \tilde{s}_i^2) \\
    & \quad \ \ + \sum_{i > j} \, \biggl[ \; \sum_{a=1,2} \Bigl[ {\rm Li}_2 \bigl( z_i^{-1}\tilde{z}_j \zeta_a^{-\ell_c} (\xi'_{ij}\xi_{\ell_c}^{-\frac{1}{4}})^{\ell_c} \bigr) - {\rm Li}_2 \bigl( z_j \tilde{z}_i^{-1} \zeta_a^{\ell_c} (\xi_{ji}'^{-1}\xi_{\ell_c}^{\frac{1}{4}})^{\ell_c} \bigr) \Bigr]  \\
    & \qquad \qquad \quad + \sum_{a=3,4} \Bigl[ {\rm Li}_2\bigl( z_j \tilde{z}_i^{-1} \zeta_a^{-\ell_c} (\xi_{ji}'^{-1}\xi_{\ell_c}^{-\frac{1}{4}})^{\ell_c} \bigr) - {\rm Li}_2\bigl( z_i^{-1}\tilde{z}_j \zeta_a^{\ell_c} (\xi'_{ij}\xi_{\ell_c}^{\frac{1}{4}})^{\ell_c} \bigr) \Bigr] \biggr] \\
    & \quad \ \ + \sum_i \biggl[ \sum_{a=1,2} {\rm Li}_2 \bigl( z_i^{-1} \tilde{z}_i \zeta_a^{-\ell_c} (\xi'_{ii}\xi_{\ell_c}^{-\frac{1}{4}})^{\ell_c} \bigr) - \sum_{a=3,4} {\rm Li}_2 \bigl( z_i^{-1}\tilde{z}_i \zeta_a^{\ell_c} (\xi'_{ii}\xi_{\ell_c}^{\frac{1}{4}})^{\ell_c} \bigr) \biggr] \,.
\end{split}
\end{equation}

Note that the arguments of the dilogarithms above contain factors of the type~$(\xi'_{ij}\xi_{\ell_c}^{-\frac{1}{4}})^{\ell_c} = \exp \Bigl( - 2\pi\ii \ell_d \bigl( \frac{\mf{m}_i - \tilde{\mf{m}}_j}{2} + \frac{1}{4} \bigr) \Bigr)$, which deserve a comment. 
Since we have~$\gcd(c,d)=1$, there are three possible cases, i.e.,~$\gcd(c,4d) = \gcd(c,4) =1$, $2$, or~$4$. When~$\gcd(c,4)=1$, all these factors in~\eqref{eq:Wstart} equals~1.  When~$\gcd(c,4)=2$, they are equal to~$\rme^{-2\pi\ii \frac{d}{2}}$. 
In this case it is clear from~\eqref{eq:Wstart} that this phase can be absorbed into a redefinition of~$\zeta_a$. 
When~$\gcd(c,4)=4$, there is no such simplification, and we will restrict to the first two cases from now on. 

The case~$c=1$ and~$d=0$, which is relevant for~the~$q \to 1$ limit of the ABJM index, has been studied in~\cite{Choi:2019zpz, Nian:2019pxj} by the saddle-point approximation in the small parameter~$\ve$. Even in this approximation, the presence of the dilogarithm functions in~$\CW$ makes it difficult to perform an exact analysis. However, the potential~$\CW$ simplifies in the large-$N$ limit. We review the main points of this analysis in Section~\ref{sec:largeN}, and use the large-$N$ method to analyze the perturbation theory in~$\ve$ to all orders in Section~\ref{sec:subleading}.

\subsection{The large \texorpdfstring{$N$}{N} saddle-point analysis \label{sec:largeN}} 

Having expressed the ABJM superconformal index in the generalized Cardy limit in terms of an effective potential $\CW$ \eqref{Zholoc}, in order to find its value in the large-$N$ limit we should extremize $\CW$. As already noticed in \cite{Choi:2019zpz}, the effective potential~$\CW$ obtained in the generalized Cardy limit is a straightforward 
generalization, at a mathematical level,  
of the Bethe potential introduced in \cite{Benini:2015eyy} to describe the topologically twisted index which is, a priori,
a different problem, and one can therefore follow the analysis developed in~\cite{Benini:2015eyy}. 

\medskip

Recall that the large-$N$ limit of a unitary matrix model can be implemented by replacing the discrete distribution of the~$N$ eigenvalues by a continuum  of eigenvalues on the interval~$[0,1]$.
Sums over the discrete label of the eigenvalues~$i$ turn into integrals over the interval, which one replaces by integrals over the space of eigenvalues by introducing a density of eigenvalues. 

In our case, we have two distributions~$s$ and~$\wt s$ for the integration variables that are both complex. 
Justified by the study of the numerics in \cite{Benini:2015eyy}, one introduces the following single-cut ansatz
\be\label{eq:ansatzN32}
s(x)\=  v(x) - \ii N^{\frac{1}{2}} x \,,\qquad
\wt{s}(x)\= \tilde{v}(x) - \ii N^{\frac{1}{2}} x\,,
\ee
where~$x \in [x_1,x_2] \in \IR$.
The assumptions in the ansatz are that~$\Im(\wt s)=\Im(s)$ and that~$x_2-x_1, v(x), \wt v(x)$ are all~$\text{O}(1)$ quantities 
as~$N \to \infty$.
The sums become integrals according to 
\be
\sum_{i=1}^N \,  \dots \; \mapsto N \int_{x_1}^{x_2} \rd x \, \rho(x)\, \dots \,, 
\ee
where the eigenvalue density $\rho$ obeys
\be
\label{eq:RhoConstraint}
\int_{x_1}^{x_2} \rd x \, \rho(x) \= 1 \,.
\ee

We split the effective action~$\mathcal{W}$ in \eqref{eq:Wstart} as a sum of three pieces
\be
\label{WDecomposition}
\mathcal{W} \= \mathcal{W}_1 + \mathcal{W}_2 + \mathcal{W}_3\,,
\ee
with
\be 
\label{Wsplit}
\begin{split}
\CW_1 &\=  N \int_{x_1}^{x_2} \rd x \, \rho(x) \, \frac{1}{2} (2\pi \ii)^2 \Bigl(s(x)^2 -\widetilde{s}(x)^2\Bigr) \,, \\[5pt]
\CW_2 &\= N^2\int_{x_1}^{x_2} \rd x \, \rho(x) \int_{x}^{x_2} \rd y \, \rho(y)
\left[\sum_{a=1,2} \left[\text{Li}_2\Bigl(\frac{\widetilde{z}(x)\,z(y)^{-1}}{\zeta^{\ell_c}_{a}} \Bigr) -
\text{Li}_2\Bigl(\frac{{z}(x)\,\widetilde{z}(y)^{-1}}{\zeta_{a}^{-\ell_c}}  \Bigr) \right] \right. \\ 
&\qquad \qquad \qquad\qquad\qquad\qquad\quad + \left. \sum_{a=3,4} \left[\text{Li}_2\Bigl(\frac{{z}(x)\,
\widetilde{z}(y)^{-1}}{\zeta_{a}^{\ell_c}} \Bigr) -
\text{Li}_2\Bigl(\frac{\widetilde{z}(x)\,z(y)^{-1}}{\zeta_{a}^{-\ell_c}}  \Bigr)\right] \right] \,, \\[5pt]
\CW_3 &\= N \int_{x_1}^{x_2} \rd x \, \rho(x)\,
\left[\sum_{a=1,2} \text{Li}_2\Bigl(\frac{\widetilde{z}(x)\,z(x)^{-1}}{\zeta^{\ell_c}_{a}} \Bigr)-
\sum_{a=3,4} \text{Li}_2\Bigl(\frac{\widetilde{z}(x)\,z(x)^{-1}}{\zeta_{a}^{-\ell_c}} \Bigr)\right] \,.
\end{split}
\ee
Here we have introduced the notation $z(x) = \rme^{2\pi\ii s(x)}$ and $\tilde{z}(x) = \rme^{2\pi\ii \tilde{s}(x)}$. Our goal is to extremize $\CW$ subject to the constraint \eqref{eq:RhoConstraint}, or extremizing the quantity
\be
\label{eq:WmuTobeExtremised}
\mathcal{W}^\mu \defeq  \mathcal{W} + N^{\frac{3}{2}}\,\mu\, \ii \left( \int_{x_1}^{x_2} \rd x \, \rho(x) - 1 \right) \, .
\ee
The resulting equations, namely
\be
\label{eq:extremize}
\delta_{\rho}\mathcal{W}^\mu\= 0 \,, \qquad \delta_{v} \mathcal{W}^\mu\=0\, , \qquad \delta_{\wt v} \mathcal{W}^\mu\=0\, , \qquad \partial_{\mu} \mathcal{W}^\mu\=0\, , 
\ee
should be solved for the distributions~$\rho$, $s(x)$ and~$\wt s(x)$.

We now evaluate the three terms~$\CW_{1,2,3}$ in~\eqref{WDecomposition} on the ansatz~\eqref{eq:ansatzN32}. As we will shortly see, each piece has a simple term scaling as~$N^\frac32$, and subleading terms scaling as~$N$, as~$N \to \infty$.\\
The first piece is
\be
\label{W1largeN}
\begin{split}
\CW_1 &\;\sim\; - N^{\frac{3}{2}} \int_{x_1}^{x_2} \rd x\,\rho(x) \, (2\pi)^2 \ii x \,\delta v(x) \,
\Bigl( 1 + \text{O}(1/N^\frac12) \Bigr)\,,
\end{split}
\ee
where~$\delta v(x) \equiv \widetilde{v}(x)\,-\,v(x)$.\\
The second piece contains terms of the following form
\be\label{eq:NonLocalTerm}
\begin{split}
&\int_{x_1}^{x_2} \rd x \, \rho(x) \int_{x}^{x_2} \rd y \, \rho(y) \, \text{Li}_2 \Bigl(\frac{\widetilde{z}(x)\,z(y)^{-1}}{\zeta_a^{\ell_c}} \Bigr)\,\\
&\qquad \qquad \qquad \qquad \=\int_{x_1}^{x_2} \rd x \, \rho(x) \int_{x}^{x_2} \rd y \, \rho(y) \,\, \text{Li}_2\Bigl(\rme^{ 2\pi N^\frac{1}{2} (x-y) +
2\pi\ii \bigl( \widetilde{v}(x) - {v}(y) - \lambda'_a \bigr)}\Bigr) \,,
\end{split}
\ee
where $\lambda_a' \equiv \ell_c \lambda_a$.
In matrix model language, these terms indicate non-local interactions between the eigenvalues at~$x$ and~$y$. However, at leading order in the large-$N$ expansion, the integrand simplifies further, and we are left only with local interactions at each $x$.
To see this, it is convenient to change integration variables from~$y$ to $\delta y \coloneqq N^{\frac{1}{2}}(y-x)$, after which the right-hand side of~\eqref{eq:NonLocalTerm} becomes
\be
N^{-\frac{1}{2}} \int_{x_1}^{x_2} \rd x \, \rho(x)\int_{0}^{N^{\frac{1}{2}}(x_2-x)} \rd \delta y \, \rho(x+\delta y/N^{\frac{1}{2}}) \,
\text{Li}_2\Bigl( \rme^{ - 2\pi \delta y + 2\pi\ii \bigl( \tilde{v}(x) - v(x+\delta y/N^{\frac{1}{2}}) - \lambda'_a \bigr) }\Bigr)\, .
\ee
At leading order in the large-$N$ expansion (at fixed~$\delta y$) this takes the asymptotic form 
\be
N^{-\frac{1}{2}} \int_{x_1}^{x_2} \rd x \, \rho(x)^2\, \int_{0}^{+\infty} \rd \delta y  \,
\text{Li}_2\Bigl(\rme^{ - 2\pi \delta y + 2\pi\ii ( \delta v(x) - \lambda'_a) }\Bigr) \,.
\ee
Then, using~$\frac{\rd \ }{\rd x} \text{Li}_3(\rme^x) \= \text{Li}_2(\rme^{x})$, one concludes that at leading order
\be\label{Li2inttoL3}
\int_{x_1}^{x_2} \rd x \, \rho(x) \int_{x}^{x_2} \rd y \, \rho(y) \,\, \text{Li}_2\Bigl(\frac{\widetilde{z}(x)\,z(y)^{-1}}{\zeta_a^{\ell_c}}\Bigr) 
\sim \frac{N^{-\frac{1}{2}}}{2\pi} \int_{x_1}^{x_2}\, \rd  x \,\rho(x)^2\,\text{Li}_3 \left( \rme^{ 2\pi\ii (\delta v(x) - \lambda_a')} \right) \,.
\ee
Upon applying the above analysis to the four terms in~$\mathcal{W}_2$, one obtains
\be \label{W2Li3}
\begin{split}
\mathcal{W}_2 \;\sim\; \frac{N^{\frac{3}{2}}}{2\pi} \int_{x_1}^{x_2} \rd x \, \rho(x)^2\,
&\left[ \sum_{a=1,2} \Bigl[  \text{Li}_3\bigl(\rme^{2\pi\ii (\delta v(x) - \lambda_a')}\bigr) - \text{Li}_3\bigl(\rme^{- 2\pi\ii (\delta v(x) - \lambda'_a)} \bigr) \Bigr] \right.  \\ 
&  \quad - \left. \sum_{a=3,4} \Bigl[ \text{Li}_3\bigl(\rme^{ 2\pi\ii (\delta v(x) + \lambda'_a)}\bigr)  - \text{Li}_3\bigl(\rme^{- 2\pi\ii (\delta v(x) + \lambda'_a)} \bigr)\Bigr] \right] \,.
\end{split}
\ee
We can simplify this expression further using the identity \eqref{eq:UsefulIdentity_PolyLogs}
\be
\label{eq:Li3Identity}
\text{Li}_3\bigl(\rme^{2 \pi\ii x}\bigr) \,-\,\text{Li}_3\bigl(\rme^{-2 \pi \ii x}\bigr)\=
\frac{4\pi^3\ii}{3}\, \overline{B}_3(x)
\ee
where~$\overline{B}_3$ is the third periodic Bernoulli polynomial that is defined in~\eqref{eq:PeriodicBernoulliDef}, obtaining
\be 
\label{W2largeN}
\mathcal{W}_2 \;\sim\; \frac{2\pi^2\ii}{3} \, N^{\frac{3}{2}}\int_{x_1}^{x_2} \rd x \, \rho(x)^2\, \left[ \sum_{a=1,2}\overline{B}_3\left( \delta v(x) - \lambda'_a \right) - \sum_{a=3,4} \overline{B}_3 \left( \delta v(x) + \lambda'_a \right) \right] \,.
\ee
Finally, the third piece~$\mathcal{W}_3$ can be written as
\be \label{W3largeN}
\mathcal{W}_3 \;\sim\; N \int_{x_1}^{x_2} \rd x \, \rho(x)\, \left[ \sum_{a=1,2} \text{Li}_2\bigl(\rme^{ 2\pi\ii( \delta v(x)- \lambda'_a)} \bigr) - \sum_{a=3,4}  \text{Li}_2\bigl(\rme^{ 2\pi\ii (\delta v(x) + \lambda'_a)} \bigr) \right] \, .
\ee
Note that the factor in front of the integral in~\eqref{W3largeN} is~$N$, instead of~$N^\frac32$ as in~$\CW_1$, $\CW_2$. This means that $\CW_3$ does not contribute to the on-shell value of the effective action $\CW$ in the \hbox{large-$N$} 
limit. However, it cannot be naively discarded, since it is relevant for the saddle point equations, as the dilogarithm~$\text{Li}_2(z)$ is not analytic at the branch point~$z=1$.
Indeed, the first derivative of the integrand in~\eqref{W3largeN} with respect to~$\delta v(x)$ can grow as~O$(N^\frac{1}{2})$, if the function approaches a branch point.
In this case, $\CW_3$ contributes at the same order as $\CW_1$ and $\CW_2$ to the equation in \eqref{eq:extremize} obtained by 
taking the derivative with respect to~$\delta {v}(x)$.

\medskip

At this stage, the degree of difficulty of the original extremization problem is substantially diminished, as the large-$N$ form of the potential~$\mathcal{W}$ given by the sum of~\eqref{W1largeN}, \eqref{W2largeN}, \eqref{W3largeN} has no non-local terms. 
Moreover, if one focuses on the contribution at order~$N^{\frac{3}{2}}$ to the potential, one notices that in terms of the field variables~$\rho(x)$ and~$\delta v(x)$ the problem is piecewise quadratic. Therefore, solutions to the variational problem~\eqref{eq:extremize} can be found using a linear ansatz~$\rho(x)= \rho_0 +x \rho_1$ for the density of eigenvalues. 
The solutions when all the~$\lambda_a$s are equal are particularly simple, and we present them below. The details for generic values of~$\lambda_a$s are discussed in Appendix~\ref{app:SaddlePointSol}.

\paragraph{Unrefined index}

We set all the chemical potentials to be equal ($\lambda_a \equiv \lambda$). The constraint~\eqref{eq:ConstraintFT} implies that~$\lambda = n_1/4$. As for $\lambda'_a$, recall from the discussion below \eqref{eq:Wstart}, that if $\gcd(c,4)=2$, a phase is absorbed in $\zeta_a$, so we write $\lambda' = (\ell_c n_1 + \ell_d)/4$, with the understanding that if $\gcd(c,4)=1$, then $\ell_d = 4d$ and thus $\ell_d/4$ can be removed from $\lambda'$ (which is defined modulo 1), and if $\gcd(c,4)=2$, then $\ell_d = 2d$ and $\ell_d/4$ is non-trivial. 
We further assume that $\delta v(x)$ does not cross a branch point of the dilogarithm, so that we can effectively ignore $\CW_3$.  This picture is also warranted by the numerics \cite{Benini:2015eyy}. To leading order in $N$, the function to be extremized $\CW^\mu$ defined in \eqref{eq:WmuTobeExtremised} takes the simple form
\begin{equation}
\begin{split}
    \CW^\mu &\= N^{\frac{3}{2}} \, \ii \int_{x_1}^{x_2} \rd x \, \rho(x) \, \left[ - 4\pi^2 x \, \delta v(x) + \frac{4\pi^2}{3} \rho(x) \left( B_3 \left( X_-(x) \right) - B_3 \left( X_+(x) \right) \right) \right] \\
    &\quad \ \ + N^{\frac{3}{2}} \, \mu \, \ii \left( \int_{x_1}^{x_2}\rd x \, \rho(x) - 1 \right) \, , 
\end{split}
\end{equation}
where
\begin{equation}
\label{eq:Xpm}
    X_\pm (x) \; \equiv \; \delta v(x) \pm \frac{\ell_c n_1 + \ell_d}{4} + w_\pm
\end{equation}
for some integers $w_\pm$ such that 
\begin{equation}
\label{eq:DefinitionXpm}
0 < X_\pm(x) < 1 \,
\end{equation}
In the following, we shall re-express these integers using $\Sigma \equiv w_+ + w_-$ and $\Delta \equiv w_+ - w_-$. The single-cut solution is defined on a single sheet of the multi-valued polylogarithms provided $w_\pm$ do not depend on $x$.

Combining the first three extremization equations \eqref{eq:extremize} gives (ignoring boundary terms)
\be
\label{eq:ExtremizationExpl}
\begin{split}
    0 &\= 4 x + \rho(x) (\ell_c n_1 + \ell_d + 2\Delta) \left( 2(\Sigma-1) + 3 \, \delta v(x) \right) \, , \\
    0 &\= 48\pi^2 x \, \delta v(x) + \pi^2 \rho(x) (\ell_c n_1 + \ell_d + 2\Delta) \big( 8 + (\ell_c n_1 + \ell_d + 2\Delta)^2 \\
    & \qquad \qquad \qquad \qquad \qquad \qquad + 12 \, \Sigma (\Sigma-2) + 48 \, \delta v(x)(\Sigma - 1) + 48 \, \delta v(x)^2 \big) - 12\mu \, , 
\end{split}
\ee
These equations are easily solved for $\rho(x)$ and $\delta v(x)$. Indeed, for fixed~$x$, the first equation is linear in $\rho$ and $\delta v$, and  upon substituting in the second equation the expression for~$\rho$ obtained from the first (and assuming that~$\rho$ is finite), we find that the resulting equation is linear in~$\delta v$. The solutions are:
\beq
\label{eq:SaddleEqual}
\begin{split}
\rho(x) &\= \frac{12 \mu + 24 \pi^2 (\Sigma -1) x }{\pi^2 \left(\ell_c n_1 + \ell_d + 2\Delta -2\right) \left(\ell_c n_1 + \ell_d + 2 \Delta\right) \left(\ell_c n_1 + \ell_d + 2 \Delta + 2\right)} \, , \\[5pt]
\delta v(x) &\= \frac{ - \pi^2 x (\ell_c n_1 + \ell_d + 2(\Delta - \Sigma))(\ell_c n_1 + \ell_d + 2(\Delta + \Sigma)) - (\Sigma -1) \left[ 6 + 8 \pi^2 x (2\Sigma-1) \right] }{12 \left( \mu + 2 \pi^2 x (\Sigma -1) \right) } \, .
\end{split}
\eeq
We focus on the solution with constant eigenvalue density, so we set $\Sigma=1$, which means that $\Delta$ must be odd.\footnote{The solution with constant eigenvalue density can also be obtained as a limit of the solutions to the saddle point equations of the refined index in Appendix~\ref{app:SaddlePointSol}.} 
The conditions~\eqref{eq:DefinitionXpm} imply
\begin{equation}
\label{eq:InequalitiesUnrefined}
\begin{split}
    \frac{6|\mu|}{\pi^2 (\ell_c n_1 + \ell_d + 2\Delta +2)(\ell_c n_1 + \ell_d + 2\Delta -2)}  \; < \; & x \; < \; \frac{- 6|\mu|}{\pi^2 (\ell_c n_1 + \ell_d + 2\Delta +2)(\ell_c n_1 + \ell_d + 2\Delta -2)} \, , \\
    -2 \; < \; & \ell_c n_1 + \ell_d + 2\Delta \; < \; 2 \,.
\end{split}
\end{equation}
In particular, the second inequality in \eqref{eq:InequalitiesUnrefined} imposes that $\ell_c n_1 + \ell_d + 2\Delta \in \{-1,0,1\}$, but the choice~0 (from the first equation in~\eqref{eq:SaddleEqual}) leads to a divergent density $\rho$ unless~$\mu=0$ (which would reduce the support to a single point), and therefore should be discarded.
Thus
\begin{equation}
    \rho \= - \sgn(\ell_c n_1 + \ell_d + 2\Delta) \, \frac{4\mu}{\pi^2} \, , \qquad \delta v \= \frac{\pi^2}{4\mu} x \, .
\end{equation}
Clearly, in order to have a positive density of eigenvalues, we need $\mu\in\R$ and $\sgn(\mu) = - \sgn(\ell_c n_1 + \ell_d+2\Delta)$. 

The first inequality in \eqref{eq:InequalitiesUnrefined} is only a necessary condition for the range of $x$ such that \eqref{eq:DefinitionXpm} holds. In fact, upon substitution of ~$\ell_c n_1 + \ell_d+2\Delta=\pm1$, \eqref{eq:DefinitionXpm} reduces to
\be
-\frac{|\mu|}{\pi^2}\,<\,x\,<\,\frac{|\mu|}{\pi^2}\, .
\ee
This is the smallest interval for $x$ such that the single-cut solution does not cross a branch cut of the polylogarithm, thus we set~$x_{1}=-\frac{|\mu|}{\pi^2}$ and~$x_2=\frac{|\mu|}{\pi^2}$. Finally, we impose the normalization of $\rho$, which is equivalent to extremizing $\CW^\mu$ with respect to~$\mu$, finding
\begin{equation}
\begin{aligned}
    \rho &\= \sqrt{2} \, , &\qquad \delta v &\= - \sgn(\ell_c n_1 + \ell_d + 2\Delta) \, \frac{x}{\sqrt{2}}  \\ 
    [x_1,x_2] &\= \left[ - \frac{1}{2\sqrt{2}}, \frac{1}{2\sqrt{2}} \right] \, , &\qquad \mu &\= - \sgn(\ell_c n_1 + \ell_d + 2\Delta) \, \frac{\pi^2}{2\sqrt{2}} \, .
\end{aligned}
\end{equation}
Overall, $\Delta$ remains a free odd integer, and
\begin{equation}
    \CW \= \mp \frac{\ii \pi^2}{3\sqrt{2}} N^{\frac{3}{2}} \, \qquad \text{if $\ell_c n_1 + \ell_d = \pm 1 \mod 4$.}
\end{equation}
We then substitute in \eqref{Zholoc} and perform the same analysis for the anti-holomorphic part \eqref{eq:ZantiHol}, reaching the following conclusion. The extremization problem for the unrefined index has $\text{O}(N^{\frac{3}{2}})$ scaling only if 
\begin{equation}
\label{eq:LargeNConstraint_Unrefined}
    \ell_c n_1 + \ell_d \= \pm 1 \mod 4 \,,
\end{equation} 
and
\begin{equation}
\label{eq:LargeNUnrefinedIndex}
    \log \CI(\tau;\lambda) \; \sim \; \mp \frac{\pi}{3\sqrt{2}} N^{\frac{3}{2}} \frac{1}{\ell_c \left( \ell_c\tau- \ell_d\right)} \qquad \text{as } \tau \to \frac{4d}{c} \text{ and } \ell_c n_1 + \ell_d = \pm 1 \mod 4 \, .
\end{equation}
We recall that this holds if $\gcd(c,4d)=1,2$.

To summarize, we have the following picture: we start with the unrefined index, which is a single-valued function of $\Introtau$
\begin{equation}
\label{eq:N2Index_v1}
    \CI(\Introtau) \= \Tr_{\CH_{\rm BPS}} (-1)^{2J} \rme^{2\pi\ii \Introtau\left( 4J + \sum_a Q_a \right)} \= \sum_{n} d_n \Introq^n \, , \quad \quad ( \Introtau\sim \Introtau+1\, , \ \ \Introq\equiv \rme^{2\pi\ii \Introtau}) 
\end{equation}
and to find $d_n$ we look at the limit where $\Introq$ goes to a primitive root of unity, as explained in Section~\ref{sec:Intro}, or $\Introtau\to \Introd/\Introc$ with $\gcd(\Introc,\Introd)=1$. In order to study the asymptotic behavior, we further introduce $4\Introtau = {\ell}_D/\ell_c$, for coprime $\ell_c$, ${\ell}_D$. In terms of these, we write the constraint \eqref{eq:LargeNConstraint_Unrefined} as $4\ell_c \Introtau = {\ell}_D = \pm 1$ mod 4. Since ${\ell}_D = 4\Introd/\gcd(\Introc, 4)$, this constraint can only be solved if $\gcd(\Introc, 4)=4$, in which case we have
\begin{equation}
\label{eq:LimitUnrefinedIndex_Ttilde}
    \log \CI(\Introtau) \; \sim \; \mp \frac{\pi}{3\sqrt{2}} N^{\frac{3}{2}} \frac{1}{\frac{\Introc}{4} \bigl( \Introc \Introtau - \Introd \bigr)} \quad \text{as $\Introtau \to \frac{\Introd}{\Introc}$ with $D=\pm 1$ mod $4$.}
\end{equation}
This leads us to the conclusion that the leading growth for the index \eqref{eq:N2Index_v1} is controlled  
by the singularity near $\Introtau = \pm \frac{1}{4}$, that is, when $\Introq$ approaches the non-trivial primitive fourth roots of unity. In terms of $\QFTtau$ and $n_1$, which is how the index has been presented in the literature, our methods allow us to reach these singularities as $\tau\to 0$ and $n_1=1,3$, $\QFTtau\to 2$ and $n_1=1,3$. We are currently unable to study the case $\tau\to 1, 3$ as $n_1=0$, but it is natural to conjecture that it gives the same result. 
Since the index is a four-sheeted function of $\QFTtau$, the singularities in the large-$N$ limit appear on the first and third sheet, or taking the Cardy limit $\QFTtau \to 0$ of the $R$-charge index defined in \eqref{eq:ABJMSCI_Shifts}.
There is a clear analogy with the case of four-dimensional $\CN=4$, which is developed in Appendix \ref{app:N4Index}.

\paragraph{Refined index}
We now move to the most refined ABJM index, where the chemical potentials $\lambda_a$ are taken to be generic. In this case, the solutions to the extremization problem~\eqref{eq:extremize} are more involved than~\eqref{eq:SaddleEqual}, and have been addressed in~\cite{Benini:2015eyy} using the techniques of~\cite{Herzog:2010hf}. Details of their construction are reviewed in Appendix \ref{app:SaddlePointSol}. Here, we notice that, assuming $\lambda_a\in\R$, a solution to the extremization problem \eqref{eq:extremize} with scaling $N^{\frac{3}{2}}$ in the large-$N$ limit exists in terms of a single-cut eigenvalue distributions only provided the chemical potentials and $\ell_c$ satisfy
\be
\label{eq:LargeNConstraint_Refined}
\sum_{a=1}^4 \{\ell_c\lambda_a + \ell_d/4\} \= 1 \text{ or } 3 \, .
\ee
Here, as in \eqref{eq:Xpm}, we have introduced $\ell_d/4$, which is either an integer or a half-integer, depending on whether $\gcd(c,4)=1,2$, respectively.
Notice that this constraint consistently reduces to \eqref{eq:LargeNConstraint_Unrefined} upon setting all $\lambda_a$ to be equal.

Combining the contributions from the extremization of the holomorphic and the anti-holomorphic parts in~\eqref{IntZZbarsep}, one finds that the leading result for the ABJM index, as~$\ve = 2\pi \ii \,({c\t - 4d})\searrow 0$, is
\begin{equation}
\label{eq:FieldTheoryResult2}
    \log \CI(\QFTtau;\lambda) \; \sim \; \mp\,  \frac{8\pi\sqrt{2} \, N^{\frac{3}{2}}}{3} \frac{ \sqrt{\{ \ell_c \lambda_1  + \ell_d/4\}_{\pm}  \{ \ell_c \lambda_2  + \ell_d/4\}_{\pm}  \{ \ell_c \lambda_3  + \ell_d/4\}_{\pm} \{ \ell_c \lambda_4  + \ell_d/4\}_{\pm} }}{c(c\tau-4d) } 
    \, . 
\end{equation}
where~$\{x\}_{+}:=\{x\}$ and~$\{x\}_{-}:=\{x\}-1\,$. The upper and lower signs correspond to the first or second case in~\eqref{eq:LargeNConstraint_Refined}.

A few comments are in order. Firstly, studying the regions associated to different asymptotic behaviors of the fully refined index $\CI(\tau;\lambda)$ is quite involved. It is clear that \eqref{eq:LargeNConstraint_Refined} in the case $\ell_c=1$ and $\ell_d=0$ reduces to $n_1 = 1, 3$, and thus we obtain the same picture to the one just described for the unrefined index: in the Cardy limit $\tau \to 0$, the non-trivial large-$N$ limit is found on the first and third sheet of the multivalued 
function~$\CI(\tau)$, which corresponds to the $R$-charge index defined in \eqref{eq:ABJMSCI_Shifts}. More generally, though, the relation of \eqref{eq:LargeNConstraint_Refined} to the shifts of $\tau$ is more involved.\footnote{Of course, the constraint \eqref{eq:LargeNConstraint_Refined} can be expressed without reference to $n_1$, as it should from its definition.}

Secondly, notice that \eqref{eq:FieldTheoryResult2} is invariant under a shift of any $\lambda_a$ by~$\frac{1}{\ell_c}$, in addition to the shift~$\lambda_a\to \lambda_a+1$ due to the quantization of the charges $Q_a$ (see~\eqref{eq:ABJMSCI_v3}). The emergent periodicity suggests that in the expansion near roots of unity the superconformal index only counts operators with charges dual to~$\lambda_a$ being an integer multiple of~$\ell_c$. That is, in such an expansion the contribution to the trace~\eqref{eq:ABJMSCI_v3} of operators with charges dual to~$\lambda_a$ not being an integer multiple of~$\ell_c$ is suppressed. Interestingly, the same phenomenon happens in four-dimensions~\cite{ArabiArdehali:2021nsx,Jejjala:2021hlt,Cabo-Bizet:2021plf}.

Finally, recall that the partition function in the microcanonical ensemble is obtained by taking the Laplace transform of \eqref{eq:LimitUnrefinedIndex_Ttilde}. The saddles near~$\Introtau \to \pm \frac{1}{4}$ 
(equivalently,  $c=1$ and $d=0$ and~$2$ for fixed~$n_1=1$)
contribute equally in magnitude to the Laplace transform, and with opposite phases. 
The logarithm of the real part
agrees with the entropy of the dual supersymmetric black hole, and the interference of the phases produces macroscopic oscillations (of order $N^{\frac{3}{2}}$) in the microcanonical index. The same phenomenon has been observed in related contexts in~\cite{Choi:2019zpz, Agarwal:2020zwm, Murthy:2020rbd, Cabo-Bizet:2020ewf, Choi:2020baw, Genolini:2021qbi}. 

\paragraph{Index with pairwise equal fugacities} 

Finally, we discuss a further case, which is relevant for the comparison with holographic duals in Section \ref{sec:U12}, namely the index \eqref{eq:Index_PairwiseEqual} with pairwise equal fugacities. We set $\lambda_{1,2} \equiv \sigma_1$ and $\lambda_{3,4} \equiv \sigma_2$. The result can be straightforwardly obtained from the fully refined case just discussed (see Appendix~\ref{par:LimitingCases}). Assuming that $\sigma_{1,2}\in\R$, the single-cut eigenvalue distribution provides a solution to the extremization problem that scales like $N^{\frac{3}{2}}$ in the large-$N$ limit provided
\begin{equation}
\label{eq:LargeNConstraint_Pairwise}
    2\left( \{ \ell_c \sigma_1 + \ell_d/4\} + \{ \ell_c \sigma_2 + \ell_d/4\} \right) \= 1 \text{ or } 3 \, .
\end{equation}
Substituting the constraint $2( \sigma_1 + \sigma_2 ) = n_1$, it is straightforward to show that the left-hand side of \eqref{eq:LargeNConstraint_Pairwise} has the same parity as $\ell_c n_1+\ell_d$, which in the cases we are interested in corresponds to that of $\ell_c n_1$ (since $\ell_d$ is even). Therefore, $\ell_c n_1$ cannot be even, which singles out the $R$-charge index among the sheets in \eqref{eq:ABJMSCI_Shifts}. More precisely, if $c$ is odd, then the condition \eqref{eq:LargeNConstraint_Pairwise} is satisfied if $cn_1$ is odd, and imposes $\{ c\sigma_1 \} \leq \frac{1}{2}$ for the right-hand side to be $1$, and $\{ c\sigma_1 \} > \frac{1}{2}$ for the right-hand side to be $3$. If instead $\gcd(c,4)=2$, then one needs $cn_1/2$ odd, but now $\{ c\sigma_1 \} \leq \frac{1}{2}$ is consistent with the right-hand side being $3$, and $\{ c\sigma_1 \} > \frac{1}{2}$ is consistent with the right-hand side being $1$. \\
Upon substitution, in the case of $c$ odd, we require $n_1$ to be odd, and the resulting value of the large-$N$ partition function is
\begin{equation}
\label{eq:LargeNPairwiseIndex}
\begin{aligned}
    \log \CI(\tau;\sigma) \; &\sim \; - \frac{8\pi \sqrt{2}}{3} N^{\frac{3}{2}} \frac{\{ c  \sigma_1 \} \{ c \sigma_2 \}  }{\ell_c \left( c\tau- 4d\right) } &\quad &\text{if } \tau \to \frac{4d}{c} \text{ and } \{ c \sigma_1 \} \leq \frac{1}{2} \, , \\
    \log \CI(\tau;\sigma) \; &\sim \; + \frac{8\pi \sqrt{2}}{3} N^{\frac{3}{2}} \frac{ ( 1- \{ c \sigma_1 \}) ( 1 - \{ c \sigma_2 \})  }{c \left( c\tau- 4d\right) } &\quad &\text{if } \tau \to \frac{4d}{c} \text{ and } \{ c \sigma_1 \} > \frac{1}{2} \, .
\end{aligned}
\end{equation}

\subsection{Subleading effects \label{sec:subleading}}

Now we consider sub-leading effects in the asymptotic expansion in~$\ve$ around any rational point~$d/c$.
Our goal is to show that the expansion terminates at order~$\ve$ i.e.~all the 
terms of order~$\ve^k$, $k \ge2$ vanish in the large-$N$ limit. The same result was found in~\cite{GonzalezLezcano:2022hcf} for the case of the black hole saddle i.e.~$(c,d)= (1,0)$. 
We use the leading order analysis of the previous 
subsection as a reference and discuss the new points that arise in the all-order analysis. 
 
We begin again with the holomorphic part of the potential~\eqref{defZhol}, expand it to all orders in~$\ve$
as given in~\eqref{IdentityGZ}, and run the steps of the large-$N$ expansion as in 
Sections~\ref{subsec:GeneralizedCardyLimites}, and~\ref{sec:largeN}.
The potential~$\CW$ splits into three pieces~$\CW_{1,2,3}$ as in~\eqref{WDecomposition}, \eqref{Wsplit}, 
exactly as at leading order. The term~$\CW_1$ is classical and does not depend on~$\ve$ and therefore remains 
the same as in~\eqref{Wsplit}. The other two pieces~$\CW_2$, $\CW_3$ have, a priori, an 
infinite expansion governed by the identity~\eqref{IdentityGZ}, with the leading-order term given by~\eqref{Wsplit}. 

\medskip

Consider a general term of order~$\ve^{k}$, $k \ge 2$, in the expansion of~$\CW_2$.  
This term has the same structure as in the leading~$\text{O}(1/\ve)$ term with~$\text{Li}_{1-k}$ 
replacing~$\text{Li}_2$, with coefficients being Bernoulli polynomials given by~\eqref{psiexp}. 

We first consider the case~$\tau \to 0$, so that~$(c,d)=(1,0)$. 
We use~\eqref{IdentityGZ}
with the value of~$w$ and~$\nu$
determined by the Pochhammer symbols in~\eqref{defZhol}. 
We end up with a double integral as in~\eqref{Wsplit} with linear combinations of terms 
\be \label{BkLikcomb}
{B}_{k+1} (-\nu) \, \text{Li}_{1-k}\Bigl(\frac{\widetilde{z}(x)\,z(y)^{-1}}{\zeta}\Bigr) \,-\,
{B}_{k+1} (1+ \nu) \, \text{Li}_{1-k}\Bigl(\frac{{z}(x)\,\widetilde{z}(y)^{-1}}{\zeta^{-1}} \Bigr) \,,
\ee
with~$\zeta = \zeta_{a}^{\ell_c}$, $a=1,\dots, 4$, 
and~$\nu=0$ for vector multiplets and~$\nu=-\frac14$ for hypermultiplets.
The large-$N$ analysis of the term~$\CW_2$ proceeds as before for every~$k$. The non-local 
interactions drop out---effectively identifying~$x$ and~$y$ in~\eqref{BkLikcomb}---and the arguments of the 
polylogarithms in the combination~\eqref{BkLikcomb}, writing~$\z=\rme^{2\pi \ii \lambda}$, 
become~$Z$ and~$1/Z$ with~$Z=\rme^{2 \pi \i(\delta v(x)\,-\,\lambda)}$.

Continuing onwards, we have, in the analog of the step~\eqref{Li2inttoL3}, that~$\text{Li}_{1-k}$ integrates to~$\text{Li}_{2-k}$. 
This gives an integral as in~\eqref{W2Li3} with the integrand consisting of linear combinations of the differences 
\be \label{Bkdifferences}
{B}_{k+1} (-\nu) \, \text{Li}_{2-k}(Z) \,-\, {B}_{k+1} (1+\nu) \, \text{Li}_{2-k}(1/Z) \,, \qquad \nu = 0 \,, -\frac14 \,,
\ee
with~$Z=\rme^{2 \pi \i(\delta v(x)\,-\,\lambda)}$ as above.
Now, using
\be \label{Bknuoneid}
{B}_{k} (1+\nu) \=(-1)^k {B}_{k} (-\nu)\,, \quad k>2 \,, \quad 0 \le -\nu \le 1 \,,
\ee
we see that the combination~\eqref{Bkdifferences} equals
\be
{B}_{k+1} (-\nu) \, \Bigl( \text{Li}_{2-k}(Z) \,-\, (-1)^{k+1} \text{Li}_{2-k}(1/Z) \Bigr) \,,
\ee
which vanishes due to the following classical 
identity satisfied by polylogarithms 
\be \label{Polylogid}
\text{Li}_{-r}(Z) + (-1)^r \, \text{Li}_{-r}(1/Z) \= 0\,, r \ge 1 \,.
\ee 

The case when~$\t \to \mathbb{Q}$ is not much different compared to when~$\t \to 0$. 
Consider~$\t \to d/c$,~$c,d \in \IZ$ with~$\text{gcd}(c,d)=1$. 
The asymptotic expansion~\eqref{psiexp} now contains~$c$ terms, and in lieu of~\eqref{Bkdifferences}
we have, with~$\nu = 0$,~$-\frac14$, 
\be \label{Bkdifferencescd}
\begin{split}
&\sum_{t=1}^c {B}_{k+1} \Bigl(1-\frac{t+\nu}{c} \Bigr) \, 
\text{Li}_{2-k} \Bigl( Z  \, \rme^{2 \pi \ii \tfrac{2d}{c}(t+\nu)} \Bigr) \\
&\,-\, \sum_{t'=1}^c {B}_{k+1} \Bigl(1+\frac{\nu+1-t'}{c} \Bigr) \, 
\text{Li}_{2-k} \Bigl( Z^{-1} \rme^{-2 \pi \ii \tfrac{2d}{c}(\nu+1-t')} \Bigr)  \,.
\end{split}
\ee
Now, for each value of~$t$ in the first term, there corresponds exactly one value of~$t'$
in the second given by
\be
t' \=  c + 1 - t \,.
\ee
Pairing up the two sums in this manner, we obtain that the coefficient of the term~$\ve^k$
is proportional to 
\be \label{Bkdifferencescd1}
\begin{split}
& \sum_{t=1}^c  \biggl(  {B}_{k+1} \Bigl(1-\frac{t+\nu}{c} \Bigr) \, 
\text{Li}_{2-k} \Bigl( Z  \, \rme^{2 \pi \ii \tfrac{2d}{c}(t+\nu)} \Bigr) 
\,-\,  {B}_{k+1} \Bigl(\frac{t+\nu}{c} \Bigr) \, 
\text{Li}_{2-k} \Bigl( Z^{-1} \rme^{-2 \pi \ii \tfrac{2d}{c}(\nu+t)} \Bigr) \biggr) \\
\= & \sum_{t=1}^c   {B}_{k+1} \Bigl(1-\frac{t+\nu}{c} \Bigr) \, \biggl( \, 
\text{Li}_{2-k} \Bigl( Z  \, \rme^{2 \pi \ii \tfrac{2d}{c}(t+\nu)} \Bigr) 
\,-\,  (-1)^{k+1} \, 
\text{Li}_{2-k} \Bigl( Z^{-1} \rme^{-2 \pi \ii \tfrac{2d}{c}(\nu+t)} \Bigr) \biggr) \\
 \= & 0 \,.
\end{split}
\ee
Here, the first equality follows from~\eqref{Bknuoneid} (and that~$0\le 1-(t+\nu)/c \le 1$ for the 
above values), and the second equality follows from the fact that each term in the sum vanishes 
due to~\eqref{Polylogid}. 

\smallskip

Finally, we turn to the analysis of~$\CW_3$. Now there is a significant difference compared to the leading term.
Recall from the discussion below~\eqref{W3largeN} that the~$1/N^\frac12$ suppression of~$\CW_3$ 
is compensated by the $N^\frac12$-growth of the  derivative of~$\text{Li}_2$. This phenomenon occured due to the fact that the derivative~$\text{Li}'_2(z)$
is large while~$\text{Li}_2(z)$ itself is small is due to the logarithmic non-analyticity of the function at~$z=0$.
For the terms~$\text{O}(\ve^{k})$, the  $\text{Li}_2$ is replaced by~$\text{Li}_{1-k}$, which, for~$k \ge 2$, are meromorphic 
functions and therefore do not have large derivative, as can be explicitly checked.  
We conclude that the~$\CW_3$ term is suppressed at large~$N$ for~$k \ge 2$. 
 
\smallskip

Collecting the above facts together, we reach the conclusion that 
the perturbation expansion of the large-$N$ index around any rational point
only contains terms multiplying~$\ve^{-1}$, $\ve^0$, and~$\ve$, and the~$\text{O}(\ve^k)$ terms vanish for all~$k \ge 2$.

\section{Black holes and supersymmetric solutions}
\label{sec:BlackHole}

In this section we move to the bulk gravity computation. We briefly review the Kerr--Newman-AdS black hole, then consider a related family of complex solutions to minimal gauged supergravity, and finally connect to the BPS black hole. Similar studies of these solutions have been made in \cite{Cassani:2019mms, Bobev:2019zmz}.

\subsection{Kerr--Newman-AdS black hole}

The Lorentzian action for Einstein--Maxwell theory with a cosmological constant is
\begin{equation}
\label{eq:LagrangianLorentzian}
\begin{split}
    S &\= \frac{1}{16\pi G_4} \int_{Y_4}\left( \bulkR + 6 - \bulkF^2 \right) \vol_{\bulkg} \, .
\end{split}
\end{equation}
Here $\bulkR$ is the Ricci scalar of the metric $\bulkg$, $\bulkF$ is the curvature of the $U(1)$ gauge field $\bulkA$, and $-3$ is the cosmological constant. A black hole solution with rotation and electric charge has been known for a long time \cite{Carter:1968ks}. In a frame that is non-rotating at infinity, which is more immediate for uses in AdS/CFT \cite{Caldarelli:1999xj, Gibbons:2004ai}, the solution is
\begin{equation}
\label{eq:Metric_Four_Dimensional_Minimal_NonRotating}
\begin{split}
\rd s^2 &\= - \frac{\Delta_r\Delta_\Theta}{B\Xi^2}  \rd \nonrotatingt^2 + \sin^2\Theta \, B \left( \rd\nonrotatingphi + a \Delta_\Theta \frac{\Delta_r - (1+r^2)(r^2+a^2)}{B W \Xi^2} \rd \nonrotatingt\right)^2  \\
& \qquad + W \left( \frac{\rd r^2}{\Delta_r} + \frac{\rd \Theta^2}{\Delta_\Theta}  \right) \, , \\[10pt]
\bulkA &\= \frac{ m r \sinh \delta}{W\Xi}\Big( \Delta_\Theta \, \rd \nonrotatingt - a \sin^2\Theta \, \rd \nonrotatingphi \Big) + \gamma \, \rd \nonrotatingt  \, .
\end{split}
\end{equation}
Here $\gamma$ is a constant, $\Theta\in [0,\pi), \nonrotatingphi\in [0,2\pi)$, and
\begin{equation}
\label{eq:Quantites_Four_Dimensional_Minimal_Rotating}
\begin{split}
& \Delta_r \= (r^2 + a^2)(1 + r^2) - 2 m r \cosh \delta + m^2 \sinh^2 \delta \, , \\
& \Delta_\Theta \= 1 -  a^2 \, \cos^2 \Theta \, , \qquad W \= r^2 + a^2 \cos^2 \Theta \, , \qquad \Xi \= 1 - a^2 \, , \\
& B \; \equiv \; \frac{\Delta_\Theta(r^2+a^2)^2 - a^2 \sin^2\Theta \, \Delta_r}{W \Xi^2} \, .
\end{split}
\end{equation}

\medskip

The black hole has an outer horizon at $r=r_+$, the largest positive root of $\Delta_r$. On a slice of constant $\nonrotatingt$ and $r$ outside the horizon, the solution is topologically a sphere, and it is easy to see that it closes off smoothly at $\Theta=0,\pi$ with $\nonrotatingphi\sim \nonrotatingphi + 2\pi$. We then perform a Wick rotation $\nonrotatingt=-\ii \nonrotatingtau$ and look near the horizon in geodesic coordinates. The space caps off smoothly at the horizon only if we identify
\begin{equation}
\label{eq:IdentificationsNonRotating}
    (\nonrotatingtau, \nonrotatingphi) \; \sim \; (\nonrotatingtau, \nonrotatingphi + 2\pi) \; \sim \; (\nonrotatingtau + \beta, \nonrotatingphi - \ii \Omega \beta) \, ,
\end{equation}
where the temperature and angular velocity of the horizon are
\begin{equation}
    \label{eq:HorizonQuantities}
\begin{split}
    \beta \= 4\pi \frac{a^2 + r_+^2}{\Delta'_r(r_+)} \, , \qquad \Omega \= a\frac{1 + r_+^2}{a^2 + r_+^2} \, .
\end{split}
\end{equation}
The metric \eqref{eq:Metric_Four_Dimensional_Minimal_NonRotating}, once Wick rotated, is real if we take $a$ to be pure imaginary and $\delta, m$ to be real, whereas the gauge field $\bulkA$ is pure imaginary provided $\gamma$ is real. This also makes $\Omega$ above pure imaginary, and the topology is that of the product of a disc and a 2-sphere~\cite{Gibbons:1976ue}. The metric \eqref{eq:Metric_Four_Dimensional_Minimal_NonRotating} has two obvious commuting Killing vectors $\partial_{\nonrotatingt}$,  
$\partial_{\nonrotatingphi}$, and the surface $\{ r=r_+\}$ is a Killing horizon of the linear combination 
\begin{equation}
\label{eq:KillingGenerator}
    V \= \frac{\partial \ }{\partial \nonrotatingt} + \Omega \frac{\partial \ }{\partial \nonrotatingphi } \, ,
\end{equation}
The electrostatic potential is\footnote{More generally, we should only pick the $\Theta$-independent part of $\iota_V \bulkA |_{r\to \infty}$.}
\begin{equation}
    \begin{split}
    \Phi_e & \defeq \iota_V \bulkA |_{r=r_+} - \iota_V \bulkA |_{r\to \infty} \\
    &\= m\sinh \delta \, \frac{r_+}{a^2 + r_+^2} \, .
\end{split}
\end{equation}
In Euclidean signature, the horizon is a bolt for $V$, and regularity of the gauge field on the disc requires that $\iota_V \bulkA$ vanishes at the origin $r=r_+$. This fixes the gauge choice in \eqref{eq:Metric_Four_Dimensional_Minimal_NonRotating} to be $\gamma = -\Phi_e$. The Bekenstein--Hawking entropy of the horizon is
\begin{equation}
    S \= \frac{\pi}{G_4} \frac{r_+^2+a^2}{1-a^2} \, .
\end{equation}

\medskip

That the black hole is asymptotically anti-de Sitter can be shown by considering the region $r\to \infty$ and applying the following coordinate change in the asymptotic region \cite{Hawking:1998kw}
\begin{equation}
    \label{eq:ChangeRotatingFrame}
\begin{aligned}
    \frac{\cos\bdrytheta}{z} &\= r \cos\Theta \, , &\qquad z^{-2} &\= \frac{r^2\Delta_\Theta + a^2\sin^2\Theta}{\Xi} \, ,
\end{aligned}
\end{equation}
obtaining, as~$z \to 0$, 
\begin{equation}
\label{eq:Boundarybehavior}
\begin{split}
    \rd s^2 &\; \sim \; \frac{\rd z^2}{z^2} + \frac{1}{z^2} \left( - \rd \nonrotatingt^2 + \left( \rd \bdrytheta^2 + \sin^2\bdrytheta \, \rd\nonrotatingphi^2 \right) \right) \,, \\
    \bulkA &\; \sim \; - \Phi_e \, \rd\nonrotatingt \, .
\end{split}
\end{equation}
The boundary metric in Euclidean signature is not just the round $S^1\times S^2$, due to the identifications \eqref{eq:IdentificationsNonRotating}. To see this explicitly, define $\bdryphi = \nonrotatingphi + \ii \Omega \nonrotatingtau$, so that the boundary metric has the fibered form
\begin{equation}
\label{eq:BdryLineElement}
    \rd s^2_{\rm bdry} \= \rd \bdrytau^2 + \left( \rd \bdrytheta^2 + \sin^2\bdrytheta \, (\rd\bdryphi - \ii \Omega \, \rd\bdrytau)^2 \right) \, ,
\end{equation}
but now with the identifications
\begin{equation}
\label{eq:IdentificationsBoundary}
    (\bdrytau, \bdryphi) \; \sim \; (\bdrytau, \bdryphi + 2\pi) \; \sim \; (\bdrytau + \beta, \bdryphi) \, ,
\end{equation}
This metric is real for pure imaginary~$a$, as consistent with the comments below~\eqref{eq:HorizonQuantities}. It is clear that the metric \eqref{eq:BdryLineElement} and the boundary gauge field $\CA \sim \ii \Phi_e \, \rd \bdrytau$ match the metric of the Euclidean background over which the index is computed as a functional integral \eqref{eq:FiberedBdryMetric}, and the background gauge field coupled to the $\mf{u}(1)_R$ symmetry generated by $H_1$, $A^{(H_1)} = \ii \Phi^{(H_1)} \, \rd\bdrytau$ \eqref{eq:FibrationParameters_AllEqual}.

\medskip

In order to compute the conserved charges, we can use the standard methods of holographic renormalization (see \cite{Papadimitriou:2005ii} for a review). We introduce a cutoff at $z = \delta \geq 0$ and consider the geometry of the hypersurface $\partial Y_\delta = \{z=\delta\}\cap Y_4 \cong M_3$, 
with induced metric~$h$. In order to make the problem well-defined and remove the divergences, we need to add to the action the Gibbons--Hawking--York term and the counterterms, so  that the renormalized on-shell action is
\begin{equation}
\label{eq:OnShellActionBH}
    I \= \lim_{\delta\to 0} \left[ S + \frac{1}{8\pi G_4} \int_{\partial Y_\delta} \left( K - 2 - \frac{1}{2}R \right) \vol_h \right] \, ,
\end{equation}
where $K$ is the trace of the extrinsic curvature of the hypersurface.
We then define the holographic stress-energy tensor and electric current
\begin{equation}
    \langle T_{ij} \rangle \= - \frac{2}{\sqrt{-\bdryg}} \frac{\delta I}{\delta \bdryg^{ij}} \, , \qquad \langle j^i \rangle \= \frac{1}{\sqrt{-\bdryg}} \frac{\delta I}{\delta A_i} \, , 
\end{equation}
where $i,j$ are labels for the boundary coordinates, and $\bdryg, \bdryA$ are, respectively, the boundary metric and gauge field given by \eqref{eq:BdryLineElement} and $A = - \Phi_e \, \rd t$. These quantities satisfy the following equations
\begin{equation}
    \nabla^i \langle T_{ij} \rangle \= \bdryF_{ji} \langle j^i \rangle \, , \qquad \nabla_i \langle j^i \rangle \= 0 \, , \qquad \langle T^i_{\ph{i}i} \rangle \= 0 \, ,
\end{equation}
and for any boundary vector $K^j$ generating a symmetry (that is $\CL_K\bdryg = 0$ and $\CL_K \bdryA=0$) we can construct a conserved current
\begin{equation}
    \nabla_i \left( ( \langle T^i_{\ph{i}j} \rangle + \langle j^i \rangle \bdryA_j ) K^j \right) \= 0 \, ,
\end{equation}
where $\nabla$ is the Levi-Civita connection of the boundary metric $\bdryg$. Let $C$ be a surface of constant $\nonrotatingt$, and $u_i$ be the future-directed unit normal to $C$. Then, the conserved charge associated to $K$ is computed by
\begin{equation}
    Q[K] = \int_{C\cap M_3} u_i ( \langle T^i_{\ph{i}j} \rangle + \langle j^i \rangle \bdryA_j ) K^j \, \vol_{C\cap M_3} \, .
\end{equation}
Notice that this definition is not invariant under gauge transformations of $A$ (as stressed in \cite{Ferrero:2020twa, Cassani:2021dwa}), but it still gives a conserved charge provided the gauge transformed $\bdryA'$ still satisfies $\CL_K\bdryA'=0$.  
The angular momentum is associated to $K=-\partial_{\nonrotatingphi}$, so in our gauge
\begin{equation}
    J \= -\int_{C\cap M_3} u_i \langle T^i_{\ph{i}\nonrotatingphi}\rangle \, \vol_{C\cap M_3} \= a \frac{m \cosh \delta}{G_4 \Xi^2} \, .
\end{equation}
We define the electric charge as
\begin{equation}
    Q_e \= \int_{C\cap M_3} u_i \langle j^i \rangle \, \vol_{C\cap M_3} \= \frac{1}{4\pi G_4}\int_{C\cap M_3} *_4 \CF \= \frac{ m\sinh \delta}{G_4 \Xi} \, .
\end{equation}
Finally, the energy is associated to $K=\partial_{\nonrotatingt}$, and in our gauge choice
\begin{equation}
\label{eq:BlackHoleEnergyMinimal}
\begin{split}
    E' &\= \int_{C\cap M_3} u_i \langle T^i_{\ph{i}\nonrotatingt} \rangle \, \vol_{C\cap M_3} - \Phi_e \int_{C\cap M_3} u_i \langle j^i \rangle \, \vol_{C\cap M_3} \= \frac{m \cosh \delta}{G_4 \Xi^2} - \Phi_e Q_e \\
    &\= E - \Phi_e Q_e \, .
\end{split}
\end{equation}
Here $E=m\cosh\delta/G_4(1-a^2)$ is the value of the energy in the gauge~$A=0$.

Moving to Euclidean signature, we can compute the holographically renormalized on-shell action, obtaining
\begin{equation}
\label{eq:OnShellAction_NonSUSY}
I \= \frac{\beta}{2G_4 \Xi}\left[ m \cosh \delta - r_+ (r_+^2 + a^2) - \frac{r_+ m^2 \sinh^2 \delta}{r_+^2 + a^2} \right] \, .
\end{equation}
This action obeys the quantum statistical relation \cite{Papadimitriou:2005ii}
\begin{equation}
\begin{split}
    I &\= - S + \beta ( Q[V] - \Phi_h \, Q_e ) \\
    &\= - S + \beta ( Q[\partial_{\nonrotatingt}] - \Omega \, Q[-\partial_{\nonrotatingphi}] - \Phi_h \, Q_e )
\end{split}
\end{equation}
where $\Phi_h = \iota_V \CA|_{r=r_+}$ is by definition a constant. Even though each term in the relation above is separately not gauge invariant, the overall relation is invariant under gauge transformations of $\bdryA$. Concretely, it reduces to the canonical form in the $A=0$ gauge
\begin{equation}
\label{eq:QuantumStatisticalRelation}
    I \= - S + \beta (E - \Omega \, J - \Phi_e \, Q_e) \, .
\end{equation}

Moreover, varying $\delta, a, m$, we find that the first law of thermodynamics holds. Written as above in terms of holographic conserved charges it reads \cite{Papadimitriou:2005ii}
\begin{equation}
    \rd Q[\partial_{\nonrotatingt}] = \beta^{-1} \, \rd S + \Omega \, \rd Q[\partial_{\nonrotatingphi}] + \Phi_h \, \rd Q_e
\end{equation}
or concretely
\begin{equation}
\label{eq:FirstLawThermodynamics}
    \rd E \= \beta^{-1}\rd S + \Omega \, \rd J + \Phi_e \, \rd Q_e \, .
\end{equation}
It follows from combining \eqref{eq:QuantumStatisticalRelation} and \eqref{eq:FirstLawThermodynamics} that $\beta$, $\Omega$, $\Phi_e$ are the chemical potentials conjugate to the conserved charges, since
\begin{equation}
\label{eq:DerivativesGibbsFreeEnergy}
    E \= \frac{\partial I }{\partial\beta}\bigg|_{\beta\Omega, \beta\Phi_e} \, , \qquad  J \= - \frac{1}{\beta}\frac{\partial I }{\partial\Omega}\bigg|_{\beta, \Omega_e} \, , \qquad  Q_e \= - \frac{1}{\beta}\frac{\partial I}{\partial\Phi_e}\bigg|_{\beta, \Omega_e} \, .
\end{equation}
This shows that the on-shell action $I=I(\beta, \Omega, \Phi_e)$ is minus the logarithm of the grand canonical partition function, the Gibbs free energy.

\subsection{Supersymmetry}
\label{subsec:SUSYBlackHole}

The Einstein--Maxwell action \eqref{eq:LagrangianLorentzian} describes the bosonic sector of four-dimensional minimal gauged supergravity \cite{Freedman:1976aw}. A solution to the equations of motion coming from \eqref{eq:LagrangianLorentzian} is supersymmetric provided there is a non-zero Dirac spinor $\epsilon$ satisfying the equation
\begin{equation}
\label{eq:SUGRAKillingSpinorEquation}
    \left( \nabla_\mu - \ii \bulkA_\mu + \frac{1}{2}\Gamma_\mu + \frac{\ii}{4} \bulkF_{\nu\rho}\Gamma^{\nu\rho}\Gamma_\mu \right) \epsilon \= 0 \, .
\end{equation}
The condition of supersymmetry on the solution~\eqref{eq:Metric_Four_Dimensional_Minimal_NonRotating} with parameters~$(a, \delta, m)$
implies~\cite{Kostelecky:1995ei, Caldarelli:1998hg}
\begin{equation}
\label{eq:SUSYCondition}
    E \= J + Q_e \qquad \Leftrightarrow \qquad a \= \coth \delta - 1 \, .
\end{equation}
Thus, the family of supersymmetric solutions can be parametrized by the two parameters $\delta, m$. Supersymmetry and global regularity requires that in presence of non-zero electric charge the angular momentum must be non-zero \cite{Caldarelli:1998hg}.\footnote{\label{footnote:MinimalMagnetic}The conclusion is different in presence of a magnetic charge \cite{Caldarelli:1998hg, Hristov:2011ye}.}

Imposing \eqref{eq:SUSYCondition} reduces $\Delta_r$ in~\eqref{eq:Quantites_Four_Dimensional_Minimal_Rotating} to a sum of squares
\begin{equation}
\label{eq:DeltarSUSY}
\begin{split}
\Delta_r|_{\rm SUSY} &\= \left( r^2- \coth \delta + 1 \right)^2 + \coth^2 \delta \left( r - m \frac{\sinh^2 \delta}{\cosh \delta} \right)^2 \, .
\end{split}
\end{equation}
Assuming reality of all parameters, both squares should vanish at the horizon, which fixes the value of the horizon radius $r_*$ and $m$:
\begin{equation}
\label{eq:BPSHorizonRadius}
r_*^2 \= \coth \delta - 1 \, , \qquad m^2 \= \frac{\cosh^2 \delta}{\rme^{\delta}\sinh^5 \delta} \,,
\end{equation}
leaving only one free parameter~$\delta$.
It is now easy to check that~$\Delta'_r|_{\rm SUSY}(r_*)=0$, 
that is,~supersymmetry and regularity of the Lorentzian metric imply extremality.

In order to define the gravitational partition function and reproduce the large-$N$ behavior of the supersymmetric index \eqref{eq:ABJMSCI_v3}, we need to consider supersymmetric solutions connected to the Euclidean solutions.
As we observed around \eqref{eq:HorizonQuantities}, the metric is real after Wick rotation provided we choose $a$ to be pure imaginary. Clearly, this can only be compatible with the supersymmetry condition \eqref{eq:SUSYCondition} if $\delta$ is complex, but this is itself incompatible with a real Euclidean metric. 
Therefore, we conclude that the Wick rotation of a real supersymmetric Lorentzian metric of the form~\eqref{eq:Metric_Four_Dimensional_Minimal_NonRotating} is generically complex.\footnote{The bulk Killing spinor equation \eqref{eq:SUGRAKillingSpinorEquation} is analytic in the supergravity fields, so it is still satisfied by the Wick-rotated solution.} 

We shall therefore focus on the family of complex metrics obtained by imposing the supersymmetry condition \eqref{eq:SUSYCondition} without requiring reality of the metric and gauge field. These solutions arise from extending to complex parameters the Euclidean ``black hole'' solution with topology~$\R^2\times S^2$. This approach was first suggested in five dimensions in \cite{Cabo-Bizet:2018ehj} and elaborated in other dimensions (including four) in \cite{Cassani:2019mms}.
Here, we have a two-parameter family, and it is convenient to trade the parameter~$\delta$ for~$r_*$ using~\eqref{eq:BPSHorizonRadius}, and~$m$ for the largest root~$r_+$ of~\eqref{eq:DeltarSUSY}, i.e.,
\begin{equation}
\label{eq:mSUSYBranches}
    m \= \frac{1}{\sinh^2 \delta} \left[ r_+ \cosh \delta \pm \ii \left( \sinh \delta (1 + r_+^2) - \cosh \delta \right) \right] \, .
\end{equation}

\medskip

The thermodynamic quantities of the Euclidean supersymmetric solutions are generically complex. The chemical potentials 
\begin{equation}
\begin{aligned}
\beta &\= \pm 2\pi \ii \frac{r_+^2 + r_*^4}{(r_+^2 - r_*^2)( 1 \pm 2\ii r_+ + r_*^2)} \, , \\
\Omega &\= r_*^2 \frac{1 + r_+^2}{r_+^2 + r_*^4} \, , \\
\Phi_e &\= r_+ \frac{r_+ r_*^2  + r_+ \pm \ii (r_+ - r_*)(r_+ + r_*)}{r_+^2 + r_*^4} \, ,
\end{aligned}
\end{equation}
and the conjugate charges
\begin{equation}
\begin{aligned}
E &\= \frac{r_+ r_*^2 + r_+ \pm \ii (r_+ - r_*)(r_+ + r_*)}{G_4(1 - r_*^4)(1 - r_*^2)} \, , \\
J &\= r_*^2\frac{ r_+ r_*^2 + r_+ \pm \ii (r_+ - r_*)(r_+ + r_*)}{G_4(1 - r_*^4)(1 - r_*^2)} \, , \\
Q_e &\= \frac{ r_+ r_*^2 + r_+ \pm \ii (r_+ - r_*)(r_+ + r_*)}{G_4(1 - r_*^4)} \, 
\end{aligned}
\end{equation}
can be computed directly or read off by imposing \eqref{eq:SUSYCondition} in the corresponding expressions for Wick-rotated non-supersymmetric solutions.
Here the different sign choices refer to the two branches of solutions to \eqref{eq:mSUSYBranches}. The conserved charges satisfy the supersymmetry condition~\eqref{eq:SUSYCondition}
by construction, and we observe that the chemical potentials also satisfy the constraint
\begin{equation}
\label{eq:ChemicalPotentialsConstraint}
\beta \left( 1 - 2 \Phi_e + \Omega \right) = \mp 2\pi\ii \, .
\end{equation}

If in addition to supersymmetry we impose extremality, we restrict to the Wick rotation of the supersymmetric Lorentzian extremal solution discussed around \eqref{eq:DeltarSUSY}, which we call the BPS locus and indicate by an underscript $*$, as in \cite{Cabo-Bizet:2018ehj}. The charges of the extremal solution
\begin{equation}
\label{eq:BPSQuantities}
E_* \= \frac{r_*}{G_4(1 - r_*^2)^2} \, , \qquad J_* \= \frac{r_*^3}{G_4(1 - r_*^2)^2} \, , \qquad Q_{e*} \= \frac{r_*}{G_4(1 - r_*^2)} \,
\end{equation}
are real, and the extremal chemical potentials 
\begin{equation}
\label{eq:BPSQuantities2}
\Omega_* \= 1 \, , \qquad \Phi_{e *} \= 1 \,  \\
\end{equation}
are real and fixed to constant values independent of the charges.
Clearly the constraint \eqref{eq:ChemicalPotentialsConstraint} stops being meaningful. It will be useful instead to define the ``reduced chemical potentials'' for the supersymmetric solutions
\begin{equation}
\label{eq:ReducedChemicalPotentials}
    \tau_g \; \equiv \; \beta\frac{\Omega - \Omega_*}{2\pi\ii} \, , \qquad \j_g \; \equiv \; \beta \frac{\Phi_e - \Phi_{e*}}{2\pi\ii} \, ,
\end{equation}
which by construction satisfy the constraint
\begin{equation}
\label{eq:ReducedChemicalPotentialsConstraint}
    \tau_g - 2 \j_g \= \mp 1 \, .
\end{equation}
This allows us to write the quantum statistical relation \eqref{eq:QuantumStatisticalRelation} as
\begin{equation}
\label{eq:QuantumStatisticalRelationSUSY}
\begin{split}
    I &\= \beta (E - J - Q_e) - S - 2\pi\ii \tau_g \, J - 2\pi\ii\j_g \, Q_e \\ 
    \Rightarrow \quad I|_{\rm SUSY} &\= - S - 2\pi\ii \tau_g \, J - 2\pi\ii \j_g \, Q_e \, ,
\end{split}
\end{equation}
the second equation following from \eqref{eq:SUSYCondition}. In the following, we shall approach the BPS locus via a limiting process taking $\beta\to \infty$, $\Omega\to \Omega_*$, $\Phi_e \to \Phi_{e*}$ keeping $\tau_g$, $\varphi_g$ fixed to a complex value.

\medskip

As shown above, the constraint \eqref{eq:ChemicalPotentialsConstraint} follows from imposing supersymmetry on the parameters of the solution. It should be the case that it follows directly from the requirement that the bulk Killing spinor satisfying \eqref{eq:SUGRAKillingSpinorEquation} is anti-periodic when transported around the circle generated by $V$ \cite{Cassani:2019mms}. The antiperiodicity of smooth Killing spinors is consistent with the topological statement that in Euclidean solutions the circle generated by $V$ shrinks in the bulk. 
It was shown in \cite{Cabo-Bizet:2018ehj} that this is indeed the case in the asymptotically AdS$_5$ context. 
However, to the best of our knowledge, an explicit expression for the bulk Killing spinor of the four-dimensional black hole is not known (see \cite{Klemm:2013eca} for a discussion of the supersymmetry of the metric). Nonetheless, we can draw some conclusions by looking near the conformal boundary (see also \cite{Bobev:2019zmz}). 
As we show in Appendix~\ref{app:KSE}, 
the anti-periodicity of the boundary Killing spinor around the 
circle that is contractible in the bulk leads to the following condition,
\begin{equation}
\label{eq:Constraint_AllEqual_v2}
    \beta \left( 1 - 2 \Phi_e + \Omega \right) \= 2\pi \ii n_0 \, ,
\end{equation}
with \textit{odd} $n_0$, or equivalently
\begin{equation}
\label{eq:ConstraintGravitationalPotentials_v2}
    \tau_g - 2\varphi_g \= n_0 \, .
\end{equation}
It should be remarked, though, that at this level the topological argument is a formal statement, since the non-extremal supersymmetric solutions are complex. In fact, the bulk solution imposes a stronger condition, as the chemical potentials satisfy \eqref{eq:Constraint_AllEqual_v2} with $n_0=\mp 1$ depending on the solution of \eqref{eq:mSUSYBranches} chosen (see \eqref{eq:ChemicalPotentialsConstraint}).

\medskip

One of the main interests in the family of complex solutions is that they allow us to define a regularization of the on-shell action of the extremal supersymmetric black hole. The Wick-rotated extremal supersymmetric black hole has an infinite throat due to an $\mathbb{H}^2$ factor in the metric, whereas the family described above is originated from an extension to complex parameters of the Wick-rotation of the non-extremal supersymmetric black hole solution, which has the topology of the product of a disc and a $2$-sphere. Thus, the on-shell action of the BPS solution can be \textit{defined} as the limit of the on-shell action of the solutions in the complex family, and this limit is well-defined \cite{Cassani:2019mms}. One can compute the on-shell action of the solutions by direct application of the holographic renormalization, as done above for the non-supersymmetric case. 
In fact, it is possible to derive the supersymmetric result in a shorter and elegant manner extending the methods in \cite{BenettiGenolini:2019jdz}, as done in \cite{Cassani:2021dwa} for accelerating black holes.

Notice that the boundary Killing vector $\xi^i = \overline{\tilde{\chi}_E}\gamma^i \chi_E$, computed from the boundary Killing spinor~\eqref{eq:BoundaryKS}, is
\begin{equation}
\label{eq:BulkKillingVector}
    \xi \= 2u\tilde{u} \left[ \frac{\partial \ }{\partial \bdrytau} + \ii \left( \Omega - 1 \right) \frac{\partial \ }{\partial \bdryphi} \right]
\end{equation}
and can thus be extended to a bulk Killing vector, which is constructed as a  bilinear of the bulk Killing spinor. 
The on-shell action of any smooth supersymmetric solution of minimal gauged supergravity with real metric and gauge field can be expressed in terms of topological data of the circle action generated by the bulk ``supersymmetric'' Killing vector field on the spacetime \cite{BenettiGenolini:2019jdz}. In this case, we are considering a family of solutions outside the reality assumptions, and yet the formula found in \cite{BenettiGenolini:2019jdz} remarkably still holds, as we now show.

Recall that the topology of the Wick-rotated black hole is $\R^2\times S^2$, parametrized by $(r,\bdrytau)-(\Theta,\bdryphi)$. In order to find the generators of the rotations with unitary weight on the two factors, we rescale the coordinates
\begin{equation}
    \varphi_1 \= \frac{2\pi}{\beta}\bdrytau \, , \qquad \varphi_2 \= \bdryphi \,, 
    \quad \qquad \varphi_{1,2} \;\sim \;\varphi_{1,2} + 2 \pi \,.
\end{equation}
In these coordinates, the Killing vector has the form $\xi = b_1 \partial_{\varphi_1}+ b_2 \partial_{\varphi_2}$ with weights
\begin{equation}
    b_1 \= 2u\tilde{u} \frac{2\pi}{\beta} \, , \qquad b_2 \= - 2 u\tilde{u} \frac{2\pi \tau_g}{\beta} \, .
\end{equation}
The Killing vector has isolated fixed points at the North and South pole of the $S^2$ factor, so the on-shell action of the solution is given by the sum of the contributions
\begin{equation}
    I|_{\rm SUSY} \= \frac{\pi}{2 G_4}\sum_{\rm nuts_\mp} \pm \frac{(b_1\pm b_2)^2}{4b_1b_2} \, ,
\end{equation}
where the label $\pm$ for the nut corresponds to the chirality of the bulk spinor there. In our case, a solution in the $\pm$ branch of solutions to \eqref{eq:mSUSYBranches} has nuts with $\pm$ chirality. Substituting in the formula the values of the weights and using the constraint \eqref{eq:ReducedChemicalPotentialsConstraint}, we find
\begin{equation}
\label{eq:OnShellActionSUSY}
    I|_{\rm SUSY} \= \pm \frac{\pi}{G_4}\frac{\j^2_g}{\tau_g} \, ,
\end{equation}
where again the $\pm$ refers to the branch of the solution. As mentioned above, this result is consistent with what is obtained by direct substitution of \eqref{eq:SUSYCondition} in \eqref{eq:OnShellActionBH}.\footnote{In four-dimensional minimal gauged supergravity there is no need to add finite counterterms in order for holographic renormalization to be consistent with supersymmetry \cite{Genolini:2016ecx}.}
The final result~\eqref{eq:OnShellActionSUSY} for the action in terms of the variables $\tau_g, \j_g$ is independent of~$\b$ and therefore the limit~$\b \to \infty$ is smooth. This is identified as the regulated on-shell action of the BPS solution.

The above result is also consistent with the results of \cite{BenettiGenolini:2019jdz}: the on-shell action should only depend on topological data of the circle action of $\xi$, which is well-defined for the complex solutions and it is independent of the deformation parameter $\beta$. Even more concretely, observe that the Killing vector \eqref{eq:BulkKillingVector}, after a Wick rotation back to Lorentzian, coincides with the null generator of the horizon of the BPS black hole (since it has the form $\partial_{\nonrotatingt} + \Omega_* \, \partial_{\nonrotatingphi}$, to be compared with~\eqref{eq:KillingGenerator}). The derivation above and the analogous result for black holes with orbifold horizons~\cite{Cassani:2021dwa} suggest that it should be possible to extend the proof of~\cite{BenettiGenolini:2019jdz} to complex solutions, finding a general way of regularizing the on-shell action for extremal black holes in minimal gauged supergravity.\footnote{Notice that for static magnetically charged black holes with horizon homeomorphic to a Riemann surface of genus higher than 1, an analogous derivation within a family of smooth \textit{real} supersymmetric solutions has been proposed in \cite{BenettiGenolini:2019jdz}.}

\medskip

Extending the principles of Euclidean quantum gravity to the family of supersymmetric complex solutions, we can obtain the entropy of the BPS black hole in the microcanonical ensemble by taking the Legendre transform of the on-shell action $I|_{\rm SUSY}(\tau_g, \j_g)$ (see also \cite{Choi:2018fdc, Cassani:2019mms, Bobev:2019zmz}). In the following we briefly review it highlighting the assumptions made at each stage.

First, we observe that $I|_{\rm SUSY}$ in \eqref{eq:OnShellActionSUSY} is a homogeneous function of degree one, so its Legendre transform vanishes unless we impose the non-homogeneous constraint \eqref{eq:ReducedChemicalPotentialsConstraint}. In order to find the constrained Legendre transform of $I|_{\rm SUSY}(\j_g, \omega_g)$, we should extremize 
\begin{equation}
    f(\tau_g, \j_g, \Lambda) \= - I|_{\rm SUSY}(\tau_g, \j_g) - 2\pi\ii\tau_g \, J - 2\pi\ii\j_g \, Q_e + \Lambda \left( \tau_g - 2 \j_g - n_0 \right) \, ,
\end{equation}
where $n_0 = \mp 1$ represents the branch of gravitational solutions arising from \eqref{eq:mSUSYBranches}.
We notice that at the critical point $(\tau_{g*}, \j_{g*}, \Lambda_*)$ the following holds
\begin{equation}
\begin{split}
    f(\tau_{g*}, \j_{g*}, \Lambda_*) &\= - I|_{\rm SUSY}(\tau_{g*}, \j_{g*}) + \frac{\partial I|_{\rm SUSY}}{\partial \tau_g}(\tau_{g*}, \j_{g*}) \tau_{g*} + \frac{\partial I|_{\rm SUSY}}{\partial \j_g}(\tau_{g*}, \j_{g*}) \j_{g*} \\
    & \quad \ \ - n_0 \Lambda_*  \, ,
\end{split}
\end{equation}
and together with Euler's theorem for $I|_{\rm SUSY}(\tau_g, \j_g)$, this leads to the (implicit) Legendre transform
\begin{equation}
    \tilde{f}(J_*, Q_{e*}) \= - n_0 \Lambda_*(J_*, Q_{e*}) \, .
\end{equation}
Concretely, we can combine the critical point equations into the quadratic equation for $\Lambda_*(J_*, Q_{e*})$
\begin{equation}
\label{eq:QuadraticEquationLambda}
    0 \= \Lambda_*^2 + \left( 2\pi\ii Q_{e*} - n_0 \frac{\pi}{G_4} \right) \Lambda_* + \left( -\pi^2 Q_{e*}^2 + n_0 \frac{2\pi^2\ii}{G_4} J_* \right) \, .
\end{equation}
We then impose the constraints 
\begin{equation}
    J_*, Q_{e*} \in \R \, , \qquad \Lambda_*(J_*, Q_{e*}) \in \R \, .
\end{equation}
The first one says that the charges are real. 
The second one leads to the entropy being real, as we see shortly.
With these constraints, we can write the real and imaginary parts of equation~\eqref{eq:QuadraticEquationLambda} separately, finding
\begin{equation}
    \Lambda_* \= - n_0 \frac{\pi J_*}{G_4 Q_{e*} } \, , \qquad J_* \= \frac{Q_{e*}}{2} \left( - n_0 \sigma_1  \sqrt{1 + 4 G_4 Q_*^2} - 1 \right) \, ,
\end{equation}
where $\sigma_1 =\pm 1$ represents the additional sign choice for the two solutions of the quadratic equation~\eqref{eq:QuadraticEquationLambda}. Substituting in $\tilde{f}$ gives
\begin{equation}
    \tilde{f}(J_*, Q_{e*}) \= \frac{\pi J_*}{G_4 Q_{e*}} \= \frac{\pi}{2 G_4} \left( 
    - n_0 \sigma_1  \sqrt{1 + 4 G_4^2 Q_{e*}^2} - 1 \right) \, .
\end{equation}
The  requirement that the entropy~$\tilde{f}(J_*, Q_{e*})$ should be positive implies that $\sigma_1= - n_0$, so that
\begin{equation}
    \tilde{f}(J_*, Q_{e*}) \= \frac{\pi J_*}{G_4 Q_{e*}} \= \frac{\pi}{2 G_4} \left( \sqrt{1 + 4 G_4^2 Q_{e*}^2} - 1 \right) \, .
\end{equation}
This corresponds to the Bekenstein--Hawking entropy of the BPS black hole
\begin{equation}
S_* \= \frac{\pi J_*}{ G_4 Q_{e *}} \, , 
\end{equation}
and to the non-linear constraint between the extremal charges imposed by supersymmetry
\begin{equation}
J_* \= \frac{Q_{e*}}{2} \left( \sqrt{1 + 4 G_4^2 Q_{e*}^2} - 1 \right) \, .
\end{equation}

\section{A family of saddles in AdS}
\label{sec:Uplift}

In this section we introduce the gravitational dual to the generalised Cardy limits of Section~\ref{sec:RationalPoints}, where~$\tau$ approaches a rational point. This gravitational construction is modelled after the analogous one for five-dimensional black holes dual to $\CN=4$ SYM introduced in \cite{Aharony:2021zkr}.

\subsection{Uplift to eleven dimensions}

In order to appeal to the AdS/CFT dictionary and compare the gravitational results to the field theory computation, we need to embed the four-dimensional gravitational solution $(Y_4, \bulkg(Y_4), \bulkA)$ in eleven-dimensional supergravity. In particular, in order to match the field theory limit taken in Section~\ref{sec:RationalPoints}, we shall uplift the four-dimensional minimal gauged supergravity on $S^7$ to a solution of eleven-dimensional supergravity $(Y_{11}, \bulkg(Y_{11}), \CC)$. The eleven-dimensional metric and gauge field can be locally written as a fibration \cite{Gauntlett:2007ma}
\begin{equation}
\label{eq:M2Ansatz}
\begin{split}
\bulkg(Y_{11})  &\= \bulkg(Y_4) +  4 \left[ \left( \rd\bdrypsi + \sigma + \frac{1}{2} \bulkA \right)^2 + \bulkg(\mathbb{CP}^3) \right]  \, , \\
\rd\CC &\= 3 \, \vol(Y_4) - 4 *_{4} \bulkF \wedge J  \, .
\end{split}
\end{equation}
Here we have used the local form of the metric on a seven-dimensional Sasaki--Einstein space (such as $S^7$): $\partial_{\bdrypsi}$ is the Reeb vector, $J$ is the K\"ahler form on the K\"ahler--Einstein base $\mathbb{CP}^3$, and is such that $\rd\sigma=2J$. Moreover, $\bulkg(\mathbb{CP}^3)$ is the Fubini--Study metric normalized with $\Ric(\bulkg(\mathbb{CP}^3)) = 6 \, \bulkg(\mathbb{CP}^3)$, and the volume of $S^7$ is ${\rm Vol}(S^7) = \pi^4/3$. The adapted coordinate $\bdrypsi$ is periodic with period $2\pi$. We see that $\rd\CC$ has an flux through the internal space quantized as
\begin{equation}
\label{eq:NM2Branes}
N \= - \frac{1}{(2\pi \ell_P)^6}\int_{S^7} *_{11}\rd\CC \=  \frac{128 \pi^4}{(2\pi\ell_P)^6} \, .
\end{equation}
Combining the above equation with the reduction of the Ricci scalar on the internal space, we find the canonical identification for dual to ABJM theory:
\begin{equation}
\label{eq:G4M2}
\frac{1}{G_4} \= \frac{2\sqrt{2}}{3} N^{\frac{3}{2}}  \, .
\end{equation}

Now we focus on the Wick rotation of the supersymmetric Kerr--Newman-AdS black hole with chemical potentials $(\beta, \Omega, \Phi_e)$. First, we observe that the non-vanishing holonomy of the gauge field at the boundary \eqref{eq:Boundarybehavior}, together with the presence of the gauge field in the fibration term in \eqref{eq:M2Ansatz}, imply that $Y_{11}$ does not have the same asymptotics as the direct product AdS$_4\times S^7$:
\begin{equation}
\begin{split}
\bulkg(Y_{11})  &\; \sim \; \frac{\rd z^2}{z^2} + \frac{1}{z^2} \left( \rd \bdrytau^2 + \rd \bdrytheta^2 + \sin^2\bdrytheta \, (\rd\bdryphi - \ii \Omega \, \rd\bdrytau)^2 \right)\\
& \quad \ \ + 4 \left[ \left( \rd\bdrypsi + \sigma + \frac{\ii}{2} \Phi_e \, \rd\bdrytau \right)^2 + \bulkg(\mathbb{CP}^3) \right] \, .
\end{split}
\end{equation}
Regularity of the solution requires
\begin{equation}
    (\bdrytau, \bdryphi, \bdrypsi) \; \sim \; (\bdrytau+\beta, \bdryphi,\bdrypsi) \; \sim \; (\bdrytau, \bdryphi+2\pi, \bdrypsi) \; \sim \; (\bdrytau, \bdryphi, \bdrypsi + 2\pi) \, ,
\end{equation}
but at the cost of having explicit fibration terms in the metric. Alternatively, we can shift the realization of the chemical potentials to the twisting of the coordinates by defining $\nonrotatingpsi=\bdrypsi + \ii \Phi_e \, \bdrytau /2$, so that the metric is explicitly asymptotically locally AdS$_4\times S^7$, but the regularity of the solution then requires
\begin{equation}
\label{eq:ElevenDimensionalIdentifications}
    (\nonrotatingtau, \nonrotatingphi, \nonrotatingpsi) \; \sim \; (\nonrotatingtau+\beta, \nonrotatingphi-\ii \Omega \beta, \nonrotatingpsi+ \ii \Phi_e \beta/2) \; \sim \; (\nonrotatingtau, \nonrotatingphi + 2\pi, \nonrotatingpsi) \; \sim \; (\nonrotatingtau, \nonrotatingphi, \nonrotatingpsi+2\pi) \, .
\end{equation}

It is clear that the supersymmetric structure at the conformal boundary of the black hole solution is the same as the one where the field theory is formulated on in Section \ref{sec:ABJMIndex}, with the same fibration parameter $\Omega$ and with $\Phi_e =\Phi^{(H_1)}$ (see \eqref{eq:FibrationParameters_AllEqual}). Furthermore, the identifications in~\eqref{eq:ElevenDimensionalIdentifications} can be combined to show that the same relations are satisfied by a solution with the potentials
\begin{equation}
\label{eq:ElevenDimensionalShifts}
    \beta' \= \beta \, , \qquad \Omega' \= \Omega + \frac{2\pi\ii}{\beta}n_\Omega' \, , \qquad \Phi_e' \= \Phi_e + \frac{2\pi\ii}{\beta} 2n_e \, .
\end{equation}
More precisely, though, if $n'_\Omega$ is odd, the periodicity of the spinors around $S^1$ changes, whereas this doesn't happen if $n'_\Omega = 2n_\Omega$ (see the discussion of the Killing spinor in the previous section and notice that $n'_\Omega$ odd would change the parity of $n_0$ in \eqref{eq:Constraint_AllEqual_v2}). It is clear that these shifts correspond to the shifts of the chemical potentials in the partition function of the dual field theory~\eqref{eq:PeriodicitiesABJM_AllEqual}. 
Observe that all the solutions $(\beta', \Omega', \Phi'_e)$ have the same boundary conditions as the starting solution $(\beta, \Omega, \Phi_e)$, and so they must be summed over in the functional integral, with the reduced chemical potentials
\begin{equation}
\label{eq:ShiftsReducedCP_Minimal}
    \tau'_g \= \tau_g + 2n_\Omega \, , \qquad \j_{g}' \= \j_g +2n_e \, .
\end{equation}
We expand on this below.
From the eleven-dimensional perspective, all the shifts~\eqref{eq:ElevenDimensionalShifts} are obtained by combining conditions from regularity of the solution. From the effective viewpoint on $Y_4$, instead, the regularity of the solution~\eqref{eq:IdentificationsNonRotating} only leads to the shift of $\Omega$ in \eqref{eq:ElevenDimensionalShifts}. The shift of $\Phi_e$ follows from imposing the boundary condition for the bulk Abelian gauge field by fixing its holonomy around the Euclidean circle, namely
\begin{equation}
\label{eq:BdryGaugeHolonomy}
    \exp \left( \frac{\ii}{2} \int_{S^1_\beta}\bdryA \right) = \exp \left( - \frac{\beta}{2} \Phi_e \right) \, .
\end{equation}
The factor of $\frac{1}{2}$ is due to the fact that operators generically have half-integer $R$-charges.\footnote{Recall that from the boundary viewpoint, $\bdryA$ couples to the current corresponding to the symmetry generated by $H_1$, which is integrated over the $S^2$.}

\subsection{AdS/CFT comparison and orbifold solutions}

In order to perform the match warranted by the AdS/CFT correspondence, we begin by matching the chemical potentials on both sides of the correspondence using the boundary conditions, given in~\eqref{eq:FibrationParameters_AllEqual} and \eqref{eq:ReducedChemicalPotentials}, respectively.  
In these equations, we take~$n_0 = \mp 1$ on the positive/negative branch of solutions, which implies
\begin{equation}
\label{eq:RelationGravityFieldTheory_Minimal}
\tau_g \; \leftrightarrow \; \QFTtau + n_1 \mp 1 \, , \qquad 2\j_g \; \leftrightarrow \; \QFTtau + n_1 \, .
\end{equation}
The gravitational action~\eqref{eq:OnShellActionSUSY} is singular as $\tau_g \to 0$ (and $\j_g$ is finite there because of \eqref{eq:ReducedChemicalPotentialsConstraint}). The same singularity is reproduced by the field theory saddle with $(c,d)=(1,0)$ if we choose $n_1 = \pm 1$ (for the positive or negative branch of solutions, respectively). 
With this choice, the on-shell action of a complex supersymmetric solution on the positive (resp. negative) branch matches the large-$N$ behavior of the unrefined index $\CI(\tau, \lambda)$ as $\QFTtau\to 0$ and $\lambda = 1/4$ (resp. $\lambda = - 1/4$). Indeed, the gravitational action~\eqref{eq:OnShellActionSUSY} expressed in terms of the field theory parameters reads
\begin{equation}
\label{eq:GravityActionInFTParameters_Unrefined}
I|_{\rm SUSY}(\tau_g = \QFTtau) \= \pm \frac{\pi}{3\sqrt{2}} N^{\frac{3}{2}}  \frac{(\QFTtau \pm 1)^2}{\QFTtau} \, ,
\end{equation}
whose singular part clearly matches (the negative of) \eqref{eq:LargeNUnrefinedIndex} when $\ell_c=1$ and $\ell_d=0$.

\medskip

Starting from the eleven-dimensional geometry uplifting a solution with chemical potentials $(\tilde{\beta}, \tilde{\Omega}, \tilde{\Phi}_e)$, we can also define the following identification of the coordinates
\begin{equation}
\label{eq:OrbifoldIdentifications}
\begin{split}
    \left( \nonrotatingtau, \nonrotatingphi, \nonrotatingpsi \right) \; &\sim \; \left( \nonrotatingtau+\frac{\tilde{\beta}}{c_g}, \nonrotatingphi - \ii \tilde{\Omega} \frac{\tilde{\beta}}{c_g} - \frac{2\pi r}{c_g}, \nonrotatingpsi + \frac{\ii}{2} \tilde{\Phi}_e\frac{\tilde{\beta}}{c_g}  - \frac{2\pi s}{c_g}\right) \; \sim \; \left( \nonrotatingtau, \nonrotatingphi + 2\pi, \nonrotatingpsi \right) \\
    &\; \sim \; \left( \nonrotatingtau, \nonrotatingphi, \nonrotatingpsi+2\pi \right) \, ,
\end{split}
\end{equation}
or equivalently
\begin{equation}
\label{eq:OrbifoldIdentifications_v2}
\begin{split}
    \left( \bdrytau, \bdryphi, \bdrypsi \right) \; &\sim \; \left( \bdrytau+\frac{\tilde{\beta}}{c_g}, \bdryphi - \frac{2\pi r}{c_g} ,\bdrypsi  - \frac{2\pi s}{c_g} \right) \; \sim \; \left( \bdrytau, \bdryphi+2\pi, \bdrypsi \right) \; \sim \; \left( \bdrytau, \bdryphi, \bdrypsi + 2\pi \right)
\end{split}
\end{equation}
where $r,s, c_g$ are integers. It is clear from the construction that $r$ and $s$ are only defined modulo $c_g$, so that the metric is a $\Z_{c_g}$ quotient of the original solution.\footnote{In fact, as we shall see, we need to allow $r=0, \dots, 2c_g-1$, and $s=0,\dots, c_g-1$ in order to preserve the periodicity conditions for the fermions along the circle. A transformation by integer shifts $(r,s)$ has order $c_g$ in $\Z_{c_g}$ if and only if $\gcd(r,c_g)=\gcd(s,c_g)=1$.\label{footnote:Ranges}} It is also clear from the above identifications that this construction is different from the standard~$\Z_k$ quotient of the internal sphere, which acts only on the Hopf fiber as $\bdrypsi \sim \bdrypsi + 2\pi/k$, and changes the dual field theory \cite{Aharony:2008ug}. 
Since the construction of the solutions \eqref{eq:OrbifoldIdentifications} crucially involves the Euclidean circle, their Lorentzian interpretation is not transparent. Instead they represent the gravitational duals to the Euclidean saddle-points of the field theory~\cite{Aharony:2021zkr}.

\medskip

In order to interpret these solutions, we first notice that the identifications \eqref{eq:OrbifoldIdentifications} can be neatly expressed in terms of the shifted potentials in \eqref{eq:ElevenDimensionalShifts}
\begin{equation}
\begin{split}
    \left( \nonrotatingtau, \nonrotatingphi, \nonrotatingpsi \right) \; &\sim \; \left( \nonrotatingtau+\frac{\tilde{\beta}'}{c_g}, \nonrotatingphi - \ii \tilde{\Omega}' \frac{\tilde{\beta}'}{c_g}, \nonrotatingpsi + \frac{\ii}{2} \tilde{\Phi}_e'\frac{\tilde{\beta}'}{c_g}  \right) \; \sim \; \left( \nonrotatingtau, \nonrotatingphi + 2\pi, \nonrotatingpsi \right) \\
    \; &\sim \; \left( \nonrotatingtau, \nonrotatingphi, \nonrotatingpsi+2\pi \right) \, .
\end{split}
\end{equation}
Comparing with \eqref{eq:ElevenDimensionalIdentifications}, we see that these are the identifications required of a solution with potentials $(\tilde{\beta}'/c_g, \tilde{\Omega}', \tilde{\Phi}_e')$. In the saddle-point approximation to the gravity path integral, we should sum over solutions with shifted chemical potentials, provided they have the same boundary conditions. Although here $\tilde{\beta} \neq \tilde{\beta}'/c_g$, the supersymmetric index is independent of the size of the thermal circle, and the boundary values of the holonomies of the gauge field and the angular velocity (see discussion around \eqref{eq:BdryGaugeHolonomy}), which control the supersymmetric index, are indeed the same.
Thus, the $\Z_{c_g}$ quotient of a supersymmetric solution labelled by $(\tilde{\beta}, \tilde{\Omega}, \tilde{\Phi}_e)$, or more appropriately for the supersymmetric locus $(\tilde{\tau}_g, \tilde{\j}_g)$, contributes to the gravitational path integral with
\begin{equation}
\label{eq:ReducedChemicalPotentialsOrbifold}
    \beta \= \frac{\tilde{\beta}}{c_g} \, , \qquad \tau_g \= \frac{\tilde{\tau}_g}{c_g} - \frac{r}{c_g} \, , \qquad \j_g \= \frac{\tilde{\j}_g}{c_g} + \frac{2s}{c_g} \, .
\end{equation}

In order to ensure that the orbifold~\eqref{eq:OrbifoldIdentifications} preserves supersymmetry, we need to check that the Killing spinor is globally defined, i.e., that it is anti-periodic around $S^1_\beta$. This requires that the chemical potentials of a gravitational solution satisfy the constraint \eqref{eq:ConstraintGravitationalPotentials_v2}, which applied to \eqref{eq:ReducedChemicalPotentialsOrbifold}, and combined with the assumption that $\tilde{\tau}_g - 2\tilde{\varphi}_g = \mp 1$ leads to
\begin{equation}
\label{eq:SUSYConditionOrbifold_Minimal}
    4s \= - r \mp 1 - c_g n_0
\end{equation}
for an appropriate odd $n_0$. For every $c_g$, there are $c_g$ combinations of $(r,s,n_0)$ solving the equation, provided $r=0, \dots, 2c_g-1$, and $s=0,\dots, c_g-1$, and $r$ and $c_g$ have opposite parity.
This is an extension of the earlier conclusions \eqref{eq:ShiftsReducedCP_Minimal}, which impose that $r$ should be even if $c_g=1$.

By construction, the on-shell action of the solution obtained via a $\Z_{c_g}$ quotient of the supersymmetric solution with $(\tilde{\tau}_g, \tilde{\j}_g)$ is $1/c_g$ the on-shell action of the original solution, that is
\begin{equation}
\label{eq:ActionOrbifoldPI}
\begin{split}
    I|_{\rm SUSY}\left(\tau_g, \j_g \right) &\= \frac{1}{c_g} I|_{\rm SUSY}(\tilde{\tau}_g, \tilde{\j}_g) \= \pm \frac{\pi}{G_4} \frac{(c_g\j_g-2s)^2}{c_g(c_g\tau_g+r)} \\
    &\= \pm \frac{\pi}{4 G_4} \frac{(c_g\tau_g + r \pm 1)^2}{c_g(c_g\tau_g+r)}
\end{split}
\end{equation}
This represents the contribution of the orbifold to the gravitational sum dual to a grand canonical partition function with parameters $(\tau_g, \j_g)$. In order to match with field theory, we explain in detail the simplest case with $c$ odd. The function $I|_{\rm SUSY}$ is singular as $\tau_g \to - r/c_g$, and now the identification of the chemical potentials relates $\tau_g$ with $\QFTtau + n_0 + n_1$ (where $n_0$ is the odd number such that \eqref{eq:SUSYConditionOrbifold_Minimal} holds). Therefore, using \eqref{eq:SUSYConditionOrbifold_Minimal}, the corresponding singularity in field theory would appear as $\QFTtau \to (4s \pm 1 - c_gn_1)/c_g$ (where we remark again that the sign refers to the branch of supersymmetric black holes that the orbifold is a quotient of). Indeed, recall that in order for the index to have a large-$N$ behavior O$(N^{\frac{3}{2}})$, we need $c n_1 = \pm 1$ mod 4, and we already established around \eqref{eq:GravityActionInFTParameters_Unrefined} that for $c=1$ the two signs are related to the choice of branches for the dual gravity solution. Consistently, we establish the dictionary $c_g \leftrightarrow c$ and $s \leftrightarrow d$, and we find that the singular behavior of the on-shell action of the $\Z_{c}$ quotient \eqref{eq:OrbifoldIdentifications} of a solution in the positive (resp. negative) branch matches the large-$N$ limit of unrefined index $\CI(\tau;\lambda)$ as $\QFTtau \to \frac{4d}{c}$ and $c n_1 = \pm 1$ mod 4
\begin{equation}
\label{eq:GravityActionInFTParameters_Unrefined_Orbifold}
    I|_{\rm SUSY}(c\tau_g = c\QFTtau + cn_0 \pm 1) \= \pm \frac{\pi}{3\sqrt{2}} N^{\frac{3}{2}} \frac{(c\QFTtau - 4d \pm 1)^2}{c(c\QFTtau -4d)} \, .
\end{equation}
For comparison, consider \eqref{eq:LargeNUnrefinedIndex}, recalling that $\ell_c=c$ and $\ell_d = 4d$ if $c$ is odd.

\medskip

The only fixed points of the identifications \eqref{eq:OrbifoldIdentifications_v2} are at the horizon, that is $r=r_+$, where the circle generated by $V$ in \eqref{eq:KillingGenerator} shrinks. The remaining space in the eleven-dimensional solution is the product of the $S^2$ transverse in the black hole, and the internal $S^7$ of which $\bdrypsi$ is the Hopf coordinate. It is not possible for both $r$ and $s$ to be both vanishing while preserving supersymmetry: the condition \eqref{eq:SUSYConditionOrbifold_Minimal} would not be satisfied for $c_g>1$, and indeed the transverse space to the fixed point would be $\C/\Z_{c_g}$, which does not supersymmetry. If $r=0$, then the quotient acts only on the Hopf fiber of $S^7$ and the disc transverse to $S^2$. Since the Hopf fiber never shrinks, there are no fixed points. Finally, if $s=0$, then we have a quotient of the black hole only, and we have fixed point sets isomorphic to the round $S^7$ at the two poles of the $S^2$ (where $\theta=0,\pi$ and the circle generated by $\partial_{\bdryphi}$ shrinks to zero size). The transverse space is $\C^2/\Z_{c_g}$, which preserves supersymmetry.

\medskip

We conclude the discussion of the minimal gauged supergravity solutions with two comments on further refinements of the gravitational path integral. First, it is straightforward to compute corrections to the on-shell action subleading in $N$ using the four-derivative corrections to the minimal gauged supergravity action proposed in \cite{Bobev:2020egg, Bobev:2021oku}. These corrections do not modify the Killing spinor equations, and the equations of motion derived from the corrected action are a consequence of the two-derivative equations of motion.\footnote{This circumstance can be explained appealing to field redefinitions \cite{Cassani:2023vsa}.} This implies that the supersymmetric solutions to the higher-derivative theory are just those of the two-derivative theory, so the on-shell action with the higher-derivative corrections follows again from a localization principle \cite{Genolini:2021urf}. For the supersymmetric family of complex solutions, we have
\begin{equation}
\begin{split}
    I_{\rm HD} &\= \pm \left[ \frac{\pi}{G_4}\frac{\j^2_g}{\tau_g} + \frac{64\pi^2}{\tau_g} \left( - \alpha_1 (1 \mp \j_g)^2 + \alpha_2 \j_g^2 \right) \right] \\
    &\= \pm \left[ \left( \frac{\pi}{2G_4} + 32\pi^2 \alpha_2 - 32\pi^2 \alpha_1 \right) \frac{(\tau_g \pm 1)^2}{2\tau_g} \pm 64\pi^2 \alpha_1 \right] \, .
\end{split}
\end{equation}
Here $\alpha_{1,2}$ are constants introduced to parametrize the higher derivative corrections, respectively a supersymmetrization of the Weyl squared action and the Gauss--Bonnet term, which would be determined by the uplifting of the higher derivative action in eleven dimensions, or comparing with the localized partition function, obtaining \cite{Bobev:2021oku}
\begin{equation}
    \frac{\pi}{2G_4} + 32\pi^2 \alpha_2 \= \frac{2}{3}\pi N^{\frac{3}{2}} - \frac{3}{8\sqrt{2}} \pi N^{\frac{1}{2}} \, , \qquad 32\pi^2 \alpha_1 \= - \frac{\pi}{\sqrt{2}} N^{\frac{1}{2}} \, ,
\end{equation}
and finally \cite{Bobev:2021oku, Bobev:2022wem}
\begin{equation}
    I_{\rm HD} \= \pm \frac{\sqrt{2}}{3} \pi \left[ \left( N^{\frac{3}{2}} + \frac{15}{16}N^{\frac{1}{2}} \right) \frac{(\tau_g \pm 1)^2}{2\tau_g} \mp \frac{1}{3}N^{\frac{1}{2}} \right] \, .
\end{equation}
According to the prescriptions described around \eqref{eq:ActionOrbifoldPI}, the higher derivative contribution of a $\Z_{c_g}$ quotient of a supersymmetric solution $(\tilde{\tau}_g, \tilde{\j}_g)$ would be $\frac{1}{c_g}I_{\rm HD}(\tilde{\tau}_g, \tilde{\j}_g)$.

The second comment concerns the saddle point approximation to the gravitational path integral. The AdS/CFT correspondence instructs us to compare the grand canonical field theory partition function in the limit of large~$N$ and fixed $k$ with a sum over gravitational saddles
\begin{equation}
\label{eq:AdSCFTMaster_1}
    Z_{\rm GCE}(\Omega, \Phi) \; \sim \; \sum_{n_\Omega, n_a \in \Z} \rme^{ - I|_{\rm SUSY}\left( \beta, \, \Omega + \frac{2\pi\ii}{\beta} 2n_\Omega + \frac{2\pi\ii}{\beta}\sum_a n_a, \, \Phi_a + \frac{2\pi\ii}{\beta}n_a \right) } \, ,
\end{equation}
where the quantity in the exponent in the right-hand side is the on-shell action of the relevant supergravity solution with the appropriate boundary conditions fixed by the radius of the circle $\beta$, the boundary metric \eqref{eq:FiberedBdryMetric}, and the holonomy of the gauge fields at the boundary fixed by the form \eqref{eq:BackgroundGaugeFields_v0} to be
\begin{equation}
    \exp \left( \frac{\ii}{2} \int_{S^1_\beta} A_a \right) \= \exp \left( - \frac{\beta}{2}\Phi_a \right) \, , \qquad a \= 1, 2, 3, 4 \, .
\end{equation}
As we have seen, supersymmetry imposes the relation~\eqref{eq:ChemicalPotentialsConstraint_v0} between the chemical potentials, and in fact it is appropriate to use the five reduced chemical potentials defined in analogy with~\eqref{eq:ReducedChemicalPotentials}
\begin{equation}
\label{eq:ReducedChemicalPotentials_AllDifferent}
    \tau_g \; \equiv \; \beta \frac{\Omega - 1}{2\pi\ii} \, , \qquad \j_a \; \equiv \; \beta \frac{\Phi_a - \frac{1}{2}}{2\pi\ii} \, ,
\end{equation}
which are constrained by
\begin{equation}
    \tau_g - \sum_{a=1}^4 \j_a \= n_0 \, .
\end{equation}
In terms of these the field theory partition function reads
\begin{equation}
    Z \= \Tr_{\CH_{S^2}}\rme^{ - \beta \{ \CQ, \CQ^\dagger\} + 2\pi\ii \tau_g J + 2\pi\ii \sum_a \j_a R_a } \= \Tr_{\CH_{S^2}}(-1)^{2J}\rme^{ - \beta \{ \CQ, \CQ^\dagger\} + 2\pi\ii \sum_a \j_a (J+R_a) } \, .
\end{equation}
and so
\begin{equation}
\label{eq:AdSCFTMaster_2}
    Z_{\rm GCE}(\j) \; \sim \; \sum_{n_a\in\Z} \rme^{ - I|_{\rm SUSY}\left( \j_a + n_a \right) } \, ,
\end{equation}
highlighting the fact that both functions only depend on four rather than five fugacities, and that neither the field theory not the gravity depends on $\beta$. An important point stressed in \cite{Aharony:2021zkr, Boruch:2022tno} is that in the grand canonical ensemble it is not allowed to restrict to the case of equal $\Phi_a = \Phi^{(H_1)}/2$: the sum over gravitational solutions on the right-hand side of \eqref{eq:AdSCFTMaster_2} will take us away from this locus. Therefore, one can either go to a mixed ensemble, as in \cite{Boruch:2022tno}, or consider solutions in bulk gauged supergravity with multiple $U(1)$ gauge vectors, as in \cite{Aharony:2021zkr}. We now move to do this.

\section{Black holes in non-minimal gauged supergravity}
\label{sec:U12}

The black hole solutions considered in the previous section are charged under a unique Abelian gauge field, that is, they are solutions of minimal gauged supergravity in four dimensions. Thus, they can only reproduce the behavior of the ABJM index when the fugacities for the $U(1)^4$ Cartan of the $R$-symmetry are set equal. However, there are also known rotating black holes with multiple electric charges.

\subsection{Supersymmetric black holes in the \texorpdfstring{$F=-\ii X^0X^1$}{X0X1} model}

There is a rotating black hole solution with two different electric charges. We shall be quite brief in reviewing its properties, and the reader can find similar discussions in \cite{Cassani:2019mms}.

\medskip

The bosonic part of the relevant Lorentzian action is
\begin{equation}
\label{eq:LorentzianLagU12}
\begin{split}
    S \= \frac{1}{16\pi G_4} \int_{Y_4} \bigg[ &\CR \, \vol_{\CG} + \frac{1}{2} \, \rd X^2_1 \wedge * \rd {X}^2_2 - \frac{1}{2}\rd (\j X^2_1)\wedge *\rd (\j X^2_1) \\
    & + (4 + X^2_1 + {X}^2_2 ) \, \vol_{\CG} - X^{-2}_1 ( \CF_1 \wedge * \CF_1 - \j X^2_1 \CF_1 \wedge \CF_1) \\
    & - {X}^{-2}_2 \left( {\CF}_2 \wedge * {\CF}_2 + \j X^2_1 {\CF}_2 \wedge {\CF}_2 \right) \bigg] \, .
\end{split}
\end{equation}
Here $\CG$ is the metric, $\CF_1$ and ${\CF}_2$ are the curvature of two Abelian gauge fields, $X_1$ and $\j$ are a scalar and a pseudo-scalar, respectively, and we have defined
\begin{equation}
    {X}^2_2 \= X^{-2}_1 + \j^2 X^2_1 \, .
\end{equation}
There is a $\Z_2$ automorphism exchanging the two Abelian gauge factors
\begin{equation}
    \CF_1 \leftrightarrow {\CF}_2 \qquad X_1 \leftrightarrow {X}_2 \qquad \j X^2_1 \leftrightarrow - \j X^2_1 \, .
\end{equation}
The action reduces to minimal gauged supergravity \eqref{eq:LagrangianLorentzian} upon setting $X_1 = {X}_2 = 1$ and ${\CF}_1 = \CF_2 = \CF$ (which explains the non-canonical normalization of the kinetic terms for the gauge fields).

The solution we are interested in, which again we present in the frame non-rotating at infinity, is \cite{Chong:2004na, Cvetic:2005zi}
\begin{equation}
\label{eq:X0X1_Metric}
\begin{split}
    \rd s^2 &\= - \frac{\Delta_\Theta \Delta_r}{B \Xi^2} \rd \nonrotatingt^2 + \sin^2\Theta \, B \, \left( \rd\nonrotatingphi + a \Delta_\Theta \frac{\Delta_r - (1+r_1r_2)(a^2 + r_1r_2)}{B W \Xi^2} \rd \nonrotatingt \right)^2 \\
    & \ \ \ + W \left( \frac{\rd r^2}{\Delta_r} + \frac{\rd \Theta^2}{\Delta_\Theta} \right) \, .
\end{split}
\end{equation}
Here
\begin{equation}
\begin{split}
r_i &\= r + 2 \, m \, s_i^2 \, , \\
\Delta_r &\= r^2 + a^2 - 2 \, m \, r + r_1 \, r_2 \left(r_1 \, r_2 + a^2 \right) \, , \\
\Delta_\Theta &\= 1 -  a^2 \, \cos^2 \Theta \, , \qquad W \= r_1 \, r_2 + a^2 \, \cos^2 \Theta \, , \qquad \Xi \= 1 - a^2 \, , \\
B &\= \frac{\Delta_\Theta (r_1r_2 + a^2)^2 - a^2 \sin^2\Theta \, \Delta_r}{W \Xi^2} \, ,
\end{split}
\end{equation}
and $s_i = \sinh \delta_i$, $c_i = \cosh \delta_i$, $i=1,2$.
The scalar fields are
\begin{equation}
\label{eq:X0X1_Scalars}
X^2_1 \= 1 + \frac{r_1 \left(r_1 - r_2 \right) }{W} \, , \qquad \j \= \frac{a \left(r_2 - r_1 \right) \cos \Theta }{r_1^2 + a^2 \cos^2 \Theta } \, , \qquad {X}^2_2 \= 1 + \frac{r_2 \left(r_2 - r_1 \right) }{W} 
\end{equation}
and the gauge fields
\begin{equation}
\label{eq:X0X1_Vectors}
\begin{split}
    \CA_1 &\= \frac{2m s_2c_2 r_1}{W \Xi} \left( \Delta_\Theta \, \rd \nonrotatingt - a \sin^2\Theta \, \rd \nonrotatingphi \right) + \gamma_1 \, \rd \nonrotatingt \, , \\ 
    {\CA}_2 &\= \frac{2m s_1c_1 r_2}{W \Xi} \left( \Delta_\Theta \, \rd \nonrotatingt - a \sin^2\Theta \, \rd \nonrotatingphi \right) + {\gamma}_2 \, \rd \nonrotatingt \, ,
\end{split}
\end{equation}
where $\gamma_{1,2}$ are constant.

There is an outer horizon at the largest positive root of $\Delta_r$, and the Wick-rotated solution ($t=-\ii t_{\rm E}$) is smooth if we identify
\begin{equation}
    (\nonrotatingtau, \nonrotatingphi) \; \sim \; (\nonrotatingtau, \nonrotatingphi + 2\pi) \; \sim \; (\nonrotatingtau + \beta, \nonrotatingphi - \ii \Omega \beta) \, ,
\end{equation}
where the temperature and angular velocity of the horizon are
\begin{equation}
\begin{split}
    \beta \= 4\pi \frac{a^2 + r_{1+}r_{2+}}{\Delta'_r(r_+)} \, , \qquad \Omega \= a\frac{1+ r_{1+}r_{2+}}{a^2 + r_{1+}r_{2+}} \, ,
\end{split}
\end{equation}
and we defined $r_{i+}\equiv r_+ + 2m s_i^2$. The horizon is a Killing horizon for $V= \partial_{\nonrotatingt} + \Omega \partial_{\nonrotatingphi}$. The entropy is computed from the area of the horizon
\begin{equation}
    S \= \frac{\pi}{G_4} \frac{r_{1+}r_{2+} + a^2}{\Xi}
\end{equation}
The electrostatic potentials are
\begin{equation}
    \Phi_{e,1} \= 2 m  s_2 c_2 \frac{r_{1+}}{a^2 + r_{1+}r_{2+}} \, , \qquad {\Phi}_{e,2} \= 2 m s_1 c_1 \frac{r_{2+}}{a^2 + r_{1+}r_{2+}} \, ,
\end{equation}
and we choose the gauge $\gamma_1 = - \Phi_{e,1}$ and ${\gamma}_2 = - {\Phi}_{e,2}$ in order to have smooth gauge fields at the horizon. Notice that the Wick-rotated metric is real provided $a$ is pure imaginary and $\delta_{1,2}, m$ are real. If these conditions are satisfied, the Euclidean metric has the topology of the product of a disc and a 2-sphere.

The black hole is asymptotically AdS, and the boundary metric has the same form as \eqref{eq:BdryLineElement} and gauge fields $\CA_1 \sim - \Phi_{e,1} \, \rd \nonrotatingt$, ${\CA}_2 \sim - {\Phi}_{e,2} \, \rd \nonrotatingt$. The standard procedure of holographic renormalization then gives us the holographic conserved charges, as described in the previous section. In principle, the presence of scalars produces various differences with the Einstein--Maxwell theory considered earlier, affecting the form of the counterterms and the interpretation of the matching to the boundary. However, we show in Appendix \ref{app:HoloRen_Details} that for the solution considered here \eqref{eq:X0X1_Metric}, \eqref{eq:X0X1_Scalars}, \eqref{eq:X0X1_Vectors} the boundary scalar sources vanish, and thus in the main text we ignore them. 
The counterterms compatible with supersymmetry are
\begin{equation}
\label{eq:IU12Counterterms}
    I \= \lim_{\delta \to 0} \left[ S + \frac{1}{8\pi G_4} \int_{\partial Y_\delta} \left( K - \frac{1}{2}R - \sqrt{ 2+ X^2_1 + {X}^2_2} \right) \vol_h \right] \, .
\end{equation}
The holographic stress-tensor and the electric current defined as
\begin{equation}
    \langle T_{ij} \rangle \= - \frac{2}{\sqrt{-\bdryg}} \frac{\delta I}{\delta g^{ij}} \, , \qquad \langle j^i_1 \rangle \= \frac{1}{\sqrt{-\bdryg}} \frac{\delta I}{\delta \bdryA_{1i}} \, , \qquad \langle {j}^i_2 \rangle \= \frac{1}{\sqrt{-\bdryg}} \frac{\delta I}{\delta {\bdryA}_{2i}} \, ,
\end{equation}
satisfy the conservation equation
\begin{equation}
    \nabla^i \langle T_{ij} \rangle \= F_{1ji} \langle j^i_1 \rangle + {F}_{2ji} \langle {j}^i_2 \rangle \, ,
\end{equation}
and we can define a conserved charge associated to any boundary vector $K$ generating a symmetry
\begin{equation}
    Q[K] \= \int_{C\cap M_3} u_i \left( \langle T^i_{\ph{i}j} \rangle + \langle j^i_1 \rangle \bdryA_{1j} + \langle {j}_2^i \rangle {\bdryA}_{2j} \right) K^j \, \vol_{C \cap M_3} \, .
\end{equation}
In particular, we have expressions for the angular momentum associated to $-\partial_{\nonrotatingphi}$
\begin{equation}
    J \= a m \frac{1+s_1^2 + s_2^2}{G_4\Xi^2} \, ,
\end{equation}
the electric charges
\begin{equation}
\begin{split}
    Q_{e,1} &\= \int_{C\cap M_3} u_i \langle j^i_1 \rangle \, \vol_{C\cap M_3} \= \frac{1}{8\pi G_4}\int_{C\cap M_3} (X_1^{-2}*\CF_1 - \j \CF_1) \= \frac{m s_2 c_2}{G_4 \Xi} \, , \\
    {Q}_{e,2} &\= \int_{C\cap M_3} u_i \langle {j}^i_2 \rangle \, \vol_{C\cap M_3} \=  \frac{1}{8\pi G_4}\int_{C\cap M_3} (X_2^{-2}*\CF_2 + \j \CF_2) \= \frac{m s_1 c_1}{G_4 \Xi} \, ,
\end{split}
\end{equation}
and finally the energy associated to $\partial_{\nonrotatingt}$ (as in \eqref{eq:BlackHoleEnergyMinimal}, we denote by $E$ the energy in the gauge where the boundary gauge fields vanish)
\begin{equation}
\begin{split}
    E' \= m \frac{1+s_1^2+s_2^2}{G_4 \Xi^2} - \Phi_{e,1} \, Q_{e,1} - {\Phi}_{e,2} \, {Q}_{e,2} \; \equiv \; E - \Phi_{e,1} \, Q_{e,1} - {\Phi}_{e,2} \, {Q}_{e,2} \, .
\end{split}
\end{equation}
These quantities, together with the Euclidean on-shell action, satisty the quantum statistical relation
\begin{equation}
\label{eq:QuantumStatisticalRelation_NonMinimal}
\begin{split}
    I &\= - S + \beta (Q[V] - \Phi_{e,1} \, Q_{e,1} - {\Phi}_{e,2} \, {Q}_{e,2} ) \\
    &\= - S + \beta( E - \Omega J - \Phi_{e,1} \, Q_{e,1} - {\Phi}_{e,2} \, {Q}_{e,2} ) \, ,
\end{split}
\end{equation}
which reduces to \eqref{eq:QuantumStatisticalRelation} upon setting equal the two gauge fields (notice that the charges are defined to be half of the electric charge in minimal gauged supergravity). The same quantities also satisfy a first law of thermodynamics
\begin{equation}
    \rd E \= \beta^{-1} \rd S + \Omega \, \rd J + \Phi_{e,1} \, \rd Q_{e,1} + {\Phi}_{e,2} \, \rd {Q}_{e,2} \, .
\end{equation}

\medskip

The Lagrangian \eqref{eq:LorentzianLagU12} is the bosonic part of an $\CN=2$ gauged supergravity model with prepotential $\mc{F} = - \ii X^0 X^1$. The condition for having a supersymmetric solution is \cite{Cvetic:2005zi, Chow:2013gba}\footnote{As in the case of minimal gauged supergravity mentioned in footnote \ref{footnote:MinimalMagnetic}, $J$ and $Q_{e,1/2}$ cannot vanish while still having a supersymmetric black hole with spherical horizon. Different instead is the case of black holes with magnetic charge \cite{Hristov:2011qr}, though a similar Euclidean gravity analysis can be applied to that case as well \cite{BenettiGenolini:2023ucp}.}
\begin{equation}
    E \= J + Q_{e,1} + {Q}_{e,2} \qquad \Leftrightarrow \qquad a \= \coth(\delta_1+\delta_2) - 1 \, .
\end{equation}
It is easy to check, studying $\Delta_r|_{\rm SUSY}$, that in Lorentzian signature, supersymmetry and regularity imply extremality. However, as in Section \ref{subsec:SUSYBlackHole}, in order to reproduce the behavior of the index we shall need to impose supersymmetry of the Wick-rotated solutions. This necessarily leads us to consider a family of complex supersymmetric metrics obtained by deforming the Euclidean metrics. We observe that imposing $\Delta_r|_{\rm SUSY}=0$ leads to a quadratic equation for $m$ in terms of $r_+, \delta_1, \delta_2$, and thus to two branches of solutions, labelled by the choice of $\pm$ in the solution of the equation. The BPS sublocus is obtained by imposing extremality in addition to supersymmetry \cite{Cassani:2019mms}.

\medskip

The chemical potentials of the complex supersymmetric solutions satisfy
\begin{equation}
\label{eq:ConstraintGravitationalPotentialsU12}
    \beta ( 1 - \Phi_{e,1} - {\Phi}_{e,2} + \Omega ) \= \mp 2\pi \ii \, ,
\end{equation}
and if we define the ``reduced chemical potential'' by taking the difference with the value of the BPS solutions, we find
\begin{equation}
\label{eq:ReducedGravitationalPotentialsU12}
    \tau_g \; \equiv \; \beta \frac{\Omega - 1}{2\pi\ii} \, , \qquad \varphi_{g1} \; \equiv \; \beta \frac{\Phi_{e,1} - 1}{2\pi\ii} \, , \qquad {\varphi}_{g2} \; \equiv \; \beta \frac{{\Phi}_{e,2} - 1}{2\pi\ii}
\end{equation}
satisfying
\begin{equation}
\label{eq:ReducedChemicalPotentialsConstraint_U12}
    \tau_g - \varphi_{g1} - {\varphi}_{g2} \= \mp 1 \, .
\end{equation}
The reduced chemical potentials for supersymmetric solutions are complex, but they remain finite as we approach the BPS locus, whereas the chemical potentials and the conserved charges of the BPS solutions become real.

The on-shell action of the supersymmetric solutions, computed using holographic renormalization, can be expressed using the reduced chemical potentials as
\begin{equation}
\label{eq:ISUSYU12}
    I|_{\rm SUSY} \= \pm \frac{\pi}{G_4}\frac{\varphi_{g1} {\varphi}_{g2}}{\tau_g} \, .
\end{equation}
Since this expression is independent of $\beta$, it remains finite on the BPS locus, which allows us to define the on-shell action of the supersymmetric extremal black holes. The relevance of this object is two-fold. Firstly, interpreted as a Gibbs free energy in the grand canonical ensemble, it allows us to obtain the entropy in the microcanonical ensemble via a Legendre transform. Secondly, it can be related to the dual field theory partition function, and its Cardy-like limits.

\medskip

To the first purpose, we point out again that for the complex supersymmetric solutions the quantum statistical relation \eqref{eq:QuantumStatisticalRelation_NonMinimal} can be written as
\begin{equation}
\label{eq:QuantumStatisticalRelationSUSY_NonMinimal}
    I|_{\rm SUSY} \= -S - 2\pi\ii \tau_g \, J - 2\pi\ii \j_{g1} \, Q_{e,1} - 2\pi\ii {\j}_{g2} \, {Q}_{e,2} \, , 
\end{equation}
and from this expression follows a Euclidean quantum gravity derivation of the extremization procedure proposed in \cite{Choi:2018fdc} that is analogous to that described in Section \ref{subsec:SUSYBlackHole}. From \eqref{eq:QuantumStatisticalRelationSUSY_NonMinimal} follows that the function to be extremized is
\begin{equation}
\begin{split}
    f(\tau_g, \j_{g1}, \j_{g2}) &\= - I|_{\rm SUSY}(\tau_g, \varphi_{g1}, {\varphi}_{g2}) - 2\pi\ii\tau_g \, J - 2\pi\ii\varphi_{g1} \, Q_{e,1} - 2\pi\ii {\varphi}_{g2} \, {Q}_{e,2} \\
    & \quad \ \ + \Lambda ( \tau_g - \varphi_{g1} - {\varphi}_{g2} \pm 1 ) \, .
\end{split}
\end{equation}
Combining the equation satisfied by $f$ at the critical point and Euler's theorem for the homogeneous function $I|_{\rm SUSY}(\tau_g, \j_{g1}, {\j}_{g2})$, we obtain the value of the Legendre transform
\begin{equation}
    \tilde{f}( J_*, Q_{e,1*}, {Q}_{e,2*}) \= \pm \Lambda_*(J_*, Q_{e,1*}, {Q}_{e,2*}) \, .
\end{equation}
Concretely, the equation to be solved to find $\Lambda_*(J_*, Q_{e,1*}, {Q}_{e,2*})$ is
\begin{equation}
    0 \= \Lambda_*^2 + \Lambda_* \left( 2\pi\ii (Q_{e,1*} + {Q}_{e,2*}) \pm \frac{\pi}{G_4} \right) + \left( - 4\pi^2 Q_{e,1*} {Q}_{e,2*} \mp \frac{2\pi^2\ii}{G_4} J_* \right) \, .
\end{equation}
We then impose the constraints that
\begin{equation}
    J_*, Q_{e,1*}, {Q}_{e,2*} \in \R \, , \qquad \Lambda_*(J_*, Q_{e,1*}, {Q}_{e,2*}) \in \R \, .
\end{equation}
Splitting real and imaginary part of the equation above leads us to the value of the Legendre transform
\begin{equation}
    \tilde{f}(J_*, Q_{e,1*}, {Q}_{e,2*}) \= \frac{\pi J_*}{G_4(Q_{e,1*} + {Q}_{e,2*})} \= \frac{\pi}{2G_4} \left( \sqrt{1+16G_4^2 Q_{e,1*}^2 {Q}_{e,2*}^2} -1 \right) \, .
\end{equation}
These values correspond, respectively, to the entropy of the extremal black hole in the $U(1)^2$ theory and to the non-linear constraint between its charges.

\subsection{Uplift to eleven dimensions and AdS/CFT}

The theory \eqref{eq:LorentzianLagU12} can be obtained from a consistent truncation of eleven-dimensional supergravity on $S^7$ as described in \cite{Azizi:2016noi} (for the bosonic sector). Geometrically, we write the metric on $S^7$ as a $S^3\times S^3$ fibered over an interval, and introduce a gauge field only along the Hopf fiber inside each $S^3$. Concretely, we write the eleven-dimensional solution as
\begin{equation}
\label{eq:UpliftU12Model}
\begin{split}
    \CG(Y_{11}) &\= (Z_1 {Z}_2)^{\frac{1}{3}} \, \CG(Y_4) \\
    & \ \ \ \ \  + 4 (Z_1 Z_2)^{\frac{1}{3}} \bigg\{ \rd \Xi^2 + \frac{\cos^2\Xi}{4Z_1} \left[ \rd\Theta^2_1 + \sin^2\Theta_1 \, \rd\Phi_1 + (\rd\hat{\Psi}_1 + \cos\Theta_1 \, \rd\Phi_1 + \CA_1)^2 \right] \\
    & \ \ \ \ \ \qquad \qquad \qquad \quad + \frac{\sin^2\Xi}{4 {Z}_2} \left[ \rd{\Theta_2}^2 + \sin^2{\Theta_2} \, \rd {\Phi_2} + (\rd\hat{\Psi}_2 + \cos{\Theta_2} \, \rd{\Phi_2} + {\CA_2})^2 \right] \bigg\} \, , \\[5pt]
    \rd \CC &\= \left( 2 + \cos^2\Xi \, X^2_1 + \sin^2\Xi \, {X}_2^2 \right) \vol(Y_4) \\
    & \ \ \ \ \ + 2 \cos\Xi \sin\Xi \left( \frac{2}{X_1} *_4 \rd X_1 - \j X^4_1 *_4 \rd \j \right) \wedge \rd \Xi + \rd \hat{A}_3 + \hat{F}'_4
\end{split}
\end{equation}
where
\begin{equation}
\begin{split}
    Z_1 &\= X_1^2 \cos^2\Xi + \sin^2\Xi \, , \\
    Z_2 &\= \cos^2\Xi + X_2^2 \sin^2\Xi \, , \\
    \hat{A}_3 &\= - 8 \j X^2_1 \left[ \frac{\cos^4\Xi}{Z_1} \Omega_1(\CA_1) - \frac{\sin^4\Xi}{Z_2} {\Omega_2}({\CA_2}) \right] \, , \\
    \hat{F}'_4 &\= \frac{2\cos\Xi}{X^2_1} \left[ \sin\Xi \, \rd \Xi \wedge \left( \rd\hat{\Psi}_1 + \cos\Theta_1 \, \rd\Phi_1 + \CA_1 \right) + \frac{\cos\Xi}{2} \sin\Theta_1 \, \rd\Theta_1 \wedge \rd\Phi_1 \right] \\
    & \qquad \wedge (*_4 \CF_1 + \j X^2_1 \CF_1) \\
    & \qquad - \frac{2\sin\Xi}{{X_2}^2} \left[ \cos\Xi \, \rd \Xi \wedge \left( \rd \hat{\Psi}_2 + \cos {\Theta_2} \, \rd {\Phi_2} + {\CA_2} \right) - \frac{\sin\Xi}{2} \sin{\Theta_2} \, \rd{\Theta_2} \wedge \rd{\Phi_2} \right] \\
    & \qquad \wedge (*_4 {\CF_2} - \j X^2_1 {\CF_2}) \, , \\
    \Omega_1(\CA_1) &\= \frac{1}{8}\sin\Theta_1 ( \rd\hat{\Psi}_1 + \cos\Theta_1 \, \rd\Phi_1 + \CA_1) \wedge \rd\Theta_1 \wedge \rd\Phi_1 \, , \\
    {\Omega}_2({\CA}_2) &\= \frac{1}{8}\sin {\Theta}_2 ( \rd \hat{\Psi}_2 + \cos{\Theta_2} \, \rd{\Phi_2} + {\CA_2}) \wedge \rd {\Theta_2} \wedge \rd {\Phi_2} \, .
\end{split}
\end{equation}
It is straightforward to see that this uplift reduces to \eqref{eq:M2Ansatz} upon setting $X_1={X}_2=1$ and ${\CA}_1=\CA_2 = \CA$, and changing coordinates to 
\begin{equation}
    \bdrypsi \= \frac{1}{4}(\hat{\Psi}_1 + \hat{\Psi}_2) \, , \qquad \Lambda \= \frac{1}{2}(\hat{\Psi}_1 - \hat{\Psi}_2) \, .
\end{equation}
The explicit expression for the $\mathbb{CP}^3$ quantities are
\begin{equation}
\begin{split}
    \sigma &\= \frac{1}{2}\left( \cos 2\Xi \, \rd \Lambda + \cos^2\Xi \cos \Theta_1 \, \rd\Phi_1 + \sin^2\Xi \cos{\Theta}_2 \, \rd{\Phi}_2 \right) \, , \\
    \CG(\mathbb{CP}^3) &\= \rd \Xi^2 + \frac{\cos^2\Xi}{4} \left( \rd\Theta^2_1 + \sin^2 \Theta_1 \, \rd\Phi_1^2 \right) + \frac{\sin^2\Xi}{4} \left( \rd {\Theta}^2_2 + \sin^2 {\Theta}_2 \, \rd{\Phi}_2^2 \right) \\
    & \ \ \ \ \ + \sin^2\Xi \cos^2\Xi \left( \rd\Lambda + \frac{\cos\Theta_1}{2} \rd\Phi_1 - \frac{\cos {\Theta_2}}{2} \rd{\Phi_2} \right)
\end{split}
\end{equation}

As in Section \ref{sec:Uplift}, we can study the regularity of the uplift in a neighbourhood of the conformal boundary, where the metric has the form
\begin{equation}
\begin{split}
    \CG(Y_{11}) \; &\sim  \; (Z_1 Z_2)^{\frac{1}{3}} \, \left[ \frac{\rd z^2}{z^2} + \frac{1}{z^2} \left( \rd \bdrytau^2 + \rd\Theta^2 + \sin^2 \Theta \left( \rd\bdryphi - \ii \Omega \, \rd\bdrytau \right)^2 \right) \right] \\
    & \ \ \ \ \  + 4 (Z_1 Z_2)^{\frac{1}{3}} \bigg\{ \rd \Xi^2 + \frac{\cos^2\Xi}{4Z_1} \left[ \rd\Theta_1^2 + \sin^2\Theta_1 \, \rd\Phi_1 + (\rd\hat{\Psi}_1 + \cos\Theta_1 \, \rd\Phi_1 + \ii \Phi_{e,1} \, \rd\bdrytau )^2 \right] \\
    & \ \ \ \ \ \qquad \qquad \qquad \quad + \frac{\sin^2\Xi}{4 {Z_2}} \left[ \rd {\Theta_2}^2 + \sin^2 {\Theta_2} \, \rd {\Phi_2} + (\rd\hat{\Psi}_2 + \cos{\Theta_2} \, \rd {\Phi_2} + \ii {\Phi}_{e,2} \, \rd\bdrytau )^2 \right] \bigg\} \, .
\end{split}
\end{equation}
Regularity now requires
\begin{equation}
\begin{split}
    (\bdrytau, \bdryphi, \hat{\Psi}_1, \hat{\Psi}_2) \; &\sim \; (\bdrytau+\beta, \bdryphi, \hat{\Psi}_1, \hat{\Psi}_2) \\
    &\sim \; (\bdrytau, \bdryphi+2\pi, \hat{\Psi}_1, \hat{\Psi}_2) \\
    &\sim \; (\bdrytau, \bdryphi, \hat{\Psi}_1 + 4\pi, \hat{\Psi}_2) \; \sim \; (\bdrytau, \bdryphi, \hat{\Psi}_1, \hat{\Psi}_2 + 4\pi )
\end{split}
\end{equation}
while having explicit fibration terms in the metric, both in the $S^1_\beta \times S^2$, and also in the internal space, due to the non-zero holonomies of the Abelian gauge fields. On the other hand, we can twist the coordinates defining ${\Psi}_{1} = \hat{\Psi}_1 + \ii \Phi_{e,1} \bdrytau$ (and analogously for $\Psi_2$), thus absorbing the chemical potentials and finding the regularity conditions
\begin{equation}
\begin{split}
    (\nonrotatingtau, \nonrotatingphi, \Psi_1, \Psi_2) \; &\sim \; (\nonrotatingtau + \beta, \nonrotatingphi - \ii \Omega \beta, {\Psi}_1 + \ii \Phi_{e,1} \beta, {\Psi}_2 + \ii \Phi_{e,2} \beta ) \\
    &\sim \; (\nonrotatingtau, \nonrotatingphi + 2\pi, {\Psi}_1, {\Psi}_2) \\
    &\sim \; (\nonrotatingtau, \nonrotatingphi, {\Psi}_1 + 4\pi, {\Psi}_2) \; \sim \; (\nonrotatingtau, \nonrotatingphi, {\Psi}_1, {\Psi}_2 + 4\pi) \, .
\end{split}
\end{equation}
Notice that we can combine the identifications above, showing that the following choice of chemical potentials satisfy the same conditions
\begin{equation}
\label{eq:IdentificationsU12}
    \beta' \= \beta \, , \qquad \Omega' \= \Omega + \frac{2\pi\ii}{\beta} n'_\Omega \, , \qquad \Phi'_{e,1/2} \= \Phi_{e,1/2} + \frac{2\pi\ii}{\beta} 2n_{e,1,2} \, .
\end{equation}
As in the minimal gauged case, we should take $n'_\Omega = 2n_\Omega$ in order to not change the periodicity of the spinors.
The identification for the electric potential follows from the boundary condition for the Abelian gauge fields, which is imposed by fixing the holonomy
\begin{equation}
    \exp \left( \frac{\ii}{2} \int_{S^1_\beta} A_{1/2} \right) \= \exp \left( - \frac{\beta}{2}\Phi_{e,1/2} \right) \, .
\end{equation}
Again, the factor of $\frac{1}{2}$ is due to the fact that the operators have half-integer charges under the relevant symmetry: compare with \eqref{eq:PeriodicitiesABJM_PairwiseEqual}. Moreover, the constraint \eqref{eq:ConstraintGravitationalPotentialsU12} corresponds to \eqref{eq:Constraint_PairwiseEqual}.

\medskip

To compare with field theory, we should first match \eqref{eq:FibrationParameters_PairwiseEqual} with the definitions \eqref{eq:ReducedGravitationalPotentialsU12}
\begin{equation}
\label{eq:IdentificationsX0X1}
\tau_g \; \leftrightarrow \; \tau + n_1 \mp 1 \, , \qquad \j_{gA} \; \leftrightarrow \;  \frac{\tau}{2} + 2\sigma_A  \, , 
\end{equation}
which of course satisfy \eqref{eq:ReducedChemicalPotentialsConstraint_U12}. Consistently with the truncation to minimal gauged supergravity, we should set $n_1 = \pm 1$ to discuss the on-shell action of the positive/negative branch of solutions, in which case the gravity action \eqref{eq:ISUSYU12} in field theory variables reads
\begin{equation}
I|_{\rm SUSY}(\tau_g = \tau) \= \pm \frac{\pi}{3\sqrt{2}} N^{\frac{3}{2}} \frac{(\tau + 4\sigma_1)(\tau + 4\sigma_2)}{\tau} \, .
\end{equation}
The singular behavior as $\tau\to 0$ matches (the negative of) the index with pairwise equal chemical potentials $\log \CI(\tau;\sigma)$ in \eqref{eq:LargeNPairwiseIndex} on the saddle $(c,d)=(1,0)$ (since the constraint \eqref{eq:LargeNConstraint_Pairwise} becomes $n_1 = \pm 1$ mod 4 if $c=1$).

\medskip

As in the case of the uplifted black holes with one electric charge, we can define a $\Z_{c_g}$ quotient of the eleven-dimensional supergravity solution analogous to \eqref{eq:OrbifoldIdentifications} and \eqref{eq:OrbifoldIdentifications_v2}. We define it by starting with an eleven-dimensional solution $(\tilde{\beta}, \tilde{\Omega}, \tilde{\Phi}_{e,1}, \tilde{\Phi}_{e,2})$, and imposing the identification
\begin{equation}
\begin{split}
    (\nonrotatingtau, \nonrotatingphi, {\Psi}_1, {\Psi}_2) \; &\sim \; \left( \nonrotatingtau + \frac{\tilde{\beta}}{c_g} , \nonrotatingphi - \ii \tilde{\Omega} \frac{\tilde{\beta}}{c_g} - \frac{2\pi r}{c_g} , {\Psi}_1 + \ii \tilde{\Phi}_{e,1} \frac{\tilde{\beta}}{c_g} - \frac{4\pi s}{c_g} , {\Psi}_2 + \ii \tilde{\Phi}_{e,2} \frac{\tilde{\beta}}{c_g} - \frac{4\pi t}{c_g} \right) \\
    &\sim \; (\nonrotatingtau, \nonrotatingphi + 2\pi, {\Psi}_1, {\Psi}_2) \\
    &\sim \; (\nonrotatingtau, \nonrotatingphi, {\Psi}_1 + 4\pi, {\Psi}_2) \; \sim \; (\nonrotatingtau, \nonrotatingphi, {\Psi}_1, {\Psi}_2 + 4\pi) \, ,
\end{split}
\end{equation}
where $r,s,t$ are integers defined modulo $c_g$.\footnote{As mentioned in footnote \ref{footnote:Ranges}, $r=0, \dots, 2c_g-1$, whereas $s,t=0,\dots, c_g-1$.} Starting from a solution with chemical potentials (i.e. boundary conditions) $(\tilde{\beta}, \tilde{\Omega}, \tilde{\Phi}_{e,1}, \tilde{\Phi}_{e,2})$, we can rewrite its orbifold solution above in terms of the primed potentials \eqref{eq:IdentificationsU12} using 
\begin{equation}
\begin{split}
    (\nonrotatingtau, \nonrotatingphi, {\Psi}_1, {\Psi}_2) \; &\sim \; \left( \nonrotatingtau + \frac{\tilde{\beta}'}{c_g} , \nonrotatingphi - \ii \tilde{\Omega}' \frac{\tilde{\beta}'}{c_g} , {\Psi}_1 + \ii \tilde{\Phi}_{e,1}' \frac{\tilde{\beta}'}{c_g}, {\Psi}_2 + \ii \tilde{\Phi}_{e,2}' \frac{\tilde{\beta}'}{c_g} \right) \\
    &\sim \; (\nonrotatingtau, \nonrotatingphi + 2\pi, {\Psi}_1, {\Psi}_2) \\
    &\sim \; (\nonrotatingtau, \nonrotatingphi, {\Psi}_1 + 4\pi, {\Psi}_2) \; \sim \; (\nonrotatingtau, \nonrotatingphi, {\Psi}_1, {\Psi}_2 + 4\pi) \, .
\end{split}
\end{equation}
Therefore, we conclude that the $\Z_{c_g}$ quotient of the solution $(\tilde{\beta}, \tilde{\tau}, \tilde{\j}_1, \tilde{\j}_2)$ contributes with
\begin{equation}
    \beta \= \frac{\tilde{\beta}}{c_g} \, , \qquad \tau_g \= \frac{\tilde{\tau}_g}{c_g} - \frac{r}{c_g} \, \qquad \j_{g1} \= \frac{\tilde{\j}_{g1}}{c_g} + \frac{2s}{c_g} \, , \qquad \j_{g2} \= \frac{\tilde{\j}_{g2}}{c_g} + \frac{2t}{c_g} \, .
\end{equation}
Thus
\begin{equation}
\label{eq:ISUSYGravOrbifoldX0X1}
    I|_{\rm SUSY}(\tau_g, \j_{g1}, \j_{g2}) \= \frac{1}{c_g} I|_{\rm SUSY}(\tilde{\tau}_g, \tilde{\j}_{g1}, \tilde{\j}_{g2}) \= \pm \frac{\pi}{G_4} \frac{(c_g \j_{g1} - 2s)( c_g \j_{g2} - 2t)}{c_g(c_g \tau_g + r)} \, .
\end{equation}

Requiring that the supersymmetry of the original solution is preserved by the $\Z_{c_g}$ orbifold imposes that
\begin{equation}
\label{eq:SUSYConditionOrbifold_Pairwise}
    \tau_g - \j_{g1} - \j_{g2} \= n_0 \qquad \Leftrightarrow \qquad 2s + 2t \= -r \mp 1 - c_g n_0 \, .
\end{equation}
As in the case of minimal supergravity \eqref{eq:SUSYConditionOrbifold_Minimal}, this constraint can only be solved provided $r$ and $c_g$ have opposite parity.
It is easy to convince ourselves that there is a fixed subset preserving supersymmetry only if we choose $s=t=0$, in which case the quotient only acts on the black hole, and the fixed point sets develop on the horizon at the two poles of the transverse $S^2$.

We discuss in some detail the match with the field theory result \eqref{eq:LargeNPairwiseIndex} for $c$ odd, the other cases follow in a slightly more involved way. The boundary conditions identify $\j_{gA}$ again with $\QFTtau/2 + 2 \sigma_A$, as in \eqref{eq:IdentificationsX0X1}, but now $\tau_g$ is $\tau + n_0 + n_1$ where $n_0$ is the odd number such that \eqref{eq:SUSYConditionOrbifold_Pairwise} holds. The on-shell action \eqref{eq:ISUSYGravOrbifoldX0X1} is singular as $\tau_g \to - r/c_g$, or $\tau \to (2s+2t \pm 1 - c_g n_1)/c$. Consistently with the case of $c=1$ and the minimal supergravity, we relate $c \leftrightarrow c_g$, and the quotient of a solution on the positive (resp. negative) branch with the field theory condition $cn_1 = 1$ mod 4 (resp. $cn_1 = -1$ mod 4). With this choices, the singular behavior of the on-shell action \eqref{eq:ISUSYGravOrbifoldX0X1}, expressed in field theory variables, is (in the minimal case)
\begin{equation}
    I|_{\rm SUSY} \= \pm \frac{8\sqrt{2} \pi}{3} N^{\frac{3}{2}} \frac{1}{c(c \tau - 2(s + t))} \left( c \sigma_1 + \frac{2t - 2s}{4} \right) \left( c \sigma_2 + \frac{2s - 2t}{4} \right) \, .
\end{equation}
Recall that $s$ and $t$ are defined modulo $c$, so once we account for the necessity of shifting $\sigma_A\in [0,1)$ by integers, this matches \eqref{eq:ISUSYGravOrbifoldX0X1}. A more detailed match requires considerations on the shifted chemical potentials as well.

Indeed, as pointed out at the end of Section \ref{sec:Uplift}, summing over gravity solutions dual to the grand canonical field theory partition function necessarily involves summing over solutions with the gauge-invariant same boundary conditions but shifted chemical potentials. However, this leads to seemingly paradoxical results, as stressed in \cite{Aharony:2021zkr} for the five-dimensional case. For instance, consider a complex supersymmetric solution and shift the reduced chemical potentials while insisting that we remain on the same branch
\begin{equation}
\tau'_g \= \tau_g + 2n_\Omega \, , \qquad \j_{g1}' \= \j_{g1} + 2n_{e,1} \, , \qquad \j_{g2}' \= \j_{g2} + 2n_{\Omega} - 2n_{e,1}
\end{equation}
The action of the shifted solution is
\begin{equation}
I|_{\rm SUSY} \= \pm \frac{2\sqrt{2}\pi}{3}N^{\frac{3}{2}} \frac{(\j_{g1} + 2n_{e,1}) ( \j_{g2} + 2n_{\Omega} - 2n_{e,1} )}{\tau_g + 2n_\Omega} \, . 
\end{equation}
Setting $n_\Omega=0$ gives a value that diverges in the shift
\begin{equation}
\Re \left( I|_{\rm SUSY}  \right) \= \mp n_{e,1}^2 \frac{8\sqrt{2}\pi}{3}N^{\frac{3}{2}} \Re\frac{1}{\tau_g} + \mc{O}(n_{e1}) \, .
\end{equation} 
Therefore, the contribution $\rme^{-I|_{\rm SUSY}}$ of the first/second branch would diverge for $\Re(\tau_g)$ positive/negative, respectively. As suggested in \cite{Aharony:2021zkr}, though, it is possible to limit the summands on the right-hand side of the AdS/CFT equation \eqref{eq:AdSCFTMaster_2} by considering the contribution of branes wrapping cycles in the eleven-dimensional geometry. We will report on this in the future.\footnote{An analogous problem has been considered in a different setting in \cite{Iliesiu:2021are}, and solved by finding the presence of zero-modes.}

\medskip

Finally, we mention that there are known BPS rotating black hole solutions to the $U(1)^4$ \textit{STU} model with four different electric charges \cite{Hristov:2019mqp}, which are dual to the fully refined field theory index. Differently from the cases considered until now, only the extremal version of these solutions is known, since by construction it is imposed that the near horizon geometry has an infinite throat. Therefore, it is not possible to perform the previous analysis and define the family of complex supersymmetric solutions away from extremality.

\section*{Acknowledgments}

\noindent We are grateful to Arash Arabi Ardehali, Davide Cassani, Zohar Komargodski, Dario Martelli, Luigi Tizzano, Chiara Toldo, and Alberto Zaffaroni for helpful discussions. We would also like to thank the organizers and participants of the SCGP workshop ``Supersymmetric Black Holes, Holography and Microstate Counting'' for many interesting comments and discussions.
This work is supported by the ERC Consolidator Grant N.~681908, ``Quantum black holes: A macroscopic 
window into the microstructure of gravity'', and by the STFC grants ST/P000258/1 and ST/T000759/1. ACB acknowledges financial support from the INFN grant GSS (Gauge Theories, Strings and Supergravity). 
PBG gratefully acknowledges support from the Simons Center for Geometry and Physics, Stony Brook University, at which some of the research for this paper was performed.

\appendix

\section{Special functions and asymptotic limit formulas} 
\label{app:Asymp}

The Pochhammer symbol is an entire function of~$z \in \IC$, for $q \in \IC$, $|q|<1$, defined as 
\be
\label{eq:PochhammerDef}
(z;q)_\infty \defeq \prod_{n=0}^{\infty} (1\,-\, z q^n) \, .
\ee
Its asymptotic expansion for when~$q$ approaches a root of unity is given as follows~\cite{Garoufalidis:2018qds}. 
Let $w\in\C$ with $|w|<1$, $q= \xi_{{m}} \rme^{- {\varepsilon}/{m}}$ where $\xi_{{m}}$ is a primitive ${m}$-th root of unity and ${\varepsilon} >0$, and $\nu$ is a complex number such that $\nu {\varepsilon} = o(1)$ as ${\varepsilon} \to 0$. Then
\beq
\label{IdentityGZ}
\begin{split}
\log \bigl( q\, w\, \rme^{-\nu {\varepsilon}/ {m} } ; q \bigr)_\infty \, &\= \, - \frac{1}{{m}{\varepsilon}} {\rm Li}_2(w^{{m}}) - 
\left( \frac{\nu}{{m}} - \frac{1}{2} \right) \log (1 - w^{{m}}) - \frac{{\varepsilon} \nu^2}{2{m}} \frac{w^{{m}}}{1-w^{{m}}} \\
& \ \ \ - \frac{1}{{m}}\log D_{\xi_{{m}}} (w^{{m}}) - \log (1-w) + \psi_{w, \xi_{{m}}}(\nu, {\varepsilon}) \, ,
\end{split}
\eeq
where
\begin{equation}
    D_{\xi_{{m}}}(x) \equiv \prod_{t=1}^{{m}-1}\left( 1- \xi_{{m}}^t x \right)^t \, , 
\end{equation}
and $\psi_{w,\xi_{{m}}}(\nu, {\varepsilon})$ has an asymptotic expansion as ${\varepsilon}\to 0$
\begin{equation} 
\label{psiexp}
    \psi_{w, \xi_{{m}} }(\nu, {\varepsilon}) \sim - \sum_{r\geq 2}\sum_{t=1}^{{m}} \left[ B_r\left( 1- \frac{t+\nu}{{m}} \right) - 
    \delta_{r,2}\frac{\nu^2}{{m}^2} \right] {\rm Li}_{2-r}(\xi_{{m}}^t w) \, \frac{{\varepsilon}^{r-1}}{r!} \, .
\end{equation}

\medskip

The Bernoulli polynomials~$B_r$ are defined by the generating function
\begin{equation}
\label{eq:BernoulliDefn}
    \frac{t \, \rme^{tx}}{\rme^t-1} \= \sum_{r=0}^\infty B_r(x) \, \frac{t^r}{r!} \,,
\end{equation}
with the first few polynomials given by
\begin{equation}
\label{eq:BernoulliFirstFew}
\begin{aligned}
    B_0 &\= 1 \, , &\qquad B_1 &\= \frac{1}{2} - x \, , \\
    B_2 &\= \frac{1}{6} - x + x^2 \, , &\qquad B_3 &\= \frac{1}{2} x - \frac{3}{2}x^2 + x^3 \, .
\end{aligned}
\end{equation}
They satisfy the relation $B_r'(x) = r B_{r-1}(x)$. 

The periodic Bernoulli polynomials~$\overline{B}_r(z)$ are defined, for~$z \in \IC$,
through their Fourier series expansion, 
\begin{equation}
\label{eq:PeriodicBernoulliDef}
   -\frac{(2\pi \i)^j}{j!} \, \overline{B}_r(z) \=  \sum_{k\in\mathbb{Z}}{}^{'}\;\frac{\rme^{2\pi \i k z}}{k^r}
    \qquad (z\in\mathbb{C}\,,\ j\ge1) \,.
\end{equation}
The prime in the above formula means that $k=0$ has to be omitted, and that in the $j=1$ 
case---where the series is not absolutely convergent---the sum is in the sense of Cauchy principal value.
For~$x\in \IR$ we have that~$\overline{B}_r(x) = B_r\left( \{ x\} \right)$. 

\medskip

The polylogarithm functions~$\text{Li}_n$, $n \in \IZ$, 
are defined as
\be
\text{Li}_n(z) \= \sum_{k=1}^\infty \frac{z^k}{k^n} \,, \qquad |z|<1 \,,
\ee
and is extended to $\IC \, \backslash \, [1,\infty)$
by analytic continuation. 
They satisfy the relation  
\begin{equation}
\label{eq:UsefulIdentity_PolyLogs}
    \text{Li}_n(\rme^{2\pi\ii x}) + (-1)^n \, \text{Li}_n(\rme^{- 2\pi\ii x}) \= - \frac{(2\pi\ii)^n}{n!} \overline{B}_n(x) \, .
\end{equation}

\section{Factorization of the integrand in the matrix integral \label{app:factorization}}

In this appendix, we review the factorization~\eqref{eq:ABJMFactorization} of the integrand  of the ABJM index~\eqref{eq:IndexABJM}. We follow the treatment of~\cite{Choi:2019zpz}, highlighting the differences due to the limit $\tau \to 4 d/c$. Recall that we have
\begin{equation}
\label{eq:VariablesForIntegration_2}
\begin{aligned}
    s_i  &\= u_i + \ii \mf{m}_i \frac{\varepsilon}{4\pi c} \, , 
    &\qquad \overline{s}_i &\= - u_i + \ii \mf{m}_i \frac{\varepsilon}{4\pi c } \, , 
    \\ 
    \tilde{s}_i &\= \tilde{u}_i + \ii \tilde{\mf{m}}_i \frac{\varepsilon}{4\pi c } \, , 
    &\qquad \overline{\tilde{s}}_i &\= - \tilde{u}_i + \ii \tilde{\mf{m}}_i \frac{\varepsilon}{4\pi c} \, ,
\end{aligned}
\end{equation}
and the corresponding exponentiated variables $z_i \equiv \rme^{2\pi\ii s_i}$ and analogous.

The classical action~\eqref{zclass} in terms of the variables \eqref{eq:VariablesForIntegration_2} is  
\be
\label{eq:ClassicalAction_App}
\begin{split}
     Z_\text{class} 
     \= \prod_i &\exp \biggl( 2\pi \ii \frac{4\pi c}{4\ii \varepsilon} \left( s_i^2 - \tilde{s}_i^2 \right) - 2\pi \ii \frac{4\pi c}{4\ii \varepsilon} 
     \Bigl( \overline{s}_i^2 - \overline{\tilde{s}}_i^2 \Bigr) \biggr) \, .
\end{split}
\ee

\medskip

Next we consider the contribution of the vector multiplets \eqref{vec1loop}, first focusing on one of the $U(N)$ gauge groups. We shall use the following relation
\begin{equation}
\begin{split}
    \prod_{i\neq j} \frac{(x_ix_j^{-1} q^{a_{(ij)}};q^b)_\infty}{(x_i^{-1}x_j q^{b+a_{(ij)}};q^b)_\infty} 
    &\= \prod_{i\neq j} \frac{ \prod_{n=0}^\infty ( 1 - x_ix_j^{-1} q^{a_{(ij)}} q^{bn}) }{ \prod_{m=0}^\infty ( 1 - x_i^{-1}x_j q^{a_{(ij)}} q^{bm}) } ( 1 - x_i^{-1} x_j q^{a_{(ij)}} ) \\
    &\= \prod_{i\neq j} ( 1 - x_i x_j^{-1} q^{a_{(ij)}} ) \, ,
\end{split}
\end{equation}
valid for any $b$ and for any coefficients $a_{(ij)}$ symmetric in~$i,j$.
Using this, we write the first line on the right-hand side of \eqref{vec1loop} as
\begin{equation}
     \prod_{i\neq j} ( 1 - x_i x_j^{-1} q^{\frac{1}{2} |\mf{m}_{ij}|}) \= \prod_{i\neq j} \frac{(x_ix_j^{-1} q^{\frac{1}{2} |\mf{m}_{ij}|};q)_\infty}{(x_i^{-1}x_j q^{1+\frac{1}{2}|\mf{m}_{ij}|};q)_\infty} \, .
\end{equation}
It simplifies the following analysis to split the zero-point energy in \eqref{eq:IndexABJM} among the vector and chiral contributions. Therefore, we write
\begin{equation}
\label{eq:vmcontrib}
    {\rm vm} \; \equiv \; \prod_{i\neq j} q^{- \frac{1}{4} |\mf{m}_{ij}|} \frac{(x_ix_j^{-1} q^{\frac{1}{2}|\mf{m}_{ij}|};q)_\infty}{(x_i^{-1}x_j q^{1+\frac{1}{2}|\mf{m}_{ij}|};q)_\infty} \, .
\end{equation}

From the identity
\be
\frac{(X q^{\frac{\mathfrak{m}+1}{2}};q)_\infty}{(X^{-1} q^{\frac{\mathfrak{m}+1}{2}} ;q)_\infty}\,=\,(-X)^{-\mathfrak{m}} \,\frac{(X q^{\frac{-\mathfrak{m} +1}{2}}  ;q)_\infty}{(X^{-1} q^{\frac{-\mathfrak{m}+1}{2}};q)_\infty}\,\qquad \mf{m} \in \mathbb{Z}\,,
\ee
valid for any $X\in \C$ (the case $X=1$ should be treated with care), taking~$X= x^{-1} y^{-1} q^{\frac{1-R}{2}}$ it follows that \cite{Dimofte:2011py, Choi:2019zpz}
\begin{equation}
\label{eq:UsefulIdentity}
\left( x^{-1} y^{-1} q^{\frac{1-R}{2}} \right)^{\frac{1}{2}|\mf{m}|} \frac{ ( x^{-1} y^{-1} q^{\frac{2-R + |\mf{m}|}{2}} ; q)_\infty }{( x y q^{\frac{R + |\mf{m}|}{2}} ; q)_\infty} \= 
   ({\color{blue}\pm}\sgn(\mf{m}))^{\mf{m}} \,
    \left( x^{-1} y^{-1} q^{\frac{1-R}{2}} \right)^{{\color{blue}\pm} \frac{1}{2}\mf{m}} \frac{ ( x^{-1} y^{-1} q^{\frac{2-R {\color{blue}\pm} \mf{m}}{2} } ; q)_\infty }{( x y q^{\frac{R {\color{blue}\pm} \mf{m}}{2} } ; q)_\infty} \, .
\end{equation}

Using the identity~\eqref{eq:UsefulIdentity} with $y=1, R=0$, the ${\color{blue}-}$ sign for $i>j$ and ${\color{blue}+}$ sign for $i<j$, 
we can rewrite~\eqref{eq:vmcontrib} as  
\begin{equation}
\label{eq:VectorMultiplet_App}
\begin{split}
    \text{vm} &\=  \prod_{i > j} \xi_{ij}^{- \frac{1}{2}} (z_j z_i^{-1})^{-\frac{1}{2}} \frac{\left( z_j z_i^{-1} \xi_{ij} ;q \right)_\infty}{\left( z_j z_i^{-1} \xi_{ij} q ; q\right)_\infty} \ \times \ \xi_{ij}^{- \frac{1}{2}} (\overline{z}_j \overline{z}_i^{-1})^{-\frac{1}{2}} \frac{\left( \overline{z}_j \overline{z}_i^{-1} \xi_{ij} ;q \right)_\infty}{\left( \overline{z}_j \overline{z}_i^{-1} \xi_{ij} q ;q \right)_\infty} \,,
\end{split}
\end{equation}
where
\begin{equation}
    \xi_{ij} \; \equiv \; \rme^{- 2\pi\ii \frac{\mathfrak{m}_{ij}}{2} \frac{4d}{c} } \, .
\end{equation}
A similar formula applies to the contribution of the other~$U(N)$ gauge group (the second line on the right-hand side of~\eqref{vec1loop}) with~$z_i,\zbar_i$ replaced by~$\wt z_i, \wt \zbar_i$, respectively.

\medskip

Finally, we consider the contributions of the $\CN=2$ chiral multiplets, i.e.~the product of the expressions for~$a=1,\dots,4$ given in~\eqref{chi1loop}. 
We first focus on the contribution of the first line in~\eqref{chi1loop}, since the second one can be obtained simply by substituting $x_i \to x_i^{-1}$, $\tilde{x}_j \to \tilde{x}_j^{-1}$. 
As in the contribution of the vector multiplet, we also introduce part of the zero-point energy, defining
\begin{equation}
    \text{$\chi$m}_1 \= \prod_{i,j} q^{\frac{1}{8} |\mf{m}_i - \tilde{\mf{m}_j}|}  \dfrac{ \bigl( x_i^{-1}\, \tilde{x}_j \, \zeta_1^{-1} \, q^{\frac{3}{4} + \frac{1}{2} | \mf{m}_i - \tilde{\mf{m}}_j | } \, ; 
\, q \bigr)_\infty }{ \bigl( x_i \, \tilde{x}_j^{-1} \, \zeta_1 \; q^{\frac{1}{4} + {\frac{1}{2}} | \mf{m}_i - \tilde{\mf{m}}_j | } \, ; 
\, q \bigr)_\infty } \, .
\end{equation}
We split the product into $i=j$ and $i\neq j$. For the first piece with $i=j$ we use \eqref{eq:UsefulIdentity} with ${\color{blue}-}$, and the result equals
\begin{equation}
\label{eq:ChiralMultiplet1_App}
\begin{split}
    \prod_{a=1}^4 \text{$\chi$m}_a \big|_{i=j} &\= \prod_{i} \xi_{ii}'^{\frac{1}{2}} ( \tilde{z}_i z_i^{-1} )^{\frac{1}{2}} \frac{ \left( \tilde{z}_i z_i^{-1} \xi'_{ii} \zeta_1^{-1} q^{\frac{3}{4}} ; q \right)_\infty }{ \left( \tilde{z}_i z_i^{-1} \xi'_{ii} \zeta_3 q^{\frac{1}{4}} ; q \right)_\infty } \frac{ \left( \tilde{z}_i z_i^{-1} \xi'_{ii} \zeta_2^{-1} q^{\frac{3}{4}} ; q \right)_\infty }{ \left( \tilde{z}_i z_i^{-1} \xi'_{ii} \zeta_4 q^{\frac{1}{4}} ; q \right)_\infty } \\
    &\quad \ \ \times \prod_{i} \xi_{ii}'^{\frac{1}{2}} ( \overline{\tilde{z}}_i \overline{z}_i^{-1} )^{\frac{1}{2}} \frac{ \left( \overline{\tilde{z}}_i \overline{z}_i^{-1} \xi'_{ii} \zeta_3^{-1} q^{\frac{3}{4}} ; q \right)_\infty }{ \left( \overline{\tilde{z}}_i \overline{z}_i^{-1} \xi'_{ii} \zeta_1 q^{\frac{1}{4}} ; q \right)_\infty } \frac{ \left( \overline{\tilde{z}}_i \overline{z}_i^{-1} \xi'_{ii} \zeta_4^{-1} q^{\frac{3}{4}} ; q \right)_\infty }{ \left( \overline{\tilde{z}}_i \overline{z}_i^{-1} \xi'_{ii} \zeta_2 q^{\frac{1}{4}} ; q \right)_\infty } \,,
\end{split}
\end{equation}
where~$\xi'_{ij} \; \equiv \; \exp\left({- 2\pi\ii \frac{\mathfrak{m}_i - \tilde{\mathfrak{m}}_j}{2} \frac{4 d}{c} }\right)$. 
We split the product $i\neq j$ into $i>j$ and $i<j$, and again use the identity \eqref{eq:UsefulIdentity} with ${\color{blue}-}$ for $i > j$ and ${\color{blue}+}$ for $i<j$. The result is 
\begin{equation}
\label{eq:ChiralMultiplet2_App}
\begin{split}
    \prod_{a=1}^4 \text{$\chi$m}_a \big|_{i\neq j} &\= \prod_{i>j} (\xi_{ij}\tilde{\xi}_{ij})^{\frac{1}{2}} \left( \frac{z_j}{z_i} \frac{\tilde{z}_j}{\tilde{z}_i} \right)^{\frac{1}{2}} \\ 
    & \qquad \times \prod_{a=1,2} \frac{\left( \tilde{z}_jz_i^{-1} \xi'_{ij} \zeta_a^{-1} q^{\frac{3}{4}} ; q \right)_\infty}{\left( \tilde{z}_i^{-1}z_j \xi_{ji}'^{-1} \zeta_a q^{\frac{1}{4}} ; q \right)_\infty} \times \prod_{a=3,4} \frac{\left( \tilde{z}_i^{-1} z_j \xi_{ji}'^{-1} \zeta_a^{-1} q^{\frac{3}{4}} ; q \right)_\infty}{\left( \tilde{z}_jz_i^{-1} \xi'_{ij} \zeta_a q^{\frac{1}{4}} ; q \right)_\infty}  \\
    & \qquad \times (\xi_{ij}\tilde{\xi}_{ij})^{\frac{1}{2}} \left(  \frac{\overline{z}_j}{\overline{z}_i} \frac{\overline{\tilde{z}}_j}{\overline{\tilde{z}}_i} \right)^{\frac{1}{2}} \\ 
    & \qquad \times \prod_{a=1,2} \frac{\left( \overline{\tilde{z}}_i^{-1} \overline{z}_j \xi_{ji}'^{-1} \zeta_a^{-1} q^{\frac{3}{4}} ; q \right)_\infty}{\left( \overline{\tilde{z}}_j \overline{z}_i^{-1} \xi'_{ij} \zeta_a q^{\frac{1}{4}} ; q \right)_\infty}  
    \times \prod_{a=3,4} \frac{\left( \overline{\tilde{z}}_j \overline{z}_i^{-1} \xi'_{ij} \zeta_a^{-1} q^{\frac{3}{4}} ; q \right)_\infty}{\left( \overline{\tilde{z}}_i^{-1} \overline{z}_j \xi_{ji}'^{-1} \zeta_a q^{\frac{1}{4}} ; q \right)_\infty} \, .
\end{split}    
\end{equation}

Finally, upon putting together the various pieces \eqref{eq:ClassicalAction_App}, \eqref{eq:VectorMultiplet_App}, \eqref{eq:ChiralMultiplet1_App}, and \eqref{eq:ChiralMultiplet2_App}, we obtain that the integrand in \eqref{eq:IndexABJM} can be written as the product of
\begin{equation}
\begin{split}
    Z_{\rm hol}(\underline{z}, \tilde{\underline{z}}; \tau, \lambda) &= \prod_{i=1}^N \exp \biggl( 2\pi \ii \frac{4\pi c}{4\ii \varepsilon} \left( s_i^2 - \tilde{s}_i^2 \right) \biggr) ( \tilde{z}_i z_i^{-1} )^{\frac{1}{2}} \xi_{ii}'^{\frac{1}{2}} \\
    & \qquad \quad \times \frac{ \left( \tilde{z}_i z_i^{-1} \xi'_{ii} \zeta_1^{-1} q^{\frac{3}{4}} ; q \right)_\infty }{ \left( \tilde{z}_i z_i^{-1} \xi'_{ii} \zeta_3 q^{\frac{1}{4}} ; q \right)_\infty } \frac{ \left( \tilde{z}_i z_i^{-1} \xi'_{ii} \zeta_2^{-1} q^{\frac{3}{4}} ; q \right)_\infty }{ \left( \tilde{z}_i z_i^{-1} \xi'_{ii} \zeta_4 q^{\frac{1}{4}} ; q \right)_\infty } \\
    &\ \ \ \times \prod_{\substack{i,j=1 \\ i > j}}^N \frac{\left( z_j z_i^{-1} \xi_{ij} ;q \right)_\infty}{\left( z_j z_i^{-1} \xi_{ij} q ;q \right)_\infty} \frac{\left( \tilde{z}_j \tilde{z}_i^{-1} \tilde{\xi}_{ij} ;q \right)_\infty}{\left( \tilde{z}_j \tilde{z}_i^{-1} \tilde{\xi}_{ij} q ;q \right)_\infty} \\
    & \qquad \quad \times \prod_{a=1,2} \frac{\left( \tilde{z}_jz_i^{-1} \xi'_{ij} \zeta_a^{-1} q^{\frac{3}{4}} ; q \right)_\infty}{\left( \tilde{z}_i^{-1}z_j \xi_{ji}'^{-1} \zeta_a q^{\frac{1}{4}} ; q \right)_\infty} \times \prod_{a=3,4} \frac{\left( \tilde{z}_i^{-1} z_j \xi_{ji}'^{-1} \zeta_a^{-1} q^{\frac{3}{4}} ; q \right)_\infty}{\left( \tilde{z}_jz_i^{-1} \xi'_{ij} \zeta_a q^{\frac{1}{4}} ; q \right)_\infty} \, ,
\end{split}
\end{equation}
and
\begin{equation}
\begin{split}
    Z_{\rm antihol}(\overline{\underline{z}}, \overline{\tilde{\underline{z}}}; \tau, \lambda) &= \prod_{i=1}^N \exp \biggl( - 2\pi \ii \frac{ 4\pi c }{4\ii \varepsilon} 
     \Bigl( \overline{s}_i^2 - \overline{\tilde{s}}_i^2 \Bigr) \biggr) ( \overline{\tilde{z}}_i \overline{z}_i^{-1} )^{\frac{1}{2}} \xi_{ii}'^{\frac{1}{2}} \\
     & \qquad \quad \times \frac{ \left( \overline{\tilde{z}}_i \overline{z}_i^{-1} \xi'_{ii} \zeta_3^{-1} q^{\frac{3}{4}} ; q \right)_\infty }{ \left( \overline{\tilde{z}}_i \overline{z}_i^{-1} \xi'_{ii} \zeta_1 q^{\frac{1}{4}} ; q \right)_\infty } \frac{ \left( \overline{\tilde{z}}_i \overline{z}_i^{-1} \xi'_{ii} \zeta_4^{-1} q^{\frac{3}{4}} ; q \right)_\infty }{ \left( \overline{\tilde{z}}_i \overline{z}_i^{-1} \xi'_{ii} \zeta_2 q^{\frac{1}{4}} ; q \right)_\infty } \\
     & \ \ \ \times \prod_{\substack{i,j=1\\ i> j}}^N \frac{\left( \overline{z}_j \overline{z}_i^{-1} \xi_{ij} ;q \right)_\infty}{\left( \overline{z}_j \overline{z}_i^{-1} \xi_{ij} q ;q \right)_\infty} \frac{\left( \overline{\tilde{z}}_j \overline{\tilde{z}}_i^{-1} \tilde{\xi}_{ij} ;q \right)_\infty}{\left( \overline{\tilde{z}}_j \overline{\tilde{z}}_i^{-1} \tilde{\xi}_{ij} q ;q \right)_\infty} \\
    & \qquad \quad \times \prod_{a=1,2} \frac{\left( \overline{\tilde{z}}_i^{-1} \overline{z}_j \xi_{ji}'^{-1} \zeta_a^{-1} q^{\frac{3}{4}} ; q \right)_\infty}{\left( \overline{\tilde{z}}_j \overline{z}_i^{-1} \xi'_{ij} \zeta_a q^{\frac{1}{4}} ; q \right)_\infty}  
    \times \prod_{a=3,4} \frac{\left( \overline{\tilde{z}}_j \overline{z}_i^{-1} \xi'_{ij} \zeta_a^{-1} q^{\frac{3}{4}} ; q \right)_\infty}{\left( \overline{\tilde{z}}_i^{-1} \overline{z}_j \xi_{ji}'^{-1} \zeta_a q^{\frac{1}{4}} ; q \right)_\infty} \, .
\end{split}
\end{equation}
Notice that
\begin{equation}
    Z_{\rm antihol}(\overline{\underline{z}}, \overline{\tilde{\underline{z}}}; \tau, \lambda) \= Z_{\rm hol}(\overline{\underline{z}}, \overline{\tilde{\underline{z}}}; \tau, \lambda)\big|_{k\to -k, z_i \to \overline{z}_i, \tilde{z}_i \to \overline{\tilde{z}}_i, \zeta_1 \leftrightarrow \zeta_3, \zeta_2 \leftrightarrow \zeta_4} \, .
\end{equation}
From these computations follow \eqref{eq:ABJMFactorization} and \eqref{defZhol}.

\section{Saddle-point analysis of the large-\texorpdfstring{$N$}{N} index}
\label{app:SaddlePointSol}

In this appendix we review the saddle-point solution to equations~\eqref{eq:extremize}  in the general case of generic~$\lambda_a$~\cite{Benini:2015eyy}.

Due to the fact that the potential $\CW^\mu$ defined in \eqref{eq:WmuTobeExtremised}
\be
\begin{split}
\mathcal{W}^\mu\,:=\,  \mathcal{W}\Bigl(\rho(x), v(x),\widetilde{v}(x)\Bigr)\,+\,N^{\frac{3}{2}}\,\mu\,\text{i}\,\Bigl(\int_{x_1}^{x_2} \rd x \, \rho(x)-1\Bigr),
\end{split}
\ee
is piecewise polynomial, solutions the first three out of the four equations~\eqref{eq:extremize}
\be\label{ExtremizationEquations}
\delta_{\rho}\mathcal{W}^\mu\,=\,\delta_{ v} \mathcal{W}^\mu\,=\,\delta_{ \widetilde{v}} \mathcal{W}^\mu\,=\,0\,,
\ee
can be found using a rather simple linear ansatz for the density of eigenvalues
\be\label{AnsatzRho}
\rho(x)= \rho_0 +x \rho_1\,.
\ee
We will assume~$k>0$, and~\cite{Benini:2015eyy}
\be\label{eq:ConditionsBranchApp}
-\,1\,<\,{\delta v}\,-\, \lambda^{\prime}_{1,2} \,<\,0\,, \qquad  0\,<\,{\delta v}\,+\, \lambda^{\prime}_{3,4} \,<\,1\,.
\ee
where~$\lambda^{\prime}_a := \ell_c \lambda_a+\frac{\ell_d}{4}\,$. 
The conditions \eqref{eq:ConditionsBranchApp} comes from demanding $\delta v$ not to cross branch points of the polylogarithms in~\eqref{W2Li3}.

From the periodicity~$\lambda_a^{\prime}\sim\lambda_a^{\prime}+1$ of the effective potential~$\mathcal{W}$ we can assume~$0\leq\lambda^\prime_a<1$ without loss of generality. This assumption and the constraint~$\lambda_1+\lambda_2+\lambda_3+\lambda_4 \in \mathbb{Z}$
forces
\be
{\lambda^\prime_1}+{\lambda^\prime_2}+{\lambda^\prime_3}+{\lambda^\prime_4} \, \in \, \{ 0,1,2,3\} .
\ee
More general cases can be obtained using the periodicity~$\lambda_a^{\prime}\sim\lambda_a^{\prime}+1\,$.

\paragraph{A first type of solutions to~\eqref{ExtremizationEquations} }

These are solutions that do not cross branch points of polylogarithms (see~\eqref{W2Li3}). Thus, they are only valid in domains of $x$
\be
x_{\text{left}} \,<\,x\,<\,x_{\text{right}}\,.
\ee
The~$x_{\text{left}}$ and~$x_{\text{right}}$ are fixed by the inequalities obtained after plugging the explicit dependence~$\delta v=\delta v(x)$ in the conditions~\eqref{eq:ConditionsBranchApp}.
These solutions are the natural generalizations of the saddle point solution found in section~\ref{sec:largeN} for the unrefined index.

First, one solves~$\delta_{\delta v}\mathcal{W}^\mu=0$ for~$\delta v(x)$ in terms of~$\rho$ (we have reinstated $k$, which is set to $1$ to obtain the results in the main text)
\be
\delta v(x)\,=\,-\frac{\left(-\lambda^{\prime 2}_1+  \lambda^\prime_1-\lambda^{\prime 2}_2+\lambda^{\prime 2}_3+\lambda^{\prime 2}_4+  \lambda^\prime_2-  \lambda^\prime_3-\lambda^\prime_4\right) \rho + k x}{ 2\left(\lambda^\prime_1+\lambda^\prime_2+\lambda^\prime_3+\lambda^\prime_4-2 \right) \rho }
\ee
Plugging this solution and the ansatz $\rho=\rho_0+x \rho_1$ in the equation~$\delta_{\rho} \mathcal{W}^\mu=0$, and solving for~$\rho_0$ and~$\rho_1\,$, one obtains, under the assumption
\be\label{AssumptionConstraint}
{\lambda^\prime_1}+{\lambda^\prime_2}+{\lambda^\prime_3}+{\lambda^\prime_4}  \,=\, 1\,,
\ee
that
\be
\begin{split}
\rho_0& \,=\,\frac{  \mu/(4\pi^2) }{\left(\lambda^\prime _1+\lambda^{\prime} _3\right) \left(\lambda^{\prime} _1+\lambda^{\prime} _3- 1 \right) \left(\lambda^{\prime} _2+\lambda^{\prime} _3\right)
   \left(\lambda^{\prime} _2+\lambda^{\prime} _3-1\right)}\,, \\ 
\rho_1& \,=\, \frac{ k \lambda^\prime _3-k \left(\lambda^\prime _1+\lambda^{\prime} _3\right) \left(\lambda^{\prime} _2+\lambda^{\prime} _3\right)}{\left(\lambda^{\prime}_1+\lambda^{\prime}_3\right)
   \left(\lambda^{\prime}_1+\lambda^{\prime}_3-1 \right) \left(\lambda^{\prime}_2+\lambda^{\prime}_3\right) \left(\lambda^{\prime}_2+\lambda^{\prime}_3-1 \right)}
\end{split}
\ee
which matches the constant unrefined solution in the family~\eqref{eq:SaddleEqual} when $\lambda'_a$ are set equal. 

\paragraph{A second type of solutions to~\eqref{ExtremizationEquations}}
There is a second type of solutions to~\eqref{ExtremizationEquations}. These are solutions that localize, at large~$N$, around the non-analyticities of the polylogarithms in~\eqref{W2Li3}. For these type of solutions the term of order~$N^{1/2}$ in the effective potential~$\mathcal{W}$ turns out to contribute non-trivially at order~$N^{3/2}$ to the saddle point equation~$\delta_{\delta v}\mathcal{W}=0$.~\footnote{This follows from the fact that~$\partial_{\delta v(x)}\text{Li}_2(\rme^{\ii(\delta v(x)\mp \lambda')}) = \ii \, \rme^{\ii (\delta v(x)\mp \lambda')}\text{Li}_1(\rme^{\ii(\delta v(x)\mp \lambda')})$ grows as~$O(N^{\frac{1}{2}})$ if~$\delta v (x)=\pm (\lambda'- \rme^{-N^{\frac{1}{2}} Y(x)})$ (in a large-$N$ limit for which the function~$Y(x)\,>\,0\,$ remains finite).}
Without loss of generality, these solutions take the form
\be\label{FourSols}
\delta v (x)\,=\,
\epsilon_a \, \bigl(\lambda^\prime_{a}\,- \frac{1}{2\pi} \rme^{-N^{\frac{1}{2}} Y_{a}(x)} \bigr)
\ee
for~$a=1$ or $2$, or $3$, or $4$ with~$\epsilon_a\,=\,(1,1,-1,-1)\,$, and the real unknown function of~$x$,~
\be\label{PositivityYa}
Y_a(x)\,>\,0.
\ee
These solutions exist in certain connected domains of~$x$,
\be
x_{\text{left}}\,<\,x\,<\,x_{\text{right}}\,.
\ee
where~$x_{\text{left}}$ and~$x_{\text{right}}$ are fixed by the positivity conditions~$\rho(x)>0$ and~\eqref{PositivityYa}.

Plugging the ansatz~\eqref{PositivityYa} in
~$\delta_{\delta v(x)}\mathcal{W}^\mu$ one can then solve for~$Y_a(x)$ as a function of~$\rho(x)$. Then, plugging the answer in~$\delta_{\rho}\mathcal{W}^\mu=0$ one can proceed to solve for~$\rho_0$ and~$\rho_1$. Under the assumption~\eqref{AssumptionConstraint}, this previous procedure leads to the solutions~\eqref{eq:LeftEq} and~\eqref{eq:RightEq} below.

\paragraph{Constructing the dominant solution}

{Patching} solutions of~\eqref{ExtremizationEquations} of the first and second type, one can construct the dominant solution to
\be\label{ExtremizationEquations2}
\delta_{\rho}\mathcal{W}^\mu\,=\,\delta_{ v} \mathcal{W}^\mu\,=\,\delta_{ \widetilde{v}} \mathcal{W}^\mu\,=\,\delta_{\mu} \mathcal{W}^\mu\,=\,0\,.
\ee

Let us assume the following ordering
\be\label{ordering}
0\,<\,\lambda^\prime_1 \,<\,\lambda^\prime_2\,<\,\lambda^\prime_3\,<\,\lambda^\prime_4\,< 1 \, .
\ee

As it will be shown below, the last condition in~\eqref{ExtremizationEquations2} together with the choice~\eqref{ordering}, implies~$\mu \in \mathbb{R}\,$. For the moment let us further assume
\be\label{ConditionMU}
\mu\,>\,0\,.
\ee
The case~$\mu\,<\,0$ can be worked out analogously.

Given the constraint~\eqref{AssumptionConstraint} and the ordering~\eqref{ordering}, out of the four possible solutions of the second type to~\eqref{ExtremizationEquations}, only two are consistent with the required positivity conditions~$\rho(x)>0\,, \,Y_a\,>\,0\,$. One of these two is
\be\label{eq:LeftEq}
\begin{split}
\delta{v}_{\text{left}}(x) &\,=\,-\lambda^\prime_3\,+\, \frac{1}{2\pi} \rme^{-N^{\frac{1}{2}}Y_{3}(x)}\,, \qquad Y_3(x)\,=\, \frac{\mu\,+\, 4\pi^2 k \lambda^\prime _4 x }{2\pi\lambda^{\prime-}_{3,4}}\\
\rho_{\text{left}}(x)& \,=\,-\frac{\mu \,+\, 4\pi^2 k \lambda^\prime_3\, x }{(2\pi)^3\lambda^{+}_{1,3} \lambda^{+}_{2,3} \lambda^{\prime-}_{3,4}}\,, \qquad \lambda^{\prime\pm}_{i,j}:= \lambda^\prime_{i} \,\pm\, \lambda^\prime_j\,.
\end{split}
\ee
 This patch is consistent with the positivity constraints~$\rho(x)\,,\,Y_3(x)\,>\,0$ in the domain
\be\label{DomainLeft}
-\frac{\mu}{4\pi^2 k \lambda^\prime_3}\,<\,x\,<\,-\frac{\mu}{4\pi^2 k \lambda^\prime_4}\,.
\ee
The other consistent solution of the second type is
\be\label{eq:RightEq}
\begin{split}
\delta{v}_{\text{right}}(x) &\,=\,\lambda^\prime_1\,-\, \frac{1}{2\pi}\rme^{-N^{\frac{1}{2}} Y_{1}(x)}\,, \qquad Y_1(x)\,=\, \frac{\mu\,-\, 4\pi^2 k \lambda^\prime _2 x }{2\pi\lambda ^{\prime-}_{1,2}}\\
\rho_{\text{right}}(x)& \,=\,\frac{-\mu\,+\, 4\pi^2 k \lambda^\prime _1 x }{(2\pi)^3\lambda^{\prime+} _{1,3} \lambda^{\prime+}_{1,4} \lambda^{\prime-}_{1,2}}\,, 
\end{split}
\ee
which is consistent with the positivity constraints~$\rho(x)\,,\,Y_1(x)\, >\, 0$ in the domain
\be\label{DomainRight}
\frac{\mu}{4\pi^2 k \lambda^\prime_2}\,<\,x\,<\,\frac{\mu}{4\pi^2 k \lambda^\prime_1}\,.
\ee
The left and right boundary of the domains~\eqref{DomainLeft} and \eqref{DomainRight}, respectively, correspond to the points~$x$ at which~$\rho(x)=0$: at every point in~\eqref{DomainLeft} and~\eqref{DomainRight},~$\rho(x)>0$. The right and left boundaries of the domains~\eqref{DomainLeft} and \eqref{DomainRight}, respectively, correspond to the points~$x$ at which~$Y_{3}(x)$ and~$Y_1(x)$ are equal to zero: at every point in~\eqref{DomainLeft} and~\eqref{DomainRight},~$Y_{3}(x)$ and $Y_1(x)$ are larger than zero.

At last, the solution of first type is such that~$\delta v$ matches the values~\eqref{FourSols} at the left and right extrema of the interval
\be\label{domainCenter}
-\frac{\mu}{4\pi^2 k\lambda^\prime_4}<\,x\,<\,\frac{\mu}{4\pi^2 k\lambda^\prime_2}.
\ee
Such a solution is
\be\label{centerSol}
\begin{split}
\delta{v}_{\text{center}}(x) &\,=\,\frac{-\left(\lambda^\prime _3 \lambda^\prime _4-\lambda^\prime _1 \lambda^\prime _2\right) \mu/4\pi^2 \,+\,k  \left(\lambda^\prime _3 \lambda^\prime _4 \lambda^\prime _{1,2}+\lambda^\prime _1 \lambda^\prime _2 \lambda^\prime
   _{3,4}\right) x}{k
   \left(\lambda^\prime _3 \lambda^\prime _4-\lambda^\prime _1 \lambda^\prime _2\right) x\,+\,\lambda^\prime_{1,2,3,4} \, \mu/4\pi^2 }\,,\\
\rho_{\text{center}}(x)& \,=\,\frac{\lambda^\prime_{1,2,3,4}  \,\mu/4\pi^2 \,+\,k \left(\lambda^\prime _3 \lambda^\prime _4-\lambda^\prime _1 \lambda^\prime _2\right) x}{2\pi\, \lambda^{\prime+}
   _{1,3} \lambda^{\prime+} _{1,4} \lambda^{\prime+} _{2,3} \lambda^{\prime+} _{2,4}}\,, 
\end{split}
\ee
again, with
\be\label{eq:ConstraintSimpl}
\lambda^\prime_{1,2,3,4}:=\lambda^\prime_1+\lambda^\prime_2+\lambda^\prime_3+\lambda^\prime_4 \,=\, 1\,.
\ee
 Patching these three solutions to~\eqref{ExtremizationEquations} one obtains a continuous solution, in the lateral sense, which is defined in
\be\label{SupportEigenvalues}
\Bigl(-\frac{\mu}{4\pi^2 k \lambda^\prime_3}\,,\,-\frac{\mu}{4\pi^2 k \lambda^\prime_4}\Bigr)\, \cup\, \Bigl(-\frac{\mu}{4\pi^2 k\lambda^\prime_4}\,,\,\frac{\mu}{4\pi^2 k\lambda^\prime_2}\Bigr)\, \cup\, \Bigl(\frac{\mu}{4\pi^2 k \lambda^\prime_2}\,,\,\frac{\mu}{4\pi^2 k \lambda^\prime_1}\Bigr)\,,
\ee
this is, in the strict large-$N$ approximation, the left and right limits of the solution at the two interior boundary points match each other. Outside~\eqref{SupportEigenvalues}, the solution vanishes which means that the support of the eigenvalue distribution is given by
\be\label{OneCutExtrema}
x_1\,=\,-\frac{\mu}{4\pi^2 k\lambda^\prime_3}\,,\, \qquad\, x_2\,=\,\frac{\mu}{4\pi^2 k \lambda^\prime_1}\,.
\ee
The normalization condition for~$\rho$ -- which follows from the fourth equation in~\eqref{eq:extremize} --, and the assumption~\eqref{ConditionMU}, fix the value of the Lagrange multiplier~$\mu$ as follows
\be
\int_{x_1}^{x_2} \rd x \, \rho(x)\,=\,\frac{\mu ^2}{2 \,k\, (2\pi)^4\lambda^\prime _{1} \lambda^\prime _2 \lambda^\prime _3 \lambda^\prime _4}=1 \implies \mu \,=\,+4\pi^2 \sqrt{2\,k\,\lambda^\prime_1\lambda^\prime_2\lambda^\prime_3\lambda^\prime_4}
\ee
A computation shows that the value of~$\mathcal{W}(\rho(x),v(x),\widetilde{v})$ at the above-found solution is
\be
\mathcal{W}\, \longrightarrow \,- \,  \text{i} \frac{2 \sqrt{2} k^{\frac{1}{2}}\,N^{\frac{3}{2}}}{3}\, \sqrt{\lambda^\prime_1\lambda^\prime_2\lambda^\prime_3\lambda^\prime_4}
\ee
This result can be extended to~$\lambda^\prime_a\,\in\,\mathbb{R}$ out of the previously-assumed domain~\eqref{ordering}. In terms of the original variables~$\lambda_a =\frac{\lambda^\prime_a-\ell_d/4}{ \ell_c}$ the answer can be written as
\begin{equation}\label{WPlus}
\mathcal{W}\, \longrightarrow \,-   \text{i} \frac{2 \sqrt{2} k^{\frac{1}{2}}\,N^{\frac{3}{2}}}{3}\, (2\pi)^2\sqrt{\{\ell_c\lambda_1+{\ell_d}/{4}\}\{\ell_c\lambda_2+{\ell_d}/{4}\}\{\ell_c\lambda_3+{\ell_d}/{4}\} \{\ell_c\lambda_4+{\ell_d}/{4}\}}
\end{equation}
again, with
\be
\{\ell_c\lambda_1+{\ell_d}/{4}\}+\{\ell_c\lambda_2+{\ell_d}/{4}\}+\{\ell_c\lambda_3+{\ell_d}/{4}\}+\{\ell_c\lambda_4+{\ell_d}/{4}\} \,=\, +\,1\,.
\ee

\paragraph{The saddle-point solution in a different domain of~$\lambda_a$'s}

The saddle-point solution takes a different form if one assumes the $\lambda_a$'s to belong to the following domain
\be
-1\,<\,\lambda^\prime_4\,<\,\lambda^\prime_3\,<\,\lambda^\prime_2\,<\,\lambda^\prime_1\,<\,0\,.
\ee
Then, assuming
\be
\mu\,<\,0
\ee
together with~\eqref{eq:ConditionsBranchApp}, the expressions of~$\delta v$ and~$\rho$ in terms of~$\mu$ and~$\lambda^\prime_i$'s are the same as in the previous case. Again, in these expressions there is one algebraic condition analogous to~\eqref{eq:ConstraintSimpl} that in this case takes the form:
\be
\lambda^\prime_{1,2,3,4}\,=\, -\, 1.
\ee
The domains of the three different linear pieces remain as in~\eqref{SupportEigenvalues}, but this time with
\be
\mu\,=\,- 4\pi^2\,\sqrt{2\,k\,\lambda^\prime_1\lambda^\prime_2\lambda^\prime_3\lambda^\prime_4}\,,
\ee
and
\be
\mathcal{W}\, \longrightarrow \,+\,   \text{i} \frac{2 \sqrt{2} k^{\frac{1}{2}}\,N^{\frac{3}{2}}}{3}\, \sqrt{\lambda^\prime_1\lambda^\prime_2\lambda^\prime_3\lambda^\prime_4}\,.
\ee
This result can be extended to~$\lambda^\prime_a\,\in\,\mathbb{R}$ out of the previously-assumed domain~\eqref{ordering}. In terms of the original variables~$\lambda_a =\frac{\lambda^\prime_a- \ell_d/4}{\ell_c}$ the answer can be written as
\begin{equation}\label{WPlus_2}
\mathcal{W}\, \longrightarrow \,+\,   \text{i} \frac{2 \sqrt{2} k^{\frac{1}{2}}\,N^{\frac{3}{2}}}{3}\, (2\pi)^2\sqrt{\{\ell_c\lambda_1+{\ell_d}/{4}\}_-\{\ell_c\lambda_2+{\ell_d}/{4}\}_{-}\{\ell_c\lambda_3+{\ell_d}/{4}\}_{-}\{\ell_c\lambda_4+{\ell_d}/{4}\}_-}
\end{equation}
where $\{x\}_{-}:=\{x\}\,-\,1$ again, and
\be
\{\ell_c\lambda_1+\ell_d/4\}+\{\ell_c\lambda_2+\ell_d/4\}+\{\ell_c\lambda_3+\ell_d/4\}+\{\ell_c\lambda_4+\ell_d/4\} \,=\, +\,3\,.
\ee

\paragraph{Particular limit cases}\label{par:LimitingCases}
Let us focus on branch one above and study the limits for which some of the~$\lambda_a$'s coincide. The simplest case is when only a single couple of~$\lambda_a$'s coincide
\be
\lambda_{1,2}^{\prime-}\,,\,\lambda_{3,4}^{\prime-}\,\neq\, 0 \,,\qquad \lambda^{\prime-}_{2,3}\,\to\,0\,.
\ee
For this case the previous discussion applies trivially. The next case is when only two couples of~$\lambda_a$'s collide
\be
\lambda_{1,2}^{\prime-}\,\neq\, 0 \,,\qquad \lambda^{\prime-}_{2,3}\,=\,\lambda^{\prime-}_{3,4}\,\to\,0\,.
\ee
In this case the domain of the left patch, the first segment in~\eqref{eq:ConstraintSimpl}, shrinks to zero. Naively, one would say then that the saddle solution in this limit can be obtained by dropping the left patch of the generic solution. Indeed, such an expectation is reinforced by the observation that in the expansion~$\lambda_3\to \lambda_4$
\be
\int_{x_1}^{-\frac{\mu}{4\pi^2 k \lambda^{\prime}_4}} \rd x \, \rho_{\text{left}}(x) \= O((\lambda_3-\lambda_4)^1)
\ee
even though~$\rho_{\text{left}}(x)$ blows up as~$\lambda_3\to \lambda_4$~(see~\eqref{eq:LeftEq}).
This means that the contribution coming from the left patch solution to the normalization condition~$\int_{x_1}^{x_2}\rd x \, \rho(x)=1$ vanishes in the limit in which the left patch shrinks to zero. Consequently, in the limit $\lambda_3 \to \lambda_4$ one can drop the left patch of the generic solution and construct a solution by gluing the center and right patches. This new solution is properly normalized as~$\int_{x_1}^{x_2} \rd x \, \rho(x) = 1\,$ as it is required. 

Last, as in the previous case, in the limit for which all~$\lambda_a$'s collide
\be
\lambda^{\prime-}_{1,2}\to\lambda^{\prime-}_{2,3}\,\to\,\lambda^{\prime-}_{3,4}\,\to\,0\,.
\ee
one has to drop not just the left but also the right patch, and as before, the new solution is given by the center patch with
\be
\lambda_1=\lambda_2=\lambda_3=\lambda_4=\frac{n_1}{4}\,,
\ee
will be properly normalized.

\section{Killing spinor equations}
\label{app:KSE}

In Section \ref{sec:BlackHole}, we remarked that the constraint \eqref{eq:Constraint_AllEqual_v2} could be derived by studying the bulk Killing spinor equation near the boundary and imposing the existence of a contractible circle. Here we expand on those comments.

We shall first make some general considerations in Lorentzian signature. 
The bulk supersymmetry equation \eqref{eq:SUGRAKillingSpinorEquation} induces a three-dimensional charged conformal Killing spinor $\chi$ at the boundary. This spinor satisfies the equation of three-dimensional off-shell conformal supergravity \cite{Klare:2012gn, Cassani:2012ri}
\begin{equation}
\label{eq:BdryConformalKillingSpinorEquation}
    \left( \nabla_i - \ii \bdryA_i \right) \chi - \frac{1}{3}\gamma_i \gamma^j \left( \nabla_j - \ii \bdryA_j \right) \chi \= 0 \, , 
\end{equation}
where we are using the connection of the boundary metric $\bdryg$, $\gamma_i$ generate the Clifford algebra Cliff$(1,2)$, and $\bdryA$ is interpreted as a background Abelian gauge field coupling to a $\mf{u}(1)_R$ $R$-symmetry. Here we have $A=-\Phi_e\, \rd\bdryt$, for the Lorentzian boundary line element obtained from \eqref{eq:BdryLineElement} we choose the frame
\begin{equation}
    \rme^0 \= \rd \bdryt \, , \qquad \rme^1 \= \rd\bdrytheta \, , \qquad \rme^2 \= \sin\bdrytheta \, (\rd\bdryphi + \Omega \, \rd\bdryt) \, ,
\end{equation}
and $\gamma_0 = \ii \sigma^1$, $\gamma_1 = \sigma^2$, $\gamma_2 = \sigma^3$ ($\sigma^i$ being the Pauli matrices). We find that the most general solution is $\chi = \chi_- + \chi_+$ where
\begin{equation}
\begin{split}
    \chi_- &\= u_1 \exp\left[ - \frac{\ii}{2} \left( \bdryphi + \bdryt \left( 1 + 2 \Phi_e + \Omega \right) \right) \right] \rme^{ \frac{\ii}{2}\bdrytheta \sigma^3 } \begin{pmatrix} 1 \\ -1 \end{pmatrix} \\
    & \qquad + v_1 \exp\left[ \frac{\ii}{2} \left( \bdryphi - \bdryt \left( 1 + 2 \Phi_e - \Omega \right) \right) \right] \rme^{ \frac{\ii}{2}\bdrytheta \sigma^3 } \begin{pmatrix} 1 \\ 1 \end{pmatrix} \, , \\[10pt]
    \chi_+ &\= u_2 \exp\left[ - \frac{\ii}{2} \left( \bdryphi - \bdryt \left( 1 - 2 \Phi_e - \Omega \right) \right) \right] \rme^{ - \frac{\ii}{2}\bdrytheta \sigma^3 } \begin{pmatrix} 1 \\ -1 \end{pmatrix} \\
    & \qquad + v_2 \exp\left[ \frac{\ii}{2} \left( \bdryphi + \bdryt \left( 1 - 2 \Phi_e + \Omega \right) \right) \right] \rme^{ - \frac{\ii}{2}\bdrytheta \sigma^3 } \begin{pmatrix} 1 \\ 1 \end{pmatrix} \, ,
\end{split}
\end{equation}
and $u_{1,2}$, $v_{1,2}$ are arbitrary complex numbers. The two spinors satisfy the stronger charged Killing spinor equation
\begin{equation}
\label{eq:BdryKillingSpinorEquationStronger}
    (\nabla_i - \ii \bdryA_i) \chi_\mp \= \mp \frac{\ii}{2} \gamma_i \gamma^0 \chi_\mp \, .
\end{equation}
The other key spinor used in the construction of the boundary rigid supersymmetric background is the spinor~$\tilde{\chi}$ that satisfies the conformal Killing spinor equation \eqref{eq:BdryConformalKillingSpinorEquation} with opposite charge. If the gauge field is real, as it is in Lorentzian signature, we can take $\tilde{\chi}=\chi^c$, where the charge conjugate spinor is $\chi^c \equiv \gamma^0 \CC^{-1} \chi^*$. From these two spinors, we can construct a geometric background preserving two supercharges of opposite $R$-charge, and in particular a bilinear vector $\xi^i = \overline{\tilde{\chi}}\gamma^i \chi$, which is generically a conformal Killing vector.\footnote{Here $\overline{\chi}\equiv \chi^T \CC$, where the charge conjugation matrix $\mc{C}$ is the intertwiner satisfying
\[
\mc{C}^T = - \mc{C} \, , \qquad \mc{C}^* = \mc{C} \, , \qquad \mc{C}^2 = - \identity \, , \qquad \gamma_i^T = - \mc{C} \gamma_i \mc{C}^{-1} \, .
\]
With our choice of basis, one can choose $\CC = \ii \sigma^2$.} 

Issues arise when we Wick rotate to Euclidean signature. As already pointed out, both the bulk Killing spinor equation \eqref{eq:SUGRAKillingSpinorEquation} and the conformal Killing spinor equation \eqref{eq:BdryConformalKillingSpinorEquation} are analytic in the supergravity fields, so the Wick-rotated spinors are still solutions. However, in Euclidean signature one is \textit{a priori} not allowed to impose reality conditions on bosonic or fermionic fields. This implies that the spinor $\tilde{\chi}$ is independent of $\chi$, and indeed it is well-known that we can have Riemannian backgrounds supporting two independent supercharges with opposite $R$-charge, such as the fibered $S^1\times S^2$ considered in Section \ref{sec:ABJMIndex} \cite{Klare:2012gn, Dumitrescu:2012ha}. To derive this condition holographically, one should then in principle consider Riemannian bulk solutions with a supersymmetry obtained by doubling the Killing spinor equations, as stressed, for instance, in \cite{Freedman:2013oja, Bobev:2020pjk}. We shall take a simpler approach, often taken in the literature, and Wick-rotate the Lorentzian spinors to $\chi \to \chi_E$ and $\chi^c \to \tilde{\chi}_E$. The resulting spinors are independent (i.e. $\tilde{\chi}_E \neq \chi_E^c$) and solve, respectively, the conformal Killing spinor equation \eqref{eq:BdryConformalKillingSpinorEquation} and that with opposite charge.

In fact, it follows from \eqref{eq:BdryKillingSpinorEquationStronger} that these spinors are also solutions to the new minimal supergravity Killing spinor equations 
\begin{equation}
\label{eq:NMKSE}
\begin{split}
0 &\= \nabla_i \chi_E - \ii A_i^{(\rm nm)} \chi_E + \ii V_i \chi_E + \frac{\ii}{2} V^j \gamma_{ji}\chi_E + \frac{1}{2} H \gamma_i \chi_E \, , \\
0 &\= \nabla_i \tilde{\chi}_E + \ii A_i^{(\rm nm)} \tilde{\chi}_E - \ii V_i \tilde{\chi}_E - \frac{\ii}{2} V^j \gamma_{ji} \tilde{\chi}_E + \frac{1}{2} H \gamma_i \tilde{\chi}_E \, , 
\end{split}
\end{equation}
with $H=0$, an appropriate choice of $V_i$ (depending on the choice $\chi_\pm$), and $A^{(\rm nm)}= \bdryA_i + \tfrac{3}{2}V_i$. Unsurprisingly, these are the Killing spinor equations of three-dimensional off-shell new minimal supergravity that would give the backgrounds considered at the end of Section \ref{sec:ABJMIndex} \cite{Klare:2012gn}. For our purposes it is consistent to choose
\begin{equation}
\label{eq:BoundaryKS}
\begin{split}
    \chi_E &\= u \exp\left[ \frac{1}{2} \left( \ii\bdryphi +  \bdrytau \left( 1 - 2 \Phi_e + \Omega\right) \right) \right] \rme^{ - \frac{\ii}{2}\bdrytheta \sigma^3 } \begin{pmatrix} 1 \\ 1 \end{pmatrix} \, , \\
    \tilde{\chi}_E &\= \tilde{u} \exp \left[ - \frac{1}{2}\left( \ii \bdryphi + \bdrytau \left( 1 - 2 \Phi_e + \Omega \right) \right) \right] \rme^{ \frac{\ii}{2}\bdrytheta \sigma_3} \begin{pmatrix} 1 \\ -1 \end{pmatrix} \, ,
\end{split}    
\end{equation}
where again $u$, $\tilde{u}$ are arbitrary complex numbers.
These spinors solve the conformal Killing spinor equation \eqref{eq:BdryConformalKillingSpinorEquation} with positive and negative $R$-charge, respectively. Moreover, they also solve \eqref{eq:NMKSE} with $V = - \ii \, \rd \bdrytau$ and $H=0$. In order to have well-defined global spinors on the fibered $S^1\times S^2$ background \eqref{eq:BdryLineElement}, we should impose a constraint on the chemical potentials. First, notice that under $\bdryphi \to \bdryphi+2\pi$, the spinors are antiperiodic as should be. However, in order for the spinors to be antiperiodic as $\bdrytau\to \bdrytau + \beta$ we should require
\begin{equation}
\label{eq:Constraint_AllEqual_v2_App}
    \beta \left( 1 - 2 \Phi_e + \Omega \right) \= 2\pi \ii n_0 \, ,
\end{equation}
with odd $n_0$. This provides us with an additional way to argue for the constraint between the chemical potentials. In Section \ref{sec:ABJMIndex}, we found that the constraint \eqref{eq:Constraint_AllEqual} followed directly from identifying the superconformal index as a thermal partition function. We also argued around \eqref{eq:TimeDependenceNonMinimalKS} that it could be derived by looking at the rigid supersymmetric background and requiring thermal boundary conditions for the Killing spinor (which is the argument just reproduced in detail to get to \eqref{eq:Constraint_AllEqual_v2}). Whilst until now the constraint has been discussed from the field theory viewpoint, we now find an argument from the regularity of the Euclidean gravity solution: $\chi_E$ is the leading order spinor in the radial expansion of the bulk spinor $\epsilon$ near the boundary, and the anti-periodicity of $\chi_E$ (and thus of $\epsilon$) around $S^1_\beta$ is needed in order to be able to have a smooth disc filling $S^1_\beta$.

\section{Holographic renormalization for the \texorpdfstring{$F=-\ii X^0X^1$}{X0X1} model}
\label{app:HoloRen_Details}

The presence of scalars in the non-minimal supergravity model considered in Section \ref{sec:U12} leads to some subtleties in the application of the standard holographic renormalization dictionary, which we glossed over in the main text. In this appendix we briefly review these details and show why they do not affect our conclusions.

\medskip

A solution of the $F=-\ii X^0X^1$ model \eqref{eq:LorentzianLagU12} is asymptotically locally AdS. Therefore, there is a conformal boundary and in a neighbourhood of the conformal boundary it is possible to introduce a Fefferman--Graham coordinate $z$ such that the conformal boundary is at $z=0$. The fields in the solution admit an analytic expansion in $z$ of the form
\begin{equation}
\begin{split}
    \mc{G} &\= \frac{\rd z^2}{z^2} + h_{ij}(x,z) \rd x^i \rd x^j \, , \\
    h_{ij} &\= \frac{1}{z^2} \left( g_{ij} + z^2 g^{(2)}_{ij} + z^3 g^{(3)}_{ij} + o(z^3) \right) \, , \\
    \mc{A}_a &\= A_a + z A^{(1)}_a + z^2 A^{(2)}_a + o(z^2) \, , \\
    X_1 &\= 1 + z X_1^{(1)} + z^2 X_1^{(2)} + o(z^2) \, , \\
    \j &= z \j^{(1)} + z^2 \j^{(2)} + o(z^2) \, .
\end{split}
\end{equation}
Here $x^i$ $i=1,2,3$ are coordinates on the boundary. As discussed around \eqref{eq:OnShellActionBH}, we introduce a cutoff $z=\delta\geq 0$ and look at the induced geometry on the hypersurface $\partial Y_\delta = \{ z=\delta\} \cap Y_4$ with metric $h$. The renormalized on-shell action is \eqref{eq:IU12Counterterms}, that is
\begin{equation}
    I \= \lim_{\delta \to 0} \left[ S + \frac{1}{8\pi G_4} \int_{\partial Y_\delta} \left( K - \frac{1}{2}R - \sqrt{ 2+ X^2_1 + {X}^2_2} \right) \vol_h \right] \, .
\end{equation}
The counterterms are chosen in order to be compatible with supersymmetry. This can be argued by realizing that this model is a truncation of the $STU$ model or $Spin(4)$ supergravity, which have been discussed, for instance, in \cite{Freedman:2013oja, Cabo-Bizet:2017xdr, Bobev:2020pjk, BenettiGenolini:2020kxj}. The resulting action, though, should not be identified with the generating functional of the SCFT. In fact, whilst $\j^{(1)}$ is the bulk source of the SCFT operator of dimension 2 $\CO_{\Delta=2}$, $X_1^{(1)}$ is not the bulk source of the SCFT operator of dimension 1 $\CO_{\Delta=1}$, but it should be identified with its VEV. The appropriate bulk source is the variable canonically conjugate to $X_1^{(1)}$
\begin{equation}
    \CX \= \frac{1}{\sqrt{-g}} \frac{\delta I}{\delta X_1^{(1)}} = - \frac{1}{8\pi G_4} \left( (X_1^{(1)})^2 - 2 X_1^{(2)} + \frac{1}{2} (\j^{(1)})^2 \right) \, ,
\end{equation}
and the SCFT generating functional is the Legendre transform of $I$ with respect to $\CX$ \cite{Klebanov:1999tb}
\begin{equation}
    \tilde{I} = I - \int_{M_3} \CX X_1^{(1)} \, \vol_g \, .
\end{equation}
The one-point functions of the dual operators
\begin{align}
\begin{split}
    \langle \tilde{T}_{ij} \rangle &\= - \frac{2}{\sqrt{-g}} \frac{\delta \tilde{I}}{\delta g^{ij}} \\
    &= - \frac{1}{8\pi G_4} \lim_{\delta\to 0} \frac{1}{\delta} \left[ K_{ij} - K h_{ij} + \sqrt{2+X_1^2+X_2^2} \, h_{ij} - \left( R_{ij}[h] - \frac{1}{2} R h_{ij} \right)  \right] \\
    & \qquad \qquad \qquad \qquad - \CX X_1^{(1)} \, g_{ij} \, ,
\end{split} \\[5pt]
    \langle \CO_{\Delta=2} \rangle &= \frac{1}{\sqrt{-g}} \frac{\delta \tilde{I}}{\delta \j^{(1)}} = - \frac{1}{8\pi G_4} \left( X_1^{(1)} \j^{(1)} + \frac{1}{2} \j^{(2)} \right) \, , \\[5pt]
    \langle \CO_{\Delta=1} \rangle &= \frac{1}{\sqrt{-g}} \frac{\delta \tilde{I}}{\delta \CX} = - X_1^{(1)} \, , \\
\begin{split}
    \langle j_1^i \rangle &= \frac{1}{\sqrt{-g}} \frac{\delta \tilde{I}}{\delta (A_1^{(1)})_i} = - \frac{1}{8\pi G_4} \lim_{\delta\to 0} \frac{1}{\delta^3} n_\mu \left( X_1^{-2} \CF_1 + \j *_{\CG} \CF_1 \right)^{\mu i} |_{z=\delta} \, , 
\end{split} \\[10pt]
\begin{split}
    \langle j_2^i \rangle &= \frac{1}{\sqrt{-g}} \frac{\delta \tilde{I}}{\delta (A_2^{(1)})_i} = - \frac{1}{8\pi G_4} \lim_{\delta\to 0} \frac{1}{\delta^3} n_\mu \left( X_1^{-2} \CF_2 - \j *_{\CG} \CF_2 \right)^{\mu i} |_{z=\delta} \, .
\end{split}
\end{align}
These quantities satisfy the holographic Ward identities \cite{BenettiGenolini:2020kxj}
\begin{equation}
\begin{split}
\label{eq:HolographicWard}
0 &= \langle \tilde{T}^i_{\ph{i}i} \rangle + 2 \langle \mc{O}_{\Delta=1} \rangle \CX + \langle \mc{O}_{\Delta=2} \rangle \j^{(1)} \, , \\
0 &= \overline{\nabla}_i \langle j_1^i \rangle = \overline{\nabla}_i \langle j_2^i \rangle \, , \\
0 &= \overline{\nabla}^j \langle \tilde{T}_{ji} \rangle + \langle \mc{O}_{\Delta=1} \rangle \overline{\nabla}_i \CX + \langle \mc{O}_{\Delta=2} \rangle \overline{\nabla}_i \j^{(1)} + (F_1)_{ij} \langle j_1^{j} \rangle + (F_2)_{ij} \langle j_2^{j} \rangle \, ,
\end{split}
\end{equation}
where $\overline{\nabla}$ is the Levi-Civita connection associated to $g$.

To complete the procedure, we are left with the task of finding the Fefferman--Graham coordinate for the solution \eqref{eq:X0X1_Metric}. This is made subtler by the presence of the rotation, as stressed in \cite{Papadimitriou:2005ii}. The change to the Fefferman--Graham coordinates in a neighbourhood of the conformal boundary is
\begin{equation}
\begin{split}
    \Theta &\= \overline{\theta} - \frac{a^2 \cos\overline{\theta} \sin\overline{\theta}}{8} \left( 1 - a^2 \cos^2\overline{\theta} \right) z^4 + o(z^5) \, , \\
    r &\= \frac{1}{z} - m(s_1^2 + s_2^2) + \frac{1}{32}\big[ 2( -2a^2 + m^2 - 4 + 2a^2 \cos 2\overline{\theta} ) \\
    & \quad \ \ + m^2 ( \cosh 4\delta_1 - 4 \cosh 2\delta_1 \cosh 2\delta_2 + \cosh 4\delta_2 ) \big] z \\
    & \quad \ \ + \frac{m( \cosh 2 \delta_1 + \cosh 2 \delta_2 )}{6} z^2 + o(z^2) \, .
\end{split}
\end{equation}
Expanding the scalar fields \eqref{eq:X0X1_Scalars} in $z$ near the boundary, we have
\begin{equation}
\begin{aligned}
    X_1^{(1)} &\= m (s_1^2 - s_2^2) \, , &\qquad X_1^{(2)} &\= \frac{m^2}{2} (s_1^2 - s_2^2)^2 \, , \\ \j^{(1)} &\= 0 \, , &\qquad \j^{(2)} &\= - a m \cos\overline{\theta} (\cosh 2 \delta_1 - \cosh 1 \delta_2) \, .
\end{aligned}
\end{equation}
Therefore, it is clear that the sources for both scalar boundary operators vanish.

\section{Four-dimensional index}
\label{app:N4Index}

In this appendix, we review the construction and limits of the four-dimensional superconformal index for $\CN=4$ SYM with $SU(N)$ gauge group, emphasising the parallels with the construction in three dimensions explained in Section \ref{sec:ABJMIndex}.

\medskip

The four-dimensional superconformal index for $\CN=4$ SYM counts the states on $S^3$ annihilated by a supercharge $\CQ$ with
\begin{equation}
\{ \CQ, \CQ^\dagger\} \= H - J_1 - J_2 - \sum_{i=1}^3R_i \, ,
\end{equation}
where $J_{1,2}$ are the generators of the Cartan of the $\mf{su}(2)\times \mf{su}(2)$ isometries on $S^3$, and $R_{1,2,3}$ are generators of the Cartan of $\mf{so}(6)_R$ in the orthogonal basis (so they have half-integer eigenvalues). The remaining bosonic subalgebra commuting with $\CQ, \CQ^\dagger$ is generated by
\begin{equation}
J_1 + \frac{1}{3}\sum_i R_i \, , \ J_2 + \frac{1}{3}\sum_i R_i \, , \ R_1 - R_3 \, , \ R_2 - R_3 \, .
\end{equation}
The $\mf{u}(1)_R$ $R$-symmetry is generated by $r \equiv \frac{2}{3}\sum_i R_i$, with eigenvalues in $\frac{1}{3}\Z$.
We define the index as
\begin{equation}
\begin{split}
\CI(\tau_1, \tau_2, \lambda_1, \lambda_2 ) &\= \Tr_{\CH_{S^3}} \bigg[ (-1)^{2J_1} \rme^{ - \beta \{ \CQ, \CQ^\dagger \} + 2\pi\ii \tau_1 \left( J_1 + \frac{1}{3}\sum_i R_i \right) + 2\pi\ii \tau_2 \left( J_2 + \frac{1}{3}\sum_i R_i \right) } \\
& \qquad \qquad \qquad \qquad \times \rme^{2\pi\ii \lambda_1 (R_1 - R_3) + 2\pi\ii \lambda_2 (R_2 - R_3) } \bigg] \, .
\end{split}
\end{equation}
States in the theory satisfy $J_{1,2}=R_{1,2,3}$ mod 1, so we find that the index is invariant under the following shifts
\begin{equation}
    \tau_{1,2} \to \tau_{1,2} + 3 \, , \qquad \lambda_{1,2} \to \lambda_{1,2} + 1 \, .
\end{equation}
Instead, observe that shifts of (say) $\tau_{1}$ by integers lead to indices graded by $r$, the $R$-charge indices
\begin{equation}
\begin{split}
    \CI(\tau_1 + 1, \tau_2, \lambda_1, \lambda_2 ) &\= \Tr_{\CH_{S^3}} \bigg[ (-1)^{r} \rme^{ - \beta \{ \CQ, \CQ^\dagger \} + 2\pi\ii \tau_1 \left( J_1 + \frac{1}{3}\sum_i R_i \right) + 2\pi\ii \tau_2 \left( J_2 + \frac{1}{3}\sum_i R_i \right) } \\
& \qquad \qquad \qquad \qquad \times \rme^{2\pi\ii \lambda_1 (R_1 - R_3) + 2\pi\ii \lambda_2 (R_2 - R_3) } \bigg] \\
& \; \equiv \; \CI_R(\tau_1, \tau_2, \lambda_1, \lambda_2 ) \, , \\[10pt]
\CI(\tau_1 + 2, \tau_2, \lambda_1, \lambda_2 ) &\= \CI_R(\tau_1, \tau_2, \lambda_1, \lambda_2 ) \, .
\end{split}
\end{equation}

It is convenient to introduce $\lambda_3$ via the constraint
\begin{equation}
\sum_{i=1}^3 \lambda_i \= n_1 \, , \qquad n_1 \in \Z \, .
\end{equation}
This leads to
\begin{equation}
\label{eq:SymmetricIndexN4SYM}
\begin{split}
\CI(\tau_1, \tau_2; \lambda) &\= \Tr_{\CH_{S^3}}  (-1)^{2J_1} \rme^{ - \beta \{ \CQ, \CQ^\dagger\} + 2\pi\ii \left[ \tau_1 \left( J_1 + \frac{1}{3}\sum_i R_i \right) + \tau_2 \left( J_2 + \frac{1}{3}\sum_i R_i \right) + \sum_i \lambda_i R_i - n_1 R_3 \right]} \\
&= \Tr_{\CH_{S^3}}  (-1)^{2J_1} \rme^{ - \beta \{ \CQ, \CQ^\dagger\} + 2\pi\ii \left[ \sum_i \left( \frac{\tau_1}{3} + \lambda_i \right) ( J_1 + R_i ) + \tau_2 \left( J_2 + \frac{1}{3}\sum_i R_i \right) \right]} \, ,
\end{split}
\end{equation}
so that the index is really a function of the four variables $\lambda_i + \tau_1/3, \tau_2$.

The functional integral formalism also works exactly in parallel to the discussion 
in Section~\ref{sec:ABJMIndex}. The index can be seen as a functional integral over a fibered $S^1\times S^3$ with background gauge fields for the Cartan of the $R$-symmetry, and thermal boundary conditions for the fields. In particular, looking at \eqref{eq:SymmetricIndexN4SYM}, we have fibration parameters
\begin{equation}
    \Omega_1 \= 1 + \frac{2\pi\ii}{\beta}(\tau_1 + n_0 + n_1) \, , \qquad \Omega_2 \= 1 + \frac{2\pi\ii}{\beta} \tau_2 \, ,
\end{equation}
and background gauge fields $A_i = \ii \Phi_i \, \rd t_E$ with
\begin{equation}
\begin{split}
    \Phi_i &\= 1 + \frac{2\pi\ii}{\beta} \left( \frac{\tau_1 + \tau_2}{3} + \lambda_i \right) \, .
\end{split}
\end{equation}
Here $n_0$ is an odd number taking care of the grading by $(-1)^{2J_1}$, and because of supersymmetry, these quantities are not all independent
\begin{equation}
    \beta \biggl( 1 - \sum_{i=1}^3 \Phi_i + \Omega_1 + \Omega_2 \biggr) \= 2\pi\ii n_0 \, .
\end{equation}
Moreover, thanks to the relation between the charges of the states in the theory, the partition function is invariant under the following shifts
\begin{equation}
    \Omega_A \to \Omega_A + \frac{2\pi\ii}{\beta}m_A \, , \quad \Phi_A \to \Phi_A + \frac{2\pi\ii}{\beta}n_A \, , \quad \Phi_3 \to \Phi_3 + \frac{2\pi\ii}{\beta} \left( 2k - \sum_{A=1}^2(m_A + n_A) \right)
\end{equation}
with $m_A, n_A, k\in\Z$.

To simplify the discussion, we reduce to the universal case by setting $\lambda_1 = \lambda_2 = \lambda_3 \equiv \lambda$, with the constraint $3\lambda = n_1$. In this case, it is 
\begin{equation}
\label{eq:I4dUnrefined}
\begin{split}
    \CI(\tau_1, \tau_2; n_1) &\= \Tr_{\CH_{S^3}} (-1)^{2J_1} \rme^{- \beta \{ \CQ, \CQ^\dagger\} + 2\pi\ii \left[ \left(\tau_1 + n_1 \right) \left( J_1 + \frac{1}{2}r \right) + \tau_2 \left( J_2 + \frac{1}{2}r \right) \right] } \, .
\end{split}
\end{equation}
Therefore, we immediately see that if $n_1 = \pm 1$ mod 3 we have the $R$-charge index, graded by the $R$-symmetry generator $r$ satisfying $2J_1 = 2J_2 = 3r$ mod 2 on the states of $\CN=4$ SYM.

We further simplify the discussion by setting $\tau_1 = \tau_2 \equiv \tau$, obtaining the simplest index receiving contributions only from states preserving two supercharges
\begin{equation}
\label{eq:SimpleUnrefinedIndex}
\begin{split}
    \tilde{\CI}(\tau;n_1) &\= \Tr_{\CH_{S^3}} (-1)^{2J_1} \rme^{ 2\pi\ii n_1 \left( J_1 + \frac{r}{2} \right) + 2\pi\ii \tau (J_1 + J_2 + r)} \\
   &\= \Tr_{\CH_{S^3}} (-1)^{2J_1} \rme^{ 2\pi\ii (\tau - n_1) (J_1 + J_2 + r)} \, ,
\end{split}
\end{equation}
where in the last equation we used the relation $2J_2=3r$ mod $2$. For the same reason, as mentioned, both $\QFTtau$ and $n_1$ are only defined modulo $3$, so the unrefined index is really a function of the $\C$-valued variable $\exp (2\pi\ii T)$, with $T\equiv (\tau-n_1)/3$, and $T\sim T+1$.

We are interested in the generalized Cardy limit in which $\Introtau \to \Introd/\Introc$ with $\gcd(\Introc, \Introd)=1$. In order to study the asymptotic behavior of the index in this limit, we further write $3\Introtau = \ell_{\Introd}/\ell_{\Introc}$ with $\gcd(\ell_{\Introc}, \ell_{\Introd})=1$. If $\Introc$ is not a multiple of $3$, then $\ell_{\Introd} = 3\Introd$ and $\ell_{\Introc} = \Introc$, whereas if $\Introc$ is a multiple of $3$, then $\ell_{\Introd} = \Introd$ and $\ell_{\Introc} = \Introc/3$. In terms of these variables, in the generalized Cardy limit, $\log\tilde{\CI}(\tau; n_1)$ has a leading O$((3\ell_{\Introc}\Introtau - \ell_{\Introd}))^{-2})$ term provided that $\Introc$ is a multiple of $3$, in which case \cite{Cabo-Bizet:2019eaf, ArabiArdehali:2021nsx} 
\begin{equation}
\label{eq:Limit_4dIndex}
    \log \tilde{\CI}(\Introtau) \; \sim \; \pm \frac{\ii\pi}{27}N^2 \frac{1}{\frac{\Introc}{3}( \Introc \Introtau - \Introd)^2} \qquad \text{if $\Introd = \pm 1$ mod 3.}
\end{equation}
Clearly, the smallest values for which the index has a leading singular behavior are $T \to \pm 1/3$, corresponding to $\exp(2\pi\ii T)$ approaching the primitive third roots of unity. 

For historical reasons, the asymptotic behavior of the index has been studied splitting $\Introtau = (\tau - n_1)/3$ in $\tau$ and $n_1$. In terms of the variable $\tau$, the index of $\CN=4$ SYM is clearly a three-sheeted function, and the leading singularities can be reached in multiple equivalent ways: e.g. $\tau \to 1,2$ and $n_1=0$, or, if we insist on the Cardy limit $\tau \to 0$, then $n_1 = 1,2$. That is, we either need to take the limits on the second/third sheet, where the behaviour is controlled by anomalies \cite{Cassani:2021fyv, ArabiArdehali:2021nsx, Ohmori:2021dzb}, or take the Cardy limit $\tau\to 0$ of the $R$-charge index.

We derived this result using generalised Cardy limits $\tau\to \Q$. However, the same behavior for the index is discovered when approaching its study using the Bethe ansatz method: in the large-$N$ limit, one finds that the leading behavior is again on the first and second sheet, whereas on the zeroth sheet the large-$N$ limit is undefined due to the competition of terms with the same absolute value \cite[(3.63)]{Aharony:2021zkr}.

\bibliographystyle{JHEP}
\bibliography{indexrefs}

\end{document}